# Effects of C and B microalloying additions on the microstructure and processability of René 41 Ni-based superalloy

## Wai Fung Wilson Tse

A thesis in fulfilment of the requirements for the degree of

Master of Science

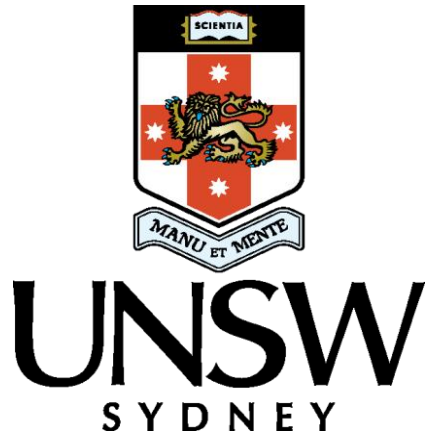

School of Material Science and Engineering

Faculty of Science

University of New South Wales

May 2022

# Thesis Title and Abstract

**Thesis submission for the degree of Master of Science (Research)**

| Thesis Title and Abstract | Declarations | Inclusion of Publications Statement | Corrected Thesis and Responses |

**Thesis Title**

Effects of C and B microalloying additions on the microstructure and processability of René 41 Ni-based superalloy

**Thesis Abstract**

René 41 is a cast and wrought Ni-based superalloy with high yield strength and stress-rupture properties contrasted with poor processability. The aim of this thesis is to systematically investigate the influence of C and B microalloying additions on processability of René 41. The first approach is an experimental one using hot compression testings and material characterisations. A second approach using machine learning methodology was also used to provide linkage for the experimental observations with industrial René 41 materials based on ultrasonic defects and chemical composition. Three René 41 variants with nominal, high C, and high B compositions were industrially fabricated and homogenized to be used in this study. The resultant flow stresses from hot compression testings were used to model hyperbolic sine constitutive equations. The activation energy for hot deformation was found to be 757, 728, and 697 kJ/mol for the nominal, high B, and high C René 41 variants respectively. Finite element method simulations based on the obtained flow curves found that effective plastic strain varied considerably through the sample geometry. Quantitative analysis via electron back-scatted diffraction found that while the three René 41 variants have nearly identical recrystallised grain size, high C contain 64 vol.% recrystallised fraction compared to that of the nominal variant with 31 vol.% at the same deformation condition. It is postulated that the presence of carbides and borides must increase the dislocation density and creates nucleation sites for particle simulated nucleation which improve grain structure during hot deformation. C-support vector machine classification model was the highest scoring supervised learning models out of the 7 chosen classification models with a F-measure of 72/100. Feature importance analysis using random forest classifier found that C, N and Ti are the highest-ranking features. Thermodynamic modelling reveals similar elements prone to macro-segregation. Thus, localised excessive precipitation of carbides may promote the formation of internal defects. Based on findings from both approaches applied in this thesis, it is proposed that the microalloying composition of René 41 should be adjusted to contain a mix population of carbides and borides to promote better hot workability without the risk of forming internal defects.

# Originality, Copyright And Authenticity Statements

# Inclusion of Publications Statements

# Table of Contents

# Acknowledgements

I would like to express my deepest gratitude to my primary supervisor Dr. Sophie Primig for offering this opportunity to me and her continuous guidance, support, and tutelage throughout my candidature. I am also equally thankful to co-supervisor Dr. Felix Theska for his mentorship, extensive knowledge, and invaluable advice he imparted to me. Their supervision has supported me through difficult times during the COVID-19 pandemic.

I would like to acknowledge industrial partner representatives Mr. Michael Lison-Pick and Dr. Steven Street for their continuous support and contributions to this project. Special thanks to everyone at Engineering Microstructure research group for all the helps and comradery for the last two years. I would also like give recognition to the administrative and technical staff at UNSW for their assistance in my candidature.

I would like to also acknowledge here the financial support from the Australian Government as a recipient of the Australian Government Research Training Program Scholarship.

Lastly, I would like to thank my friends and family for supporting me.

# List of Figures

# List of Tables

# Chapter 1 – Introduction

Ni-based superalloys originated from a binary Ni-20Cr alloy in the early 1900s as high-temperature oxidation-resistant material for electrical heating applications [1]. The term 'superalloys' was later adopted by metallurgists in the 1950s to describe a group of alloys with group VIIIB element base metals such as Ni, Co, and Fe [2]. These superalloys exhibit excellent high-temperature properties including high-temperature strength, high creep-rupture properties, and hot-corrosion resistance in aggressively oxidising environments. Therefore, they are ideal materials for elevated temperature applications such as gas turbine engines (GTEs) [3].

The energy efficiency of a GTE depends heavily on the temperature of the gas-fuel mixture entering the turbine section known as the turbine entry temperature (TET). The compression ratio defined as the volumetric changes before and after passing the gas-fuel mixture through the compressor of the GTE is an important consideration for aerospace manufacturers. This is due to the TET of gas-fuel mixture being heavily dependent on the maximum compression ratio a GTE can withstand with higher compression ratio produces higher TET. Therefore, materials with improved high-temperature mechanical properties are sought after by manufacturers to withstand higher compression ratios and TETs. In particular, the high temperature mechanical properties of precipitation-hardened Ni-based superalloys outperform other common superalloys such as solid solution hardened Ni-based superalloys or the carbide-hardened Co-based superalloys by significant margins [4]. Hence, over the last 50 years, the development of precipitation-hardened Ni-based superalloys has seen major technological advancements in terms of process design, alloy design, and advanced understanding of metallurgical phenomena [3].

Modern commercially available precipitation-hardened Ni-based superalloys have branched into many sub-categories as shown in Figure 1.1. This is due to the different manufacturing techniques required to satisfy the specifications of a modern GTE. A typical GTE is made up of 40-50 wt.% of polycrystalline Ni-based superalloy components [5]. Directionally solidified components eliminate grain boundary structures, and thereby are suitable to withstand high-temperature dynamic loading conditions. However, this requires complex manufacturing design and rare earth alloying elements, which make such components economically attractive only for turbine blade applications where creep is prevalent. On the other hand, polycrystalline-equiaxed Ni-based superalloys are more suitable to be produced in large forms such as billets and ingots while obtaining high yield strength and low-cycle fatigue strength. This is ideal for rotor disc and static applications at low to intermediate temperatures ranging from 540 to 760ºC [4].

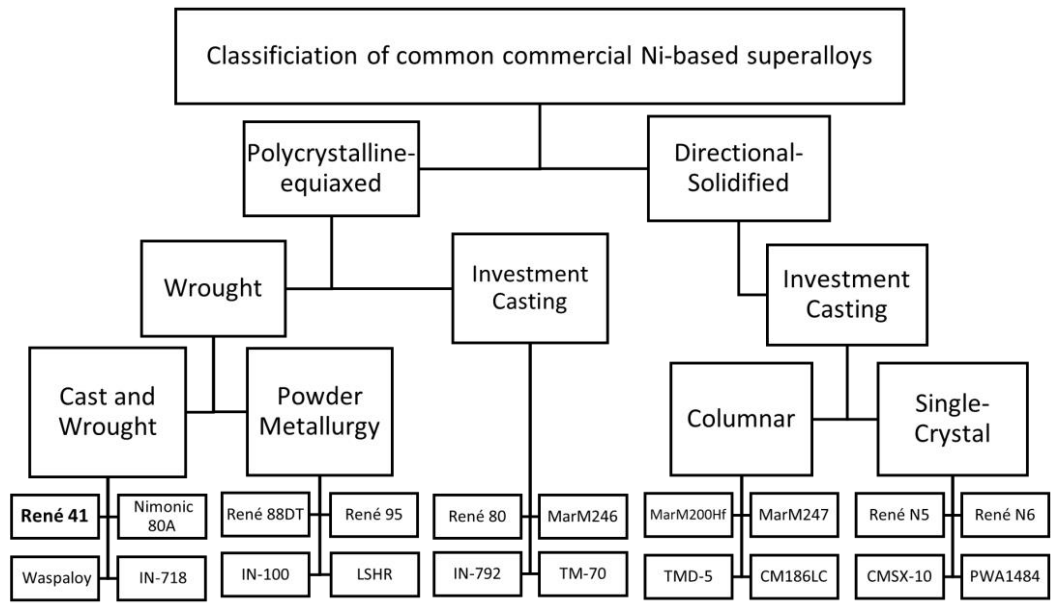

Figure 1.1 Overview of commercial Ni-based superalloys classifications and René 41 [4].

Balancing the high-temperature performance and processability of superalloys has been a major concern for superalloy developers. Improvements in mechanical and creep-rupture properties often oppose the processability of superalloys [6]. Hence, the development of cast and wrought superalloys requires both factors to be taken into consideration.

René 41 is a high-strength cast and wrought Ni-based superalloy developed in the 1950s to be selectively used for static GTE components exposed to intermediate to high temperatures (600-870ºC), such as afterburner parts, exhaust nozzle flaps, and turbine castings [4]. At 760ºC, René 41 has an ultimate tensile strength, yield strength and creep rupture strength after 1000 h of 1,105 MPa, 940 MPa, and 345 MPa, respectively [4]. This is at least 25% higher than achieved by other traditionally used commercial cast and wrought superalloys such as Inconel 718, Waspaloy, or Nimonic 80A at comparable temperatures [4]. However, the narrow processing window and sensitivity to die chill effects lower the forgeability of René 41 significantly [7]. Other concerns associated with René 41 include susceptibility to strain-age cracking which results in poor weldability and low machinability due to rapid work hardening. The combination of these factors ultimately deterred superalloy developers from exploring René 41 further for a diverse range of applications for GTE components over decades [8].

However, now the emergence of the aviation industry in the Asia-Pacific region in recent years is projected to exceed the combined air traffic of both Europe and North America by 2030. This rapid growth is expected to be met with increased demand for both the development and production of GTE-related materials, especially Ni-based superalloys. Hence, the market for Ni-based superalloys has correspondingly reported the highest compound annual growth rate (CAGR) in the Asia-Pacific region for the foreseeable future [9].

However, improvements to the high-temperature mechanical properties are often detrimental to the hot workability of the material. Thus, modifications in micro-alloying element additions such as B and C have become of interest in the design of advanced Ni-based superalloys [10,11]. Alloys contents of only up to 0.01 wt.% can have drastic effects on the fabricability and creep rupture performance [12–15]. The availability of thermomechanical processing on a laboratory scale is attractive for probing into industrial viability of such modifications. Via technologies such as physical simulations, constitutive modelling, and finite element modelling (FEM), precise prediction and modelling of flow-stress behaviour during industrial thermomechanical processes can be done [16–18]. The advent of high-resolution material characterisation techniques such as election back-scattered diffraction pattern (EBSD) acquisition [19] have also provided resources for René 41 as a candidate to be revisited for exploration as GTE material. Overcoming its shortcomings in processability will allow for higher efficiency GTE components to be made due to its superior performance in high-temperature applications.

The focus of this thesis is the influence of C and B microalloying additions on the microstructure and processability of René 41 Ni-based superalloy. The first section of this thesis explores the effects of C, and B additions on the processing-microstructure-property relationships following an integrated approach. Both thermomechanical processing and computational simulation will be used to study the influence of these so-called microalloying additions on the hot deformation behaviour, and hence the fabricability of René 41. Specially designed René 41 variants with modified C and B additions are provided by Western Australia Specialty Alloys (WASA) the industrial partner for this project. The second section of this thesis explores the utilisation of machine learning algorithms for analysis of industrial René 41 processing data also provided by WASA. This is achieved by exploring each dataset using conventional means such as data visualisation and statistics in a first step. In the second step, a predictive model for the anomaly detection based on ultrasonic testing inspection outcomes and chemical compositions of René 41 billets is achieved. The model analysis gives insights into the relationship between manufacturing defects and processing parameters without the need of extensive experimental work. This is also important for validating experimental results against industrial production.

# Chapter 2 – Literature Review

## 2.1 Phases and microstructure of Ni-based superalloys

### 2.1.1 Matrix phase

The matrix phase (denoted as γ-matrix) is a face-centric cubic (FCC) Ni-based solid solution and the main constituent of any Ni-based superalloy. The γ-matrix accommodates solute alloying elements, thus, providing solid solution strengthening [20]. In the case of wrought Ni-based superalloys, grain boundary strengthening (also known as Hall-Patch strengthening) is also an important mechanism to improve strength without sacrificing ductility unlike other strengthening mechanisms [21]. The γ-phase is inherently tough and ductile due to multiple slip systems of the FCC crystal structure and can accommodate more than ten alloying elements from the d-block transition metals such as Cr, Co, Mo, V, W, Ta, and Hf [3]. The formation of annealing twin boundaries is also a common occurrence in the γ-matrix [22]. Twin boundary volume fractions can be increased via grain boundary engineering to achieve superior mechanical properties in Ni-based superalloys. This is demonstrated by Qian and Lippold [23], where the tendency to intergranular cracking was observed to reduce via the formation of annealing twin boundaries in Waspaloy and Alloy 718 welds.

### 2.1.2 Geometrically close-packed structures

Contemporary precipitation hardened Ni-based superalloys contain a significant total volume fraction of up to 30 vol.% of geometrically closed packed (GCP) structures [6]. These are nano-to-micro scale precipitates with a structural formula of $Ni_3Al$. The most common GCP structures in superalloys are the γ' or γ" precipitates which provide anti-phase boundary (APB) or coherency strengthening [3] to the superalloys. Other GCP structures found in superalloys includes the η phase [24,25] and the δ phase [26], with various impact to the mechanical properties.

γ' precipitates have the chemical formula of $Ni_3Al$ and a $L1_2$ ordered FCC crystal structure as shown in Figure 2.1 [3]. The lattice misfit between γ-matrix and γ' precipitates is very small allowing for full lattice coherency between γ/γ' interfaces. APB strengthening is a well-researched mechanism for the origin of strengthening in Ni-based superalloys containing ordered precipitates [27]. This strengthening mechanism arises from the disruption of the chemical ordering in γ' precipitates caused by the passage of coupled partial dislocations. The result is the alignment of energetically unfavourable neighbouring atoms in the form of APB. The interfacial energy that arises from this APB formation reduces the free energy in the system. Hence, their presence increases the resistance against dislocations passing through ordered precipitates. The APB energy will be minimized by combining two coupled partial dislocations passing through the same slip plane, to maintain the chemical ordering of the γ' precipitate, thus, forming a superlattice dislocation. The FCC crystal structure adds a layer of complexity to the APB strengthening mechanism by breaking down the superlattice dislocation structure into four separate Shockley partial dislocations [28]. An illustration of one of the dislocation dissociation mechanisms and the associated complex stacking fault is shown in Figure 2.2. An in-depth breakdown of the various dislocation dissociation configurations has been made by Pope and Ezz [29]. The APB and stacking fault energy associated with the superlattice dislocation movements in Ni-based superalloys has been reported in the literature by Ahmadi et al. [30] and Kozar et al. [31] to be the most important strengthening mechanism in Ni-based superalloys.

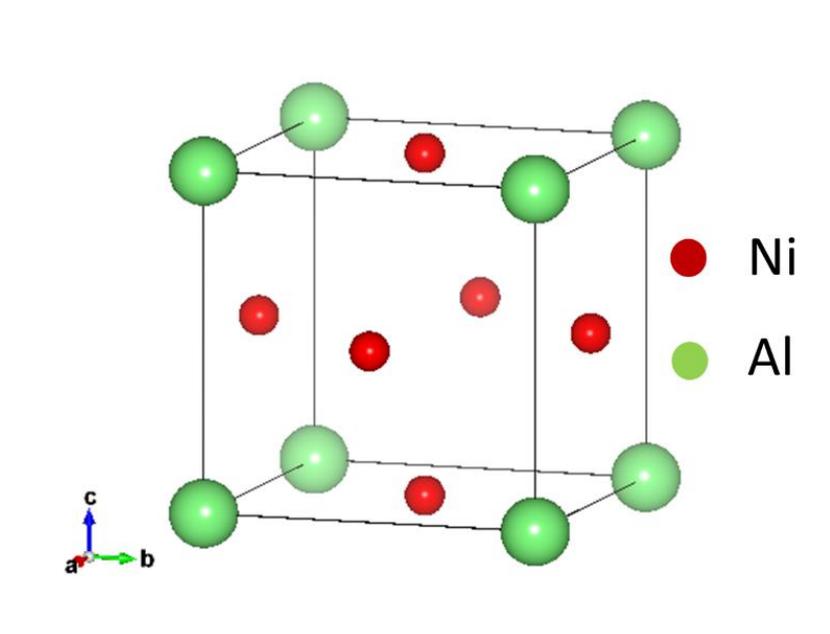

Figure 2.1 Unit cell of the Ni3Al L1$_2$ ordered FCC crystal structure [32,33].

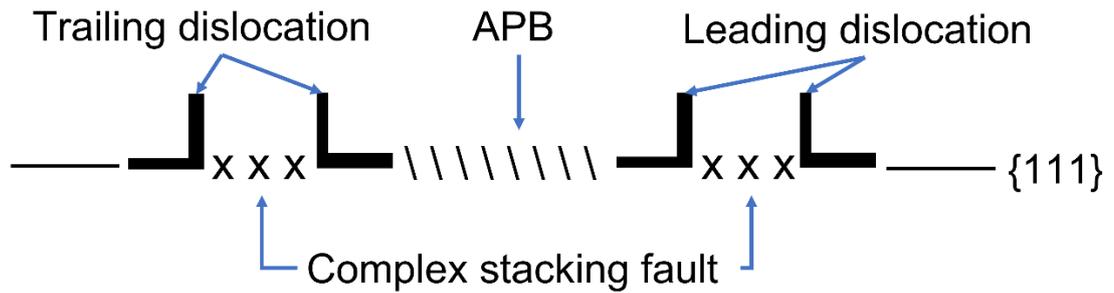

Figure 2.2 Dislocation dissociation mechanism in γ' precipitate in the {111} plane. Adapted from [29].

Superalloys with significant volume fractions of γ' precipitates exhibit an increased yield strength at elevated temperature regions as shown in Figure 2.3 when compared to FCC Ni. This phenomenon known as the high-temperature yield strength anomaly appears to contradict traditional physical metallurgy as higher temperatures tend to increase dislocation mobility, hence, lowering strengthening effects. The most widely accepted model for the high-temperature yield strength anomaly is the formation of Kear-Wilsdorf (K-W) locks [34]. A K-W lock is a cross-slip mechanism in γ', where at high temperatures high energy partial dislocations cross-slip from the {111} to the {010} planes. Here, the activation energy for slip is much greater for {010} planes than on {111} planes. As temperature increases, the likelihood of cross-slipping increases, hence, resulting in the high-temperature yield strength anomaly [28].

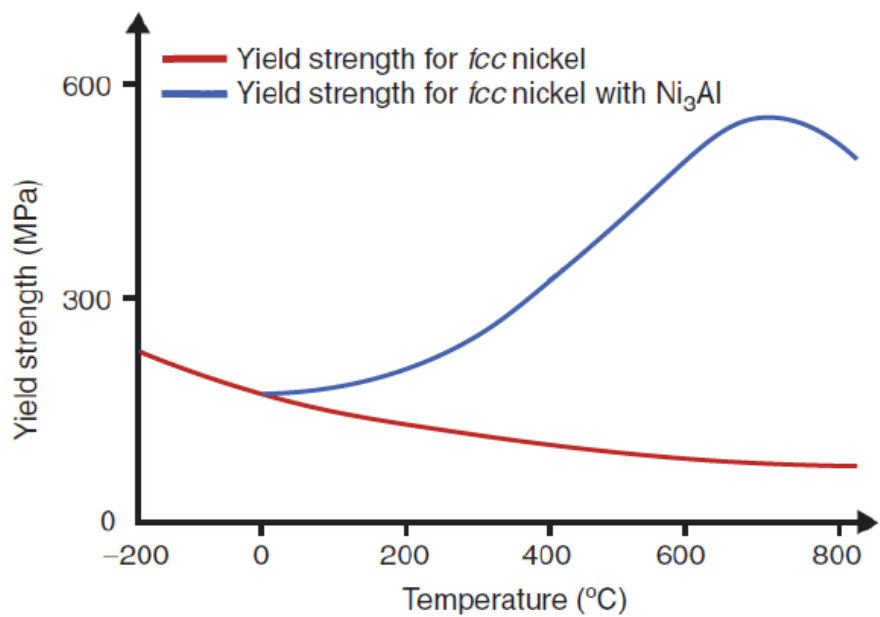

Figure 2.3 High-temperature yield strength anomaly in superalloys. Adapted from [35].

The morphology and size distribution of γ' precipitates strongly affect mechanical properties which depend on several factors such as local chemical composition, cooling rate, starting processing temperature, and strain rate etc. The microstructural evolution of γ' precipitate morphology starts as a spheroid which grows into cuboids followed by arrays of cuboids and eventually reaches solid-state dendrites. In general, as shown in Figure 2.4 there are three classes of γ' precipitates – primary γ', secondary γ' and tertiary γ'. These classes are separated based on which stages of the processing route these particles nucleated from. The classes dictate the general morphology and size distribution of the γ' precipitates, hence, providing different properties. Primary γ' are micron-scale precipitates formed in the hot working stage. These precipitates allow for grain growth pinning during sub-solvus thermo-mechanical processes, hence, reducing the overall grain size [36]. Secondary and tertiary γ' precipitates are nano-scale precipitates formed during cooling in the heat treatment schedule to provide much of the high-temperature strengthening effects to the superalloy [31].

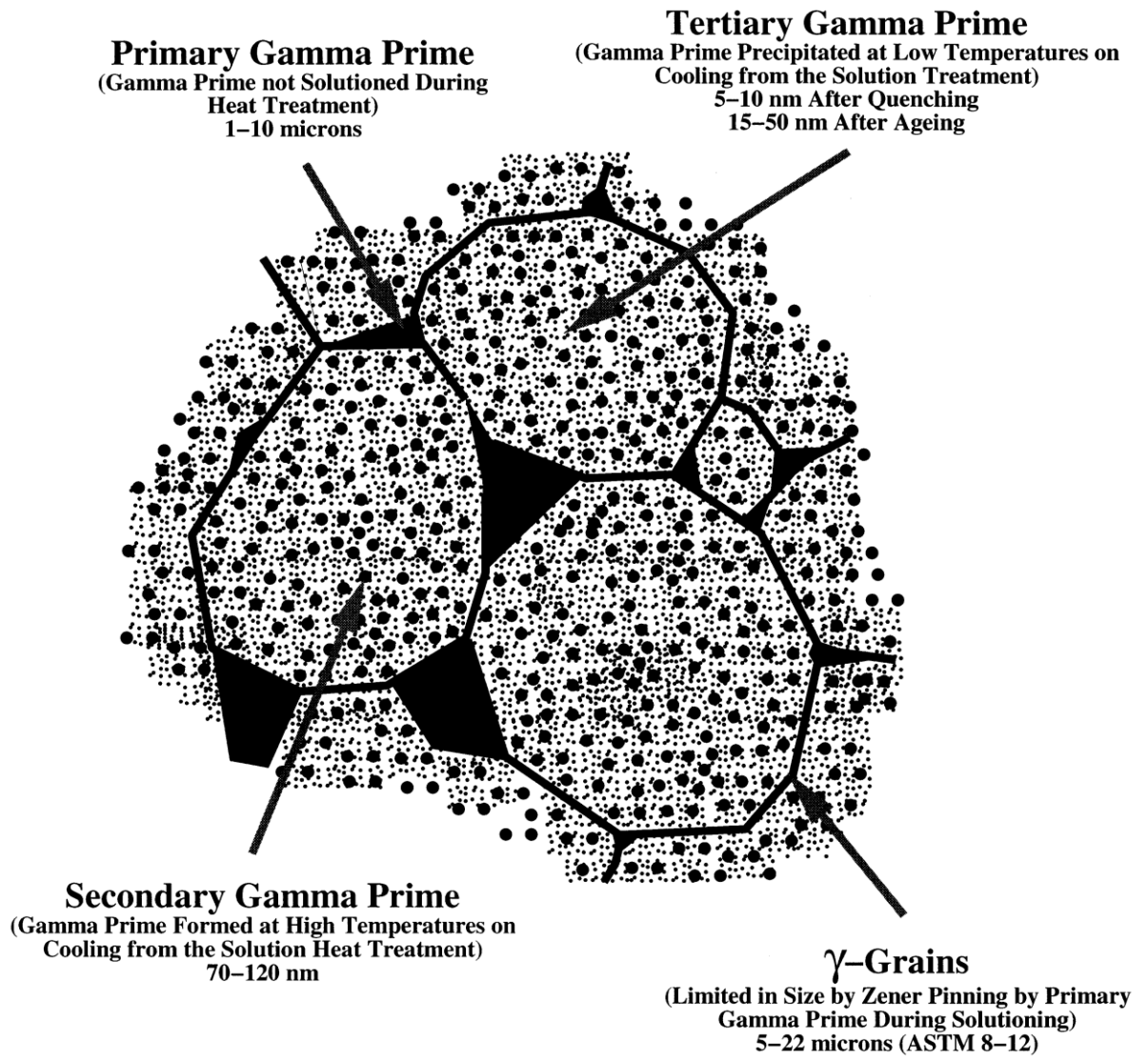

Figure 2.4 Schematic diagram of the size distributions and forming conditions of the three γ' populations in Udimet 720Li Ni-based superalloy [37].

2.1.3 Grain boundary precipitates

In polycrystalline Ni-based superalloys, many solute species can segregate to the grain boundaries due to their lattice incompatibility where they will form grain boundary precipitates. Carbides and borides are the most common intermetallic compounds that can be found at grain boundaries of superalloys. There are four carbides observed in the Ni-based superalloy microstructure – i) Primary MC carbides, ii) secondary $M_6C$ carbides, iii) secondary $M_{23}C_6$ carbides, and iv) secondary $M_7C_3$ carbides [3].

M represents the metal solute atoms depleted from the γ-matrix to form carbides with C. Primary MC carbides are formed during the solidification process of superalloys [38]. The MC carbides are comprised of mostly TiC with the NaCl (B1 in Strukturbericht designation) crystal structure as illustrated in Figure 2.5a [32]. M can also represent Ta,

Hf, Zr, Nb, V, Cr, Mo, and W for MC carbides. Most of the MC carbides share the same crystal structure as TiC except for MoC and WC having a hexagonal crystal structure [36]. MC carbides have high thermal stability, especially when combined with refractory metals such as Ti, Mo, and W [39].

Secondary carbides are formed primarily via solid-state phase transformations during heat treatments or thermomechanical processes [3]. The $M_6C$ carbides form at a temperature range of 815 to 980°C [4]. The $M_6C$ carbides have an $A_3B_3C$ crystal structure as shown in Figure 2.5b, with an $E9_3$ Strukturbericht notation. They will predominantly form in Ni-based superalloys with a combined (Mo + W) content of over 6 wt.% [38]. Depending on the alloying chemistry, 'A' atoms represent Ni, Co, Cr, Fe, Mn, Al, and V while 'B' atoms represent mainly Mo or W atoms but Nb, Ti, Ta, and Zr are also possible. There are also variants of $M_6C$ carbides in the form of $A_2B_4C$ and $A_4B_2C$ [40,41].

The $M_{23}C_6$ carbides are the most common secondary carbides in Ni-based superalloys. They are formed in the typical temperature range of 760 to 820°C [4]. The $M_{23}C_6$ carbides have a $Cr_{23}C_6$ complex cubic structure as shown in Figure 2.5c with a $D8_4$ Strukturbericht notation. While $M_{23}C_6$ carbides are mainly composed of Cr, depending on the bulk chemistry other alloying elements such as Al, Ti, Co, Ni, Mo, Ta, and W have been found in the $M_{23}C_6$ chemical composition of Ni-based superalloys [42,43].

Lastly, the Cr-rich $M_7C_3$ carbides rarely occur in Ni-based superalloys and form in the temperature range of 980 to 1093°C [38]. $M_7C_3$ carbides have a $Cr_7C_3$ complex hexagonal structure as shown in Figure 2.5d with a $D10_1$ Strukturbericht notation [32,33].

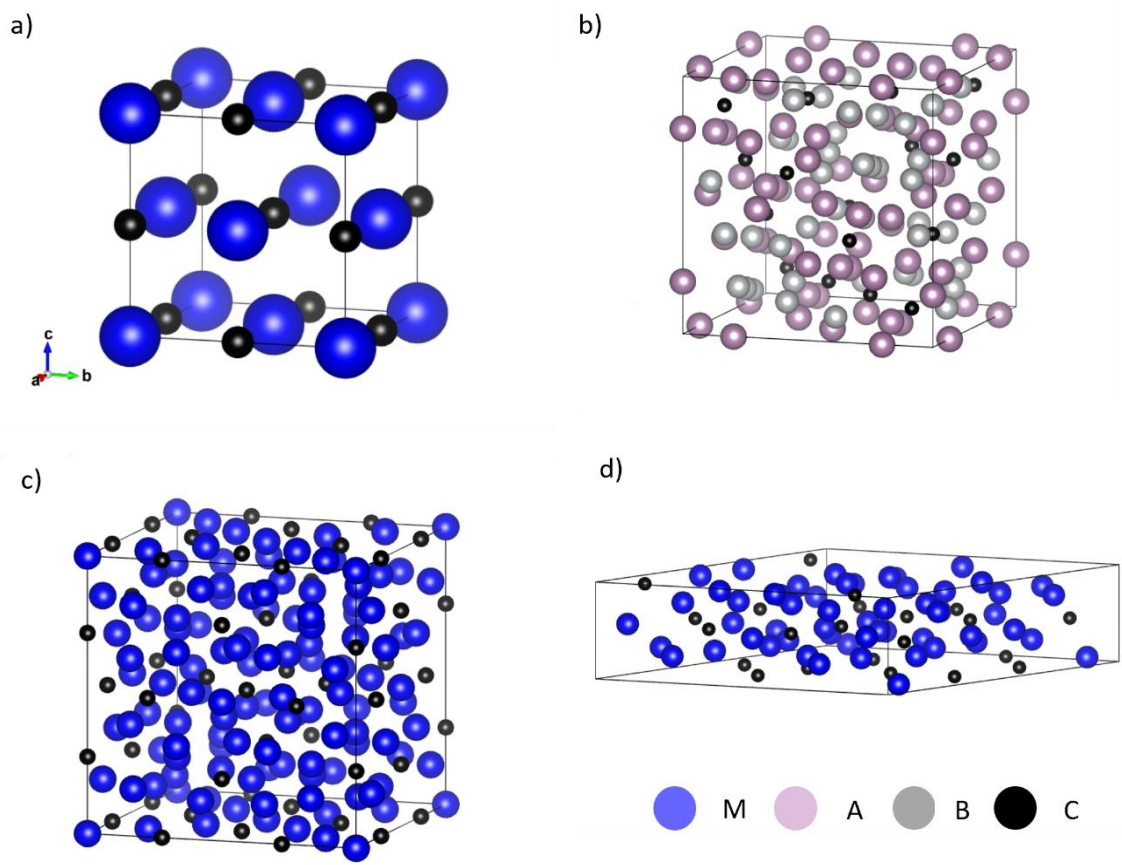

Figure 2.5 Schematics of carbide crystal structures (a) MC (b) $M_6C$ (c) $M_{23}C_6$ (d) $M_7C_3$ created from [32,33].

Borides are refractory precipitates with common metal formers such as Mo, Ti, Cr, Co, and Ni [39]. Borides in Ni-based superalloys have been observed in three forms of $M_2B$, $M_5B_3$, or $M_3B_2$ [44–47]. While $M_2B$ is rarely observed in Ni-based superalloys [48], both $M_3B_2$ and $M_5B_3$ borides are commonly found in Ni-based superalloys with B additions [49,50]. Two $M_2B$ variants form consistently either a body-centred tetragonal crystal structure as shown in Figure 2.6a, with a C16 Strukturbericht notation or a face-centred orthorhombic crystal structure as shown in Figure 6b, with a $C_b$ Strukturbericht notation. The M in $M_2B$ has been observed to be either Cr, Mo, or W in a model Ni-based superalloy [45]. The $M_3B_2$ borides have a tetragonal crystal structure as shown in Figure 2.6c, with a $D5_a$ Strukturbericht notation. The $M_3B_2$ borides observed in Ni-based superalloys can be constituted by the general $L_2SB_2$ structure where L represents heavy elements such as W, Mo, and Ti, whilst S represents lighter elements such as Cr, Co, and Ni. Similarly, the $M_5B_3$ borides have a tetragonal crystal structure as shown in Figure 2.6d, with a $D8_1$ Strukturbericht notation. The $M_5B_3$ chemical composition can also be represented as the general $L_4SB_3$ structure [46].

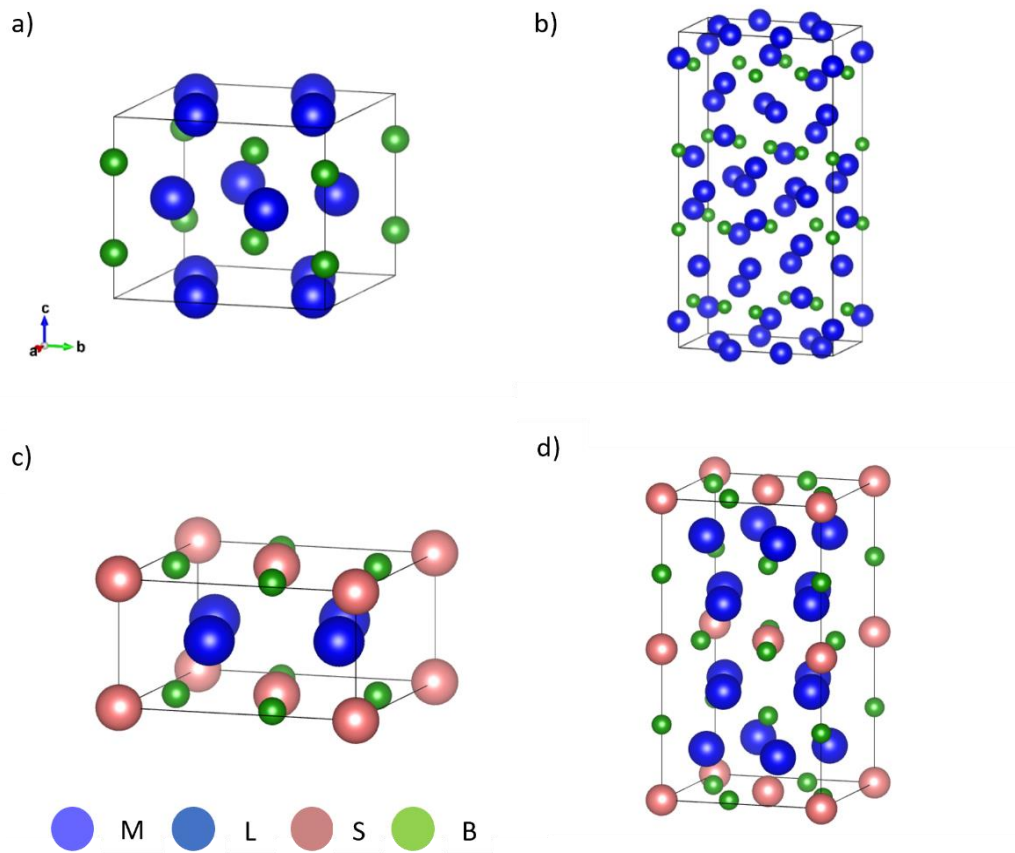

Figure 2.6 Schematics of boride crystal structures (a) $M_2B$ (C16) (b) $M_2B$ ($C_b$) (c) $M_3B_2$ (d) $M_5B_3$ created from [32,33,46].

### 2.1.5 Phases and microstructure of René 41

The main constituent phases of René 41 are the γ-matrix phase, γ'-precipitates, primary MC carbides, and secondary $M_{23}C_6$ and $M_6C$ carbides. René 41 is a cast and wrought Ni-based superalloy with a typical composition as shown in Table 2.1. With relatively high Ti/Al content, René 41 contains high γ' volume fractions in comparison to other cast and wrought Ni-based superalloys. This contributes to the superior performance at elevated temperature applications. As the Mo content in René 41 is typically around 9.75 wt.%, hence, exceeding the 6 wt.% (Mo + W) content threshold for $M_6C$ carbide formation, $M_6C$ carbides are abundant in comparison to $M_{23}C_6$ carbides [51]. The most observed morphology of $M_6C$ and $M_{23}C_6$ carbides in René 41 are in the form of discrete particles decorating high angle grain boundaries (HAGBs) [38].

Table 2.1 Typical composition of René 41 in wt.% [4].

| Cr | Co | Mo | Ti | Al | Fe | C | B | Ni |
|---|---|---|---|---|---|---|---|---|
| 19.00 | 11.00 | 9.75 | 3.15 | 1.50 | 0.50 | 0.08 | 0.006 | Bal. |

## 2.2 Effects of microalloying elements of Ni-based superalloys

### 2.2.1 Influence of Carbon

Carbon is a multi-purpose minor alloying element commonly added in cast and wrought Ni-based superalloys at a typical range of 0.01-0.4 wt.% to address a multitude of problems from reducing processing defects [52,53] to enhancing high temperature mechanical properties such as creep resistance and stress-rupture to failure [39].

C additions improve the purity of the melt by aiding in the volatilisation of low melting point tramp elements such as S via CO boil during vacuum induction melting [53]. This acts as a deoxidization and desulphurisation process to reduce grain boundary embrittlement in Ni-based superalloys. Evidence of desulphurisation by C addition has been highlighted by Holt and Wallace [53] where high residual S contents were observed in a series of experimental cast Ni-based superalloys containing low-carbon contents. In their study, the authors attributed C additions as instrumental to the removal of S during vacuum induction melting. With trace amount of S within the composition, $M_2SC$ sulfocarbides can form as grain boundary precipitates [39]. These precipitates are considered the less detrimental alternative than S solute segregation to the grain boundaries, but nonetheless cause grain boundary decohesion to a lesser degree [52]. This weakens the intergranular strength of the Ni-based superalloy, hence, promotes grain boundary embrittlement [54].

The influence of C additions on processing parameters has been difficult to evaluate in cast and wrought Ni-based superalloys. C decreases the incipient melting point and has two opposing effects: While fluidity and castability of the melt are improved, lower solution treatment temperatures are required to avoid incipient melting during solution treatment [55]. This may lead to incomplete dissolution of primary γ' and subsequently lower volume fractions of secondary and tertiary γ' precipitates, hence, decreasing mechanical properties [55]. Grodzki et al. [12] have also reported that C additions lead to a decrease in the γ/γ' eutectic volume fraction in the last solidification stage, increasing the likelihood of hot tearing in directionally solidified superalloys. Other effects of C additions on the processing of Ni-based superalloys include the reduction of common solidification defects such as freckles and shrinkage porosity [55,56]. Details of freckle formation will be discussed in section 2.3.1 as part of processing routes of cast and wrought Ni-based superalloy.

The influence of C on the microstructure of Ni-based superalloys can be divided into its influence as solute atoms within the γ-matrix, and its influence as grain boundary precipitates. The large misfit energy between the C atoms in Ni interstitial sites causes C atoms to segregate to the grain boundaries [57]. If C is not precipitated as carbides with strong carbide formers such as Ti, Cr and Mo, formation of graphite-like structures will occur when there is C content exceeding the solubility limit in the microstructure. This acts similarly to S segregation at the grain boundaries and significantly weakens the grain boundary cohesion of the alloy, leading to grain boundary embrittlement [57,58]. Therefore, the significance of C can be attributed to the different types of carbides, rather than its effect as a solute segregator.

The two commonly observed MC carbide morphologies in cast and wrought Ni-based superalloys as sown in Figure 2.7 are the blocky and Chinese-script (short 'script') morphologies. Script MC carbides are mostly found in cast Ni-based superalloys as interconnected, needle-shaped, eutectic dendrites along grain boundaries [38]. The script morphology has deleterious effect on the mechanical properties of Ni-based superalloys, as fatigue-related fractures are often initiated at MC carbides with script morphology [59]. Wrought Ni-based superalloys tend to form discrete, blocky particles nucleating inter and intra-granularly. Blocky MC carbides are mostly inert against heat-treatment schedule, thus, typically remain in the final microstructure [38].

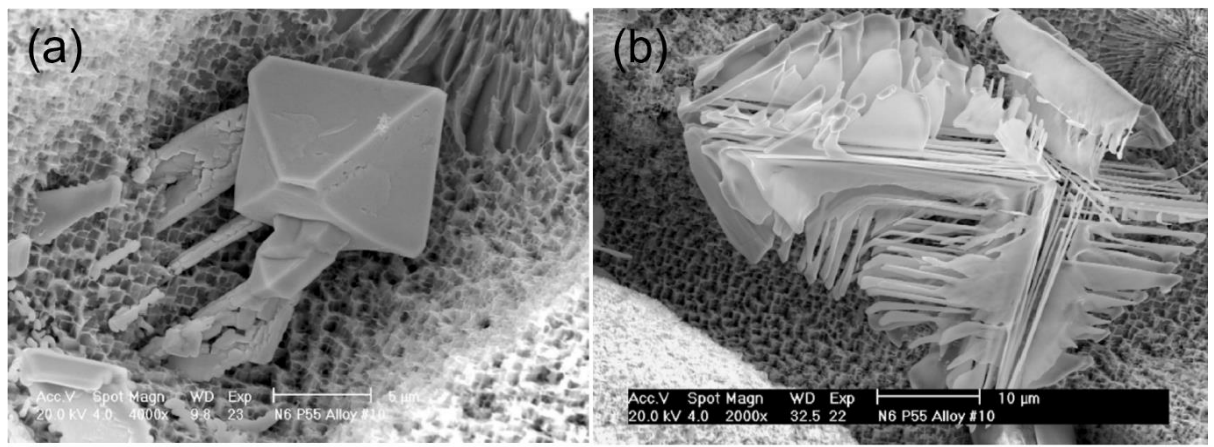

Figure 2.7 MC carbide morphologies (a) blocky (b) script in carbon-containing Ni-based superalloy [36].

The most important attribute brought by MC carbides, however, is to serve as sources of C for the formation of secondary carbides such as $M_{23}C_6$ or $M_6C$ carbides. While MC carbides are stable at high temperatures, heat treatment at specific temperature ranges can cause the decomposition of primary MC carbides into secondary $M_6C$ and $M_{23}C_6$ carbides [39]. As shown in equation 2.1 & 2.2 [60], the decomposition of MC carbides into secondary carbides is simultaneously accompanied by a phase transformation of the nearby γ-matrix into γ' precipitates. Hence, finely dispersed γ' can precipitate adjacent to $M_{23}C_6$ carbides at the grain boundaries [38]. The resulting grain boundary γ' precipitates inhibit directional growth reducing the coarsening rate of $M_{23}C_6$ carbides [61].

$$MC + \gamma \rightarrow M_{23}C_6 + \gamma' \tag{2.1}$$

$$MC + \gamma \rightarrow M_6C + \gamma' \tag{2.2}$$

$M_{23}C_6$ carbides are known for improving creep-rupture properties by inhibiting grain boundary sliding [14,62]. $M_{23}C_6$ carbides are usually found along HAGBs as discrete particles, although, there have also been observations of $M_{23}C_6$ carbides nucleating on incoherent and coherent twin boundaries [38]. The traditionally ideal microstructure for maximizing creep-rupture properties has been theorized to be a discrete dispersion of fine $M_{23}C_6$ carbides enveloped by grain boundary γ' precipitates at the HAGBs [2]. On the other hand, carbide network morphologies such as cellular or continuous film-like formations have been linked to weaker grain boundaries, and thus, reduced creep-rupture properties [38]. Two mechanisms have been proposed by Decker and Freeman [15]: Firstly, continuous $M_{23}C_6$ carbide networks act as obstacles to dislocation transmission across grains, leading to dislocation and stress accumulation. Ultimately, the pile-up of strain energy within the grain will lead to crack initiation. Secondly, coarsening of $M_{23}C_6$ leads to coarsening of grain boundary γ' precipitates. The resulting depletion of γ' formers in the nearby γ-matrix creates subsequent γ' depletion zones during heat treatment. γ' depletion zones are susceptible to crack initiation.

Blocky $M_6C$ carbides have been used extensively in the industry to limit grain growth during solution treatment [4]. This phenomenon is supported by the study of Song et al. [63] on the grain boundary pinning effect of $M_6C$ carbides during supersolvus heat treatments. This arises from the Zener-Smith pinning forces exerted by the $M_6C$ carbides to resist grain boundary migrations during grain growth [63].

$M_7C_3$ has been observed to precipitate in the γ-matrix as discrete, intragranular particles. Owing to their instability within the microstructure, $M_7C_3$ carbides decompose into $M_{23}C_6$ carbides upon exposure to temperatures below 937°C. The subsequent phase transformation into intragranular $M_{23}C_6$ carbides form in an acicular morphology which is known to be detrimental to the stress-rupture properties of Ni-based superalloys [38].

### 2.2.2 Influence of Boron

The tendency for B to segregate to the grain boundaries instead of retaining within the matrix as solid solution has been well documented in the literature [3]. Direct evidence of B segregation can be dated back to 1975 as reported by Walsh and Kear using secondary ion mass spectrometry to detect B concentration in a low carbon Udimet 700 nickel-base superalloy [64].

B additions are usually around 0.05 wt.% in Ni-based superalloys as higher B contents lead to agglomeration of brittle borides along grain boundaries. Similar to C additions, B additions influence the processability of Ni-based superalloys by decreasing the incipient melting point [65]. This promotes both an increase in the fluidity of cast Ni-based superalloys [65] and the susceptibility to hot tearing if the heat treatment temperature is above the incipient melting point [39]. Sharma et al. [66] reported the effects of B in reducing the tendency for powder metallurgy Ni-based superalloys to form prior to particle boundary precipitation. These are known processing defects that lead to lower creep-rupture properties.

B segregation has also been linked to improvements in hot workability and creep-rupture properties in a range of Ni-based superalloys [39,52]. This was first reported by Decker et al. [67]. In their study, Ni-based superalloys experienced enhanced mechanical properties when trace amounts of B were introduced via contamination of magnesia in the melting chamber crucibles during VIM. Decker and Freeman [68] later expanded upon this finding by proposing the mechanism in which B and Zr provide a stabilising effect on the grain boundaries. Thereby, the tendency of rapid agglomeration of $M_{23}C_6$ carbides and γ' precipitates at the grain boundaries is reduced as well. This leads to lower chances of brittle, precipitate related fracture, which is initiated by micro-cracking. Thereby, stress-rupture life and ductility were improved. This stabilizing effect has been studied in literature and attributed to many factors including increased grain boundary cohesion [58], minimization of embrittling agents [13], and boride formation [69].

Grain boundary cohesion is associated with failure due to grain boundary embrittlement in Ni-based superalloys [54]. Grain boundary weakness means that intergranular fracture is more likely to occur than intragranular fracture. A study by Cottrell et al. [57] highlighted a change in fracture mode when B was added to a model $Ni_3Al$ alloy, suggesting an increase in grain boundary cohesion with B additions. Furthermore, a first principles study by Sanyal et al. [58] reported an increase in grain boundary cohesion after doping with B, solidifying the beneficial effect that B can have on grain boundary cohesion.

A study by Floreen and Davidson [13] suggests that the primary effect of B is the reduction of grain boundary embrittling elements, such as O. This can make micro-cracking at the grain boundaries less prevalent. Sanyal et al. [70] also reported lower susceptibility to oxygen-induced embrittlement at Ni/boride interfaces compared to other common interfaces in Ni-based superalloys. This has been attributed to a suppression of charge localisation by O atoms in the presence of Mo atoms in borides.

Lastly, boride precipitation has also been reported with added benefits of improving grain boundary characteristics. Kontis et al. [50] reported improved creep-rupture properties from the suppression of $M_{23}C_6$ carbides with the formation of Cr-rich $M_5B_3$ borides in a model polycrystalline Ni-based superalloy. Another boride-related mechanism for increasing the creep resistance has been proposed by Wang et al. [49]. Here, a stress-induced phase transformation of $M_{23}(C,B)_6$ borocarbides to $M_5B_3$ have been proposed as the mechanism for increased creep resistance. The ideal boride morphology is proposed to be a fine dispersion at the grain boundaries to avoid grain boundary embrittlement [39], and grain boundary liquidation [47].

## 2.3 Processing route of cast and wrought Ni-based superalloys

### 2.3.1 Melting & remelting processes

An overview of a common processing route for cast and wrought Ni-based superalloys is shown in Figure 2.8. Modern cast and wrought Ni-based superalloys are either 'double-melted' or 'triple-melted' before applying thermo-mechanical processes. This refers to the application of two-step casting or three-step casting via vacuum induction melting (VIM), vacuum arc remelting (VAR) or electro-slag remelting (ESR) [3]. Schematics diagrams of these processes can be found in common superalloy literature [3,4].

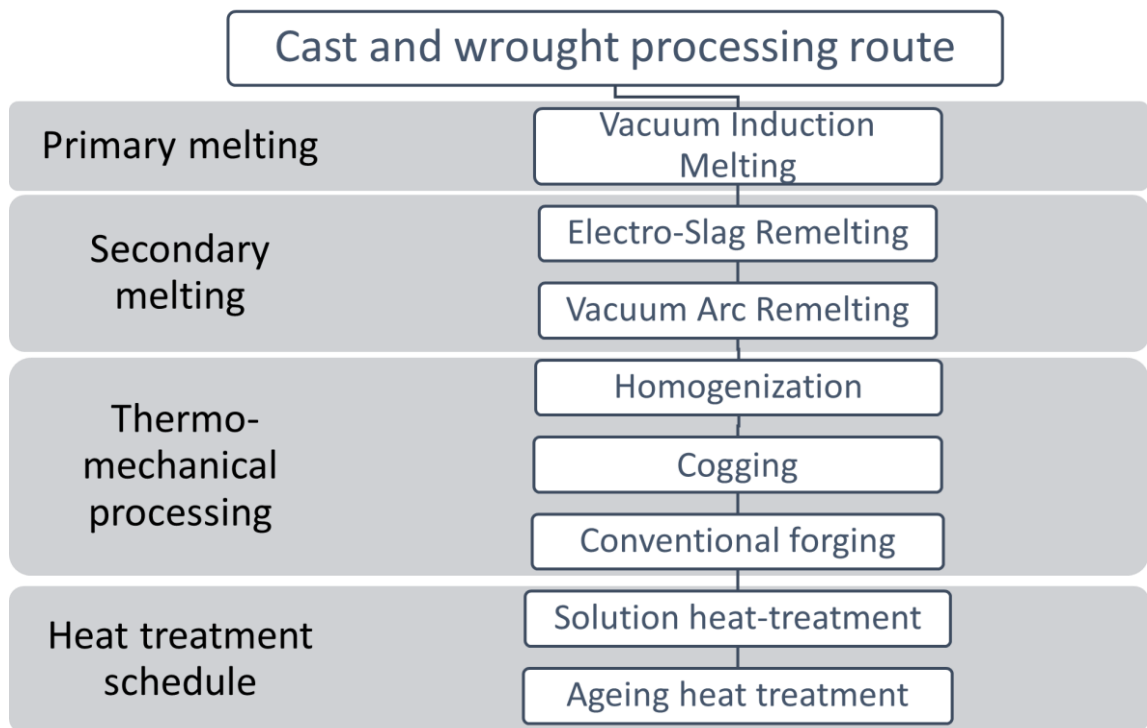

Figure 2.8 Overview of the typical cast and wrought processing route [3,4]

VIM is the standard melting practice for all superalloy stock. It is comprised of a double vacuum chamber system with the superalloy stock prepared and melted in the melting chamber. The subsequent melt is filtered through a tundish or a ceramic $ZrO_2$-based filter to remove slag, before pouring into a copper mould. VIM has three primary functions: i) combine the necessary raw materials together and adjust the superalloy chemistry to appropriate specifications, ii) remove tramp elements such as Pb and S which are detrimental to the mechanical properties and processability of the superalloy, iii) and prevent environmental contaminations from reacting with reactive elements such as Al and Ti present in the superalloy melt. The VIM superalloy castings are generally not ready to be processed by hot working due to having severe compositional segregation, inhomogeneous grain structures and major casting defects such as shrinkage cavity and solidification pipes. Therefore, additional melting practices are used to resolve these issues [3,4].

VAR is a secondary vacuum remelting technique aimed at improving the homogeneity within the casting structure while further improving the micro-cleanliness of the superalloy. The VIM ingots are placed in the centre of the remelting chamber as an electrode, and then are slowly consumed by the heat generated from a direct current (DC) electric arc running along the electrode to the solid ingot underneath. A water-cooled copper crucible is needed for rapid solidification to control the melt pool size during solidification. The molten layer formed from the surface of the electrode then drips down to the molten pool, promoting further degassing of $N_2$ and $O_2$. This also pushes impurities such as Pb and Bi to the outer rim of the ingot. This rim can be machined off to further enhance the micro-cleanliness of the ingot. The size of VAR melt pool is much smaller than VIM ingot considering only a limited surface area is available for melting in VAR process. This greatly limits the size of dendritic arms and reduces chemical segregation, hence, the overall grain structure achieved by VAR ingot is superior to VIM ingot for subsequent processes [3,4]. However, the formation of solute-lean local segregation regions known as white spots are well-documented to occur in VAR process [71]. There are three types of white spots as categorised by the ASM international White Spot Committee – discrete white spots, dendritic white spots, and solidification white spots [71]. The three types of white spots correspond to the difference formation mechanisms of the white spots as shown in Figure 2.9. White spots mostly originate from fall-in material insufficiently remelted in the melt pool allowing for locally segregated regions within the VAR ingot after solidification. In the case of discrete white points, these fall-in materials are from the outer-rim region of either the electrode or the VAR ingot specifically in the shelf, crown, and torus areas. These outer-rim areas are usually depleted in solute elements and contain inclusions such as oxides and nitrides from reacting with the crucible wall lining material forming 'dirty' white spots. Dendritic white spots are formed from the fall-in of electrode material. This material arises from porosity due to casting defects causing large dendrite clusters to dislodge from the electrode. These dendrites are solute-lean and have a higher solidus than bulk metal composition, hence, remains in the final solidification structure. Lastly, the solidification white spots can be found from the surface to the mid-radius of the VAR ingot. There have been two proposed mechanisms for the formation of solidification white spots. One of the mechanisms is the fall-in of splash beads deposited on the surface of the crucible or electrode. These splash beads cause local depletion by diffusing solute elements out of the solid-liquid boundary within the melt pool. Another mechanism is localised changes in solidification rate. Slower solidification rate allows for larger interdendritic arm spacing to form, hence, changes in local solute composition. These localised changes can be due to thermal fluctuation from fluid flow activity within the melt pool. Of the three white spots mechanisms, discrete white spots are known to be the most detrimental to the mechanical properties [71–74].

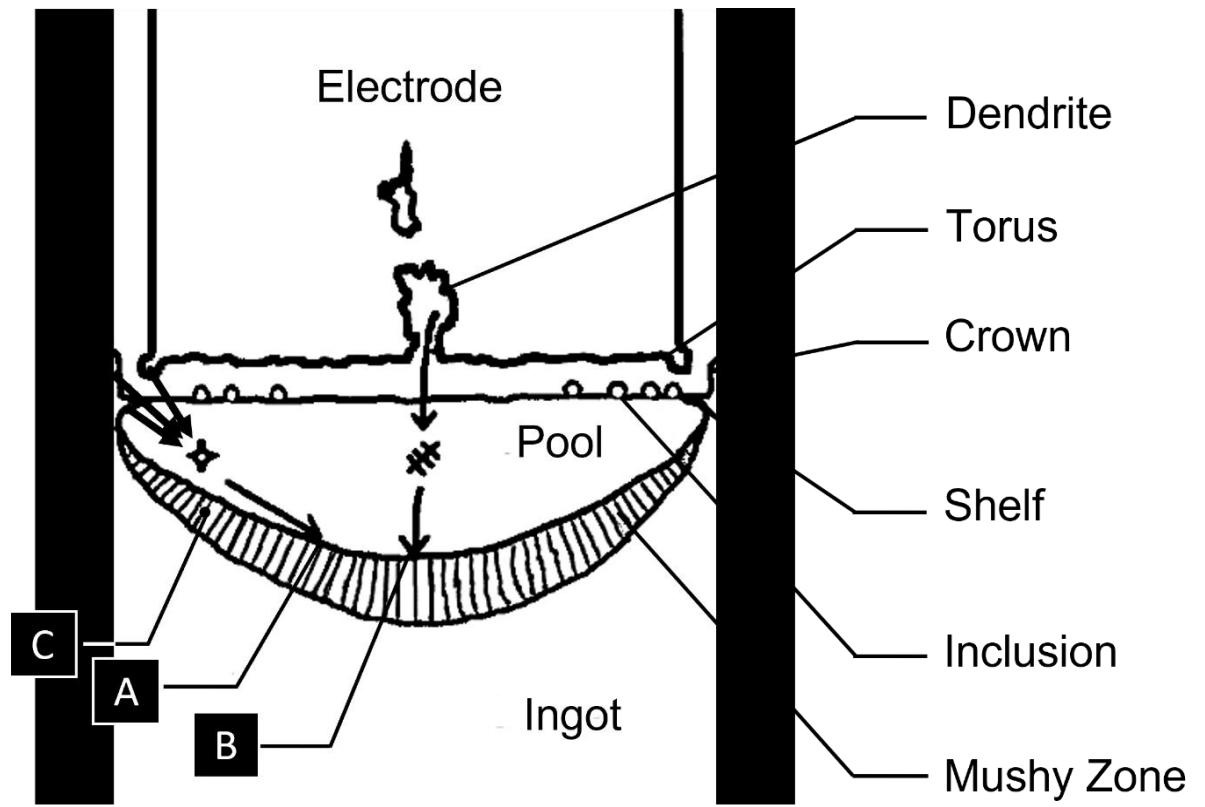

Figure 2.9 Schematic diagram of the formation mechanism of white spots (A) discrete white spot (B) Dendritic white spot (C) solidification white spot. Redrawn from [75].

ESR is characterised by the slag formation which allows for in-air condition instead of vacuum condition during the remelting process. During the initiation of the process, an alternating current (AC) is passed through the electrode into the ingot providing the heat necessary to initiate the slag reaction with the ingot. A subsequent heat source is provided by the heat generated from the exothermic reactions of the molten slag composed of mainly $CaF_2$ and oxide additions. The slag formation allows ESR droplets to precipitate out impurities much more effectively than the VAR process. Another key difference between ESR and VAR ingots are the generally deeper and steeper ESR melting pools at corresponding melt rates and crucible diameters. This means that ESR is more prone to chemical heterogeneity than VAR ingots [3,4]. Another consequence of the deeper melting pool profile is the formation of freckles which are solute-rich local segregation regions. Inter-dendritic fluid flow due to density gradient within the melt pool allows freckles to be formed. Freckles can severely comprise the mechanical properties of the superalloy due to the possible formation of TCP phases within these regions [75]. Hence, ESR always precedes VAR in triple-melting processes as it maximises the quality of the final ingot [3,4].

### 2.3.2 Thermomechanical processes

Since the microstructural evolution of Ni-based superalloys is time-temperature-strain history dependent, one major function of the thermo-mechanical processing stage is to prepare the ideal microstructure for optimising mechanical properties during heat treatment schedule. Homogenization whilst classified as a heat treatment step is typically applied before commencing any thermo-mechanical processes. This is done to minimise solute segregation across the ingot especially for high alloyed Ni-based superalloys with severe solute segregation, such as Waspaloy, René 41 and IN718. The homogenization duration can be up to four days at typically 1,200°C depending on the ingot size and alloying content. During homogenization, C tends to segregate into interdendritic regions and promote the agglomeration of primary interdendritic carbides. When the primary interdendritic carbides are broken down via subsequent hot working processes, localised carbide stringers can form. Such carbide stringers are known to cause hot tearing during forging via grain boundary embrittlement [4].

Ingot conversion is commonly used to breakdown the eutectic-dendritic grain structure into a more uniform, fine grain structure. For large ingots, this step is done by the cogging process via open-die hot forging to convert into smaller diameter billet usually by a factor of 2. Multiple heating and deformation cycles are necessary during this process as single-pass forging results in low strain penetration for most Ni-based superalloys. Hot rolling and hot extrusion can be viable alternatives to upset smaller dimension cast ingots [4]. It is critical for wrought Ni-based superalloys to achieve uniform fine grain structure in cogged billets so that subsequent forgings, often associated with large strain variation to shape into GTE components will not have a large impact on the grain structure in the final microstructure [6].

The consideration for cogging temperature is important as grain refinement requires dynamic recrystallisation to be activated. For γ'-hardened Ni-based superalloys, this temperature is dependent on the γ' solvus temperature. Therefore, the recommended starting temperature tends to be just above γ' solvus temperature for recrystallisation to occur without extensive grain growth [3]. The finishing cogging temperature is often selected just below the γ' solvus temperature to retain a fine grain structure. In the case of René 41, the typical cogging temperature ranges from a starting temperature of around 1150ºC to a finishing temperature of around 1100ºC [4]. The unusually high finishing temperature is to avoid $M_6C$ carbide precipitation which allows for more $M_{23}C_6$ carbides precipitation in subsequent heat treat schedule, hence, the creep-rupture properties of final René 41 billets can be maximised [76]. However, manufacturers may also opt into maximising ductility and stress-rupture properties instead by prioritising $M_6C$ carbides precipitation via the usage of a lower finishing temperature [7]. The cogged billets will be passed on to the forging processes converting the billets into forgings. The main objective is to produce the desired geometries suitable for their respective applications, whilst also maintaining uniform fine grain structure. The hot working temperatures are usually kept as close as possible to the γ' solvus, but not exceeding it to promote further grain refinement [3]. For René 41, it is recommended that for upset and ingot breakdown the forging temperature should start at 1150ºC and finish before 1040ºC [4]. At this temperature range, recrystallised grains are retained due to the finishing temperature at just below 1060ºC which is the reported γ' solvus for René 41 [7].

In forging, forging laps (buckling of material under compressive stress) and internal cracking can occur due to improper deformation parameter control leading to excessive work hardening within the material [3,4]. Hence, ultrasonic inspection is routinely used as the final wrought superalloys production route to detect internal defects such as cracks and cavities within the billet. Traditionally, any flaws detected exceeding 2 mm are considered not acceptable and the billet will have to be removed from wrought processing and reused as scrap material in new VIM charge [74].

Flow localization is the accumulation of strain within a narrow zone, and an important factor in influencing the materials response to hot deformation. The accumulated strain can be highly strain-hardened to the point of premature failure of components. Several factors, if not accounted for, can promote severe flow localization. The most common cause of flow localization is the poor lubrication between workpiece and forging die contact leading to barrelling of the workpiece from friction. This barrelling effect causes strain concentration at the workpiece cross-section, leading to shear band formation and a dead zone directly at the forging surface [7]. Another similar effect is the die-chill effect, where rapid heat extraction occurs at the workpiece and die contact leading to differentiate in flow stress across the cross-section profile of the workpiece leading to flow localization. Furthermore, localised flow softening due to dynamic recrystallisation (DRX) processes can also lead to preference to shear band formations, hence, promote flow localization [77,78].

### 2.3.3 Heat treatment schedule

Heat treatments of precipitation hardened Ni-based superalloys consist of two main procedures – solution treatment and ageing treatment. Although mill annealing, stress relief annealing, and surface treatment can also be classified as heat treatment steps, the utilisation of these heat treatment procedures varies between manufacturers. In general, annealing procedures are used to relieve work hardening effects and dissolve strengthening precipitates [4] while surface treatments tend to be used to increase fatigue properties and corrosion resistance [79]. Heat treatment cycles of René 41 typically consist of a solution treatment at 1,065ºC for 0.5 h followed by air cooling, and an ageing treatment at 760ºC for 16 h, followed by air cooling as well [4].

The role of the solution treatment is to dissolve secondary-phase precipitates formed from the thermo-mechanical stage and prepare the microstructure for the subsequent ageing heat treatment. Precipitation hardened Ni-based superalloys are recommended to have dry Ar atmosphere for the solution treatment due to the risk of in-air oxidation for strong oxide formers such as Al and Ti [4]. The solution treatment temperature is determined with the upper limit always kept below the incipient melting temperature of the superalloy. The soak time for solution treatment varies between 1 to 8 hours for typical wrought Ni-based superalloys [38]. Depending on the grain size requirement and precipitation population, manufacturers can opt into either supersolvus or subsolvus solution treatments. Supersolvus solution treatment dissolves primary γ' population, thus, increases γ' volume fraction of secondary and tertiary γ' populations due to increased γ' formers within the γ-matrix during subsequent ageing schedule [35]. This comes at a cost of increased grain size due to the lack of primary γ' grain boundary pinning during solution treatment. This has an adverse effect on both tensile strength and creep properties as shown in Figure 2.10 [35]. On the other hand, subsolvus solution treatment retains the fine grain structure from forging but suffers from reduced secondary and tertiary γ' population due to the presence of primary γ'. Figure 2.11 highlights the detrimental effect of increased precipitate size to the mechanical properties of Ni-based superalloys [35]. Hence, it is a balancing act between the contribution of grain size distribution and γ' precipitate size distribution for maximising mechanical properties of Ni-based superalloys. For René 41, the typical solution treatment schedule consists of a supersolvus heat treatment with ½ h at 1065 ºC followed by air cooling [4].

The main purpose of the ageing treatment is to precipitate the correct type, size, and morphology of the strengthening phases in precipitation hardened Ni-based superalloys. Since the aged microstructure will be the final microstructural evolution step before service, the in-service microstructure and mechanical properties are strongly dependent on this ageing treatment procedure. Isothermal between 612 ºC to 1038ºC is typical of the ageing temperature with a holding period of up to 24h, depending on the specific composition of Ni-based superalloys. Ageing treatment schedule is depending on the requirements for adequate size distribution of secondary and tertiary γ' precipitates to prioritise either creep resistance or creep-rupture properties. Overageing or underageing can lead to accelerated TCP phase formation and deleterious morphologies of $M_{23}C_6$ carbides. For René 41, typical ageing schedule consist of 16 h at 760ºC followed by air cooling [4].

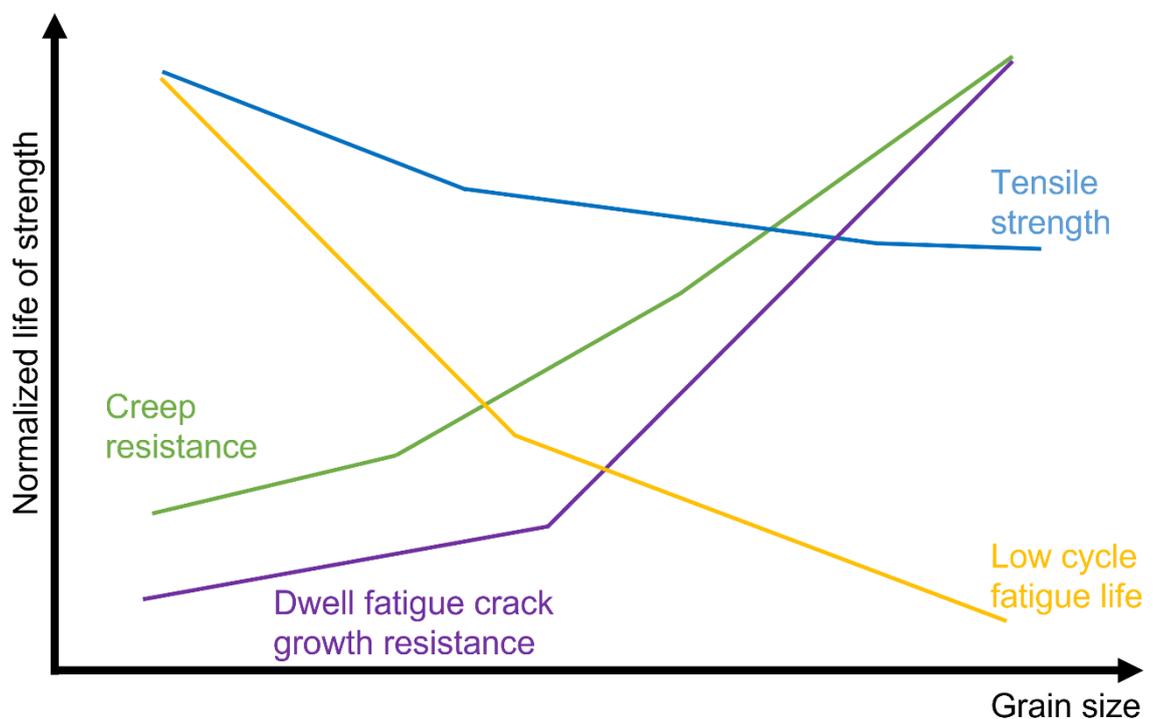

Figure 2.10 Relationship of mechanical properties vs grain size in polycrystalline Ni-based superalloys. Reproduced from [35].

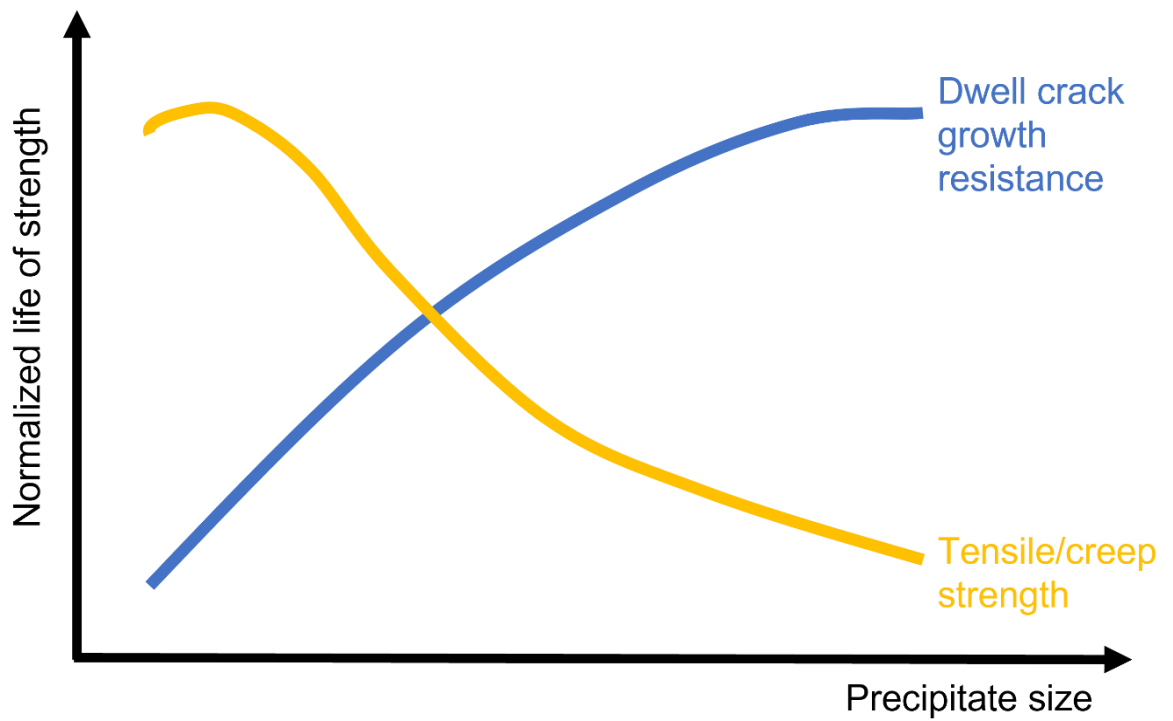

Figure 2.11 Relationship of mechanical properties vs γ' precipitate size in polycrystalline Ni-based superalloys. Reproduced from [35].

## 2.4 Hot workability of wrought Ni-based superalloys

### 2.4.1 Recrystallisation phenomena in Ni-based superalloys

Plastic deformation is stored in a deformed material mostly as accumulations of dislocations [80]. During hot deformation, this deformation energy can be released via three main dynamic restoration mechanisms – recovery, recrystallisation and grain growth [80]. Recovery refers to the arrangement and annihilation of dislocations into low energy structures. Recrystallisation can be defined as the "formation and migration of HAGBs driven by the stored energy" [81]. Lastly, grain growth is the increase of mean grain size driven by lowering the interfacial energy between grains through the reduction of total grain boundary area. In hot working, dynamic recovery (DRV) controlled processes and dynamic recrystallisation (DRX) controlled processes reflect vastly different stress-strain behaviours, as shown in Figure 2.12. For DRV driven process, there is a continuous increase of the flow stress σ until a saturation point ($σ_{sat}$). While for DRX to occur, a critical strain ($ε_c$) value is required to be achieved in the deformed material. The critical stress ($σ_c$) is the true stress value corresponding to the critical strain value. $\Delta\sigma_s$ is the net softening described as the difference between the theoretical DRV curve and the measured DRX curve. During the early stage of a DRX driven process, the flow stress is characterised by the increase in flow stress due to work hardening. This increase will slow down once it crosses the $\varepsilon_c$ where dynamic softening mechanism such as DRX and DRV occurs. The slowing will eventually reach the saturation point known as the peak stress ($\sigma_P$) where the work hardening rate is equal to the flow softening rate in the material. Then, DRX starts to take over when the peak strain ($\varepsilon_p$) is reached. Now, the driving force for DRX is the greatest. This is followed by a flow softening phenomenon gradually decreasing the flow stress into a steady state flow stress ($\sigma_{ss}$) [82,83].

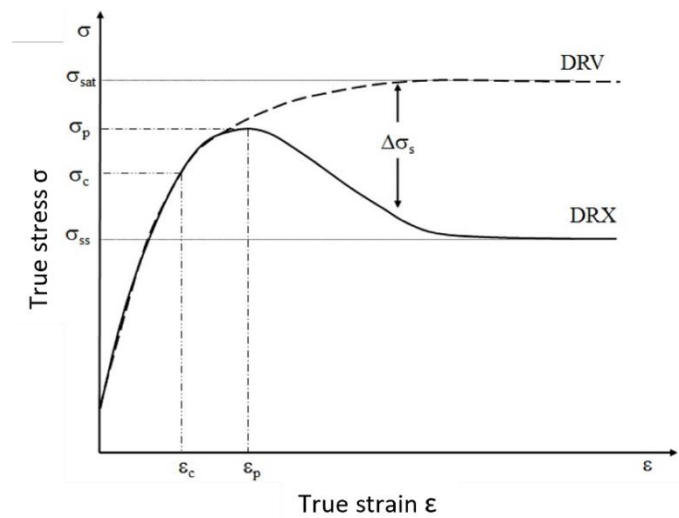

Figure 2.12 Typical dynamic recovery and dynamic recrystallisation curves [82].

There are three main classifications of DRX – discontinuous dynamic recrystallisation (DDRX), continuous dynamic recrystallisation (CDRX), and geometric dynamic recrystallisation (GDRX) as shown in Figure 2.13. Conventional DRX commonly refers to the DDRX mechanism with necklace structure at grain boundaries observed in low stacking fault energy (denoted as $\gamma_{SFE}$) metals. DDRX is characterised by a clear separation between nucleation and growth stage. Large wrought products usually have a slow temperature dissipation rate, allowing post-dynamic recrystallisation to develop after strain has been stopped. This can either be in the form of meta-dynamic recrystallisation (MDRX) or discontinuous static recrystallisation (DSRX). MDRX is a rapid softening mechanism directly following DDRX exclusively to grow newly nucleated DRX grains. Since the prior DDRX process eliminate the need for an incubation time (i.e., critical strain has been reached) to nucleate new DRX grains, MDRX can occur at the order of $10^{-2}$ s. DSRX is generally much slower as new nucleation and growth events occurs. MDRX and DSRX serve to anneal the microstructure after hot working is done to metals [84]. CDRX is characterised by the progressive lattice rotation of subgrain cell structures near grain boundaries, and the homogeneous increases in grain boundary misorientation of LAGBs to form HAGBs. GDRX is typically found in hot rolling, hot torsion, and plane strain compression processes where there is a clear rolling/ deformation texture applied to the deformed materials [84].

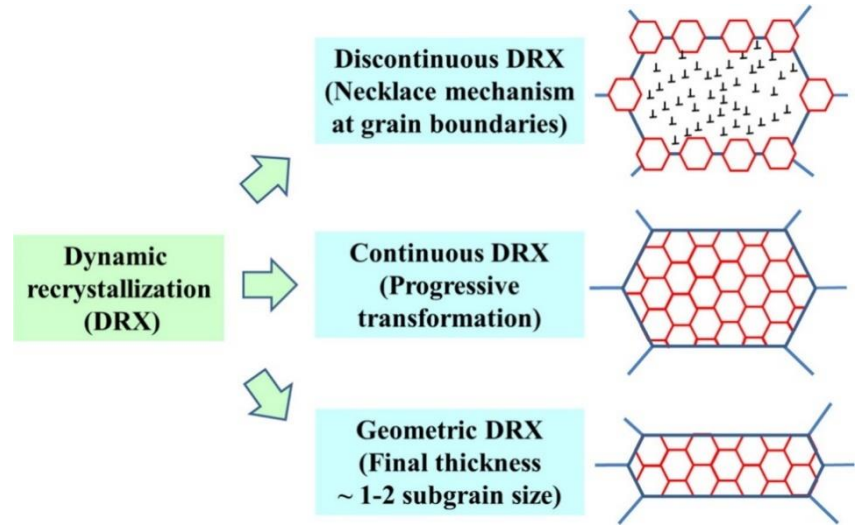

Figure 2.13 Schematics of DRX mechanisms (a) DDRX (b) CDRX (c) GDRX [83].

There are numerous studies investigating the DRX behaviour of Ni-based superalloys during hot deformation to determine the optimal processing parameters for a uniform grain structure. There have been many instances reporting conventional DDRX as the primary flow softening mechanism in several Ni-based superalloys [85–88]. Many studies attributed this phenomenon to the low $\gamma_{SFE}$ of Ni-based superalloys [89,90]. Since DDRX is greatly dependent on the grain boundary area fraction of the initial microstructure, the coarse grain structure of as-cast Ni-based superalloy poses a challenge in achieving fine grain structures, especially for supersolvus hot deformation where DDRX is reported as the sole DRX mechanism [89].

In recent years, attention has been focused on subsolvus isothermal forging of as-cast Ni-based superalloys to observe the role of primary γ' influencing the DRX process [82,91]. Xie et al. [89] observed CDRX in conjunction with DDRX in sub-solvus hot deformation of a model as-cast Ni-based superalloy. The formation of CDRX was attributed to the pinning effect of primary γ' at HAGBs against the growth of DDRX grains, allowing dislocations to accumulate into subgrain structures. Wan et al. [92] reported similar findings in Udimet 720Li and described the phenomenon as particle-induced continuous dynamic recrystallization (PI-CDRX). Another type of DRX mechanism was observed in low strain sub-solvus hot deformation of γ- γ' Ni-based superalloy as reported by Charpagne et al. [93]. These authors describe the DRX mechanism as heteroepitaxial recrystallisation (HERX) with recrystallisation nuclei forming at the interface of primary γ' precipitates. HERX grains are characterised by sharing the same crystallographic orientation with a parent primary γ' precipitate. HERX is a distinctively different DRX mechanism than conventional particle stimulated nucleation (PSN) since the recrystallised grains do not share any crystallographic relationship with the parent secondary-phase particle in PSN.

Liu et al. [90] reported significant changes in the DDRX behaviour of Udimet 720Li in response to changes of initial grain size. Due to coarsening of γ' precipitates, growth of recrystallisation nuclei has been restricted in coarse grain structures which resulted in non-uniform grain structures. Abnormal recrystallisation phenomena have also been reported by Miller et al. [94] and Charpagne et al. [93] in different processing conditions. Abnormal recrystallised grains significantly affect the fatigue life of Ni-based superalloy components and, therefore, should be avoided as much as possible during manufacturing.

The role of C and B additions in influencing the DRX behaviour of Ni-based superalloys have not been comprehensively studied. Wang et al. [95] reported findings of classical PSN events and impingement of recrystallised grain boundaries occurring at MC carbides. While B additions have been associated with improved hot workability, little has been reported on its relation to DRX behaviour in Ni-based superalloys [52]. Hosseini et al. [96] reported findings of increased γ/γ' lattice misfit of ATI 718Plus Ni-based superalloy. This could affect the interaction between γ' and DRX process as increases in γ/γ' misfit have been known to changes in γ' morphology [97]. However, this has not been further explored in literature. Hence, in this thesis further examination into these additions will be discussed in Chapter 3.

### 2.4.2 Hot deformation studies on Ni-based superalloys

Hot workability is a broad term describing the relative ease of a material to be shaped through hot deformation processes. There are both processing and materials considerations when evaluating the hot workability of a material. Intrinsic workability encompasses a material's response to processing parameters such as strain, temperature, and strain rate, and it is influenced by the material's deformation modes, initial microstructure, bulk and local chemistry, and microstructural evolution during hot deformation [77]. Physical simulation studies are means to evaluate the hot workability of a thermo-mechanically processed material using various experimental methods such as hot compressing testing, hot tensile testing, and hot torsion testing. The basic principle of these experimental methods involves quantifying intrinsic workability of a wrought material through recorded flow stress data from a series of controlled deformation done to the samples with varying strain rates and temperatures, and then convert it into mathematical expressions known as constitutive equations. These constitutive models in conjunction with microstructural analysis are used to predict flow behaviour in a range of hot working windows for industrial-scale thermo-mechanical processes by evaluating microstructural stability regions which are regions with uniform recrystallization, no flow localization, or defects [77].

On the other hand, computer-aided modelling and processing design have improved allowing for a more detailed understanding of the material during hot deformation process. Chamanfar et al. [98] applied finite element method (FEM) modelling to simulate the isothermal hot forging of wrought superalloy. The resultant flow stress generated from the FEM closely matches experimentally derived flow stress in the study. Another common use of FEM modelling is the generation of effective strain maps at the forging cross-section. This can be beneficial for understanding the level of strain gradient within the forging. While FEM allows for simulation of complex forging conditions, they require accurate materials data and correct model descriptions which can only be verified by experimental studies [4].

There have been many physical simulation studies on different wrought Ni-based superalloys to model the flow stress response of Ni-based superalloys. The most appropriate hot workability testing technique has been the hot compression testing which simulates the compression behaviour of forging processes commonly used for GTE components such as turbine discs [77]. Lin and Chen [99] have provided a critical summary of the existing constitutive equations for different hot working applications and materials with three distinctive groups – phenomenological-based models, physical-based models, and deep learning-based models. The most common constitutive model for modelling the hot deformation behaviour of Ni-based superalloys is the Sellars-McTegart hyperbolic sine equation [100]. Despite the phenomenological nature of the equation, the models established by many studies using the hyperbolic sine equation have shown close agreement with experimental data with different commercial Ni-based superalloys [85,87,90,101–105]. The hyperbolic sine equation as shown in equation 2.3 relates both the deformation parameters and intrinsic material variables into a single equation allowing the model to adjust for any changing variables. Equation 2.3 can be converted into stress-strain response via the Zener-Hollomon $Z$ parameter as shown in the form equation 2.4. The Zener-Hollomon parameter as highlighted in equation 2.5 encompass the stress-strain response with hot deformation parameters such as strain rates and temperatures.

$$\dot{\epsilon} = A(\sinh \alpha \sigma_p)^n \exp\left(-\frac{Q}{RT}\right) \tag{2.3}$$

$$\sigma = \frac{1}{\alpha} \ln\left[\left(\frac{Z}{A}\right)^{\frac{1}{n}} + \left(\left(\frac{Z}{A}\right)^{\frac{2}{n}} + 1\right)^{\frac{1}{2}}\right] \tag{2.4}$$

$$Z = \dot{\epsilon} \exp\left(\frac{Q}{RT}\right) \tag{2.5}$$

Where A, $\alpha$ are materials constants, $n$ is the stress exponent, Q is the activation energy of hot deformation, R is the universal gas constant, T is the deformation temperature in Kelvin, stress is represented by the $\sigma_p$ of the stress-strain curve, and $\dot{\epsilon}$ is the true strain rate [100]. The utilisation of constitutional modelling is essential for understanding hot deformation behaviours of Ni-based superalloys. Hence, the establishment of constitutional models for this thesis will be discussed in Chapter 3.

## 2.5 Machine learning in physical metallurgy

### 2.5.1 Overview of machine learning in material science

Material informatics is an interdisciplinary field that combines both traditional material science and information technology and has received increasing attention in publications most noticeably in the last decade [106,107]. Increasing complexity in multivariable constraints in experiments have incentivised the use of many computer science tools such as machine learning algorithms for the prediction of material properties. Ruiz et al. [108] used machine learning algorithms to predict the mechanical properties of steel rods via regression modelling. Lin et al. [109] have utilised artificial neural networks (ANN) to formulate a predictive constitutive model for the hot deformation behaviour of a commercial Ni-based superalloys.

On the other hand, selecting critical forming parameters traditionally requires a long and expensive process of physical trials. With machine learning, process optimization can be done with improved accuracy and less time investment. Vecchio et al. [110] used support vector machine algorithm to optimize superalloy investment casting process. Huang and Li [111] found processing-defect linkage during additive manufacturing processing via machine learning model analysis. Hamouche and Loukaides [112] found the optimal sheet forming processing configuration for a given final shape via deep learning classification algorithms.

### 2.5.2 Machine learning fundamentals

The basis of machine learning (ML) models is to provide generalised statements (i.e., outputs) about a given dataset that reflect true relationships in the dataset. These dataset consist of observations consisting of features either collected from real world applications or artificially created [113]. A feature is a characteristic associated with the observation and can be of varied sizes extending into a feature vector where each observation has multiple features. There are four types of machine learning models – these are i) supervised learning, ii) unsupervised learning iii) semi-supervised learning and iv) reinforcement learning [114]. Only supervised learning will be discussed in this section as this approach will be used in this thesis.

In supervised learning, a labelled dataset is used as the input of the model. A label is the outcome of the input needed for the ML model. These labels are paired with the feature vectors and used to generate a supervised learning ML model. Figure 2.14 illustrates the general framework for developing a supervised learning model. The workflow consists of splitting the original dataset into training-test sets conventionally in a 75/25 ratio but can vary depending on sample size [115]. Pre-processing is often necessary as many machine learning algorithms require strict requirements for data structure. Feature scaling is one commonly used technique for the pre-processing of data [113]. The goal is to normalise the range of the values between features while preserving the variances between each feature [113]. The three main feature scaling techniques with formulas as shown in Equation 2.6 to 2.8 are the standardisation, normalisation, and robust scaling, respectively. The training dataset is used to generate a model that can be tested against using a testing dataset. Model evaluation is performed and if results are unsatisfactory, fine tuning of the model can be done to increase the performance of the model. An output model is established when the target evaluation metrics are met. This output model can be used to predict future observations while giving insights into the original dataset [113]:

$$\hat{x}_{standardisation} = \frac{x - \bar{x}}{\sigma} \tag{2.6}$$

$$\hat{x}_{normalisation} = \frac{x - \min(x)}{\max(x) - \min(x)} \tag{2.7}$$

$$\hat{x}_{robust} = \frac{x - \tilde{x}}{Q3 - Q1} \tag{2.8}$$

where $\bar{x}$ represents the mean value of the sample population, $\sigma$ is the standard deviation of the sample population, $\min(x)$ is the minimum value in the sample population, $\max(x)$ is the maximum value in the sample population, $\tilde{x}$ is the median value of the sample population, Q1 is the first quantile of the sample population, and Q3 is the third quantile of the sample population [113].

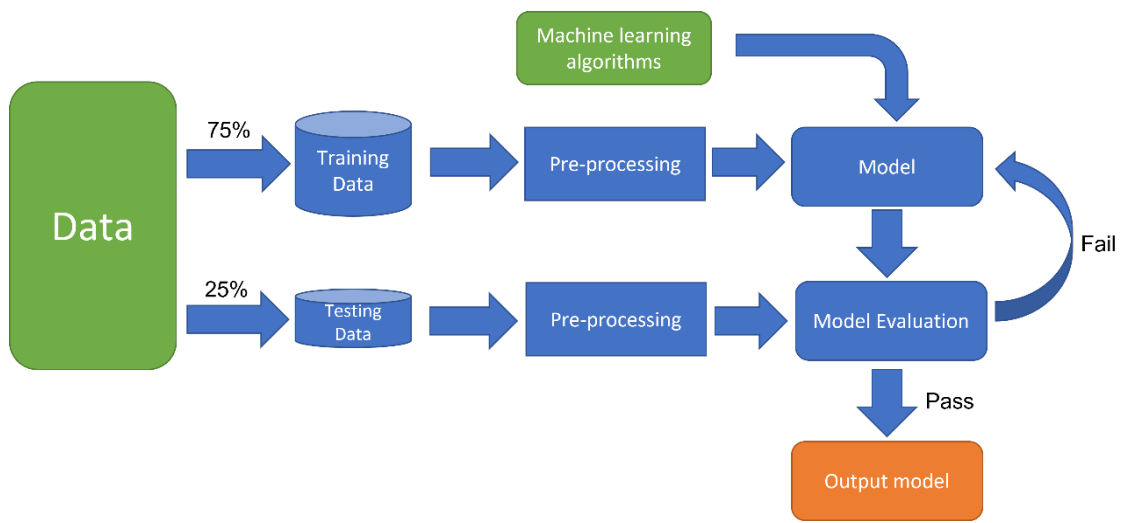

Figure 2.14 General framework of a supervised machine learning model development.

Classification problems are a subset of supervised learning models where the goal is to identify which categories a set of inputs belongs to. These can be binary (i.e., one of two classes) or multi-class in nature. In the case of binary classification, the model generates a decision boundary/surface to distinguish new observations into one of the two classes. These decision boundaries stem from the model which comprise of decision function, cost function and the optimization pathway for minimising the cost function [113]. The decision function uses the training dataset and a set of instructions usually mathematically to generate a decision boundary that classifies new observations into predicted classes. The cost function (also known as lost or error function) is the expected cost for classifying new observations in the model. The goal is to always minimise the cost function in order to minimise the total error within the model, hence, improve the performance of the model. Sometimes, instead of the cost function an objective function is used where the goal is to maximise the function to lower the error of the model. Regularization is sometimes used in conjunction with a cost function to avoid overfitting. Overfitting occurs when model overcompensating for each observation used to train the model causing the predictability of new observations to be poor [113].

2.5.3 Classification algorithms

Classification algorithms (also known as classifiers) are models which can be used to train dataset for classification problem. While there are many classifiers available, the main classifiers used in different programming environments are the logistic regressions, K-nearest neighbours, support vector machine, decision trees, and ensemble learning classifiers such as gradient boosting and random forest classifier [113,116]. Hence, these are the classification models selected to be used in this thesis as will be discussed further in Chapter 3.

Logistic regression is one of the two baseline models commonly used in classification problems (the other being linear regression for regression problem). The model is a probabilistic model based on the logistic function in the form of a sigmoid curve where the curve acts as the decision boundary for satisfying the condition to distinguish a binary classification variable. The logistic regression classifier is dependent on the decision function as shown in equation 2.9 [116].

$$f(x) = \frac{L}{1+e^{-k(x-x_0)}}$$

(2.9)

where $x_0$ is the value at the sigmoid midpoint position of the curve, L is the maximum value of the sigmoid curve, and k is the logistic growth rate of the curve.

K-nearest neighbour, more commonly known as KNN, is a classification model best suited for bivariate problems but does have the capacity to extend to multivariate problems albeit resulting in lesser degree of reliability. The k in KNN represents the number of neighbouring data points to be considered in the classifier. The model determines to which class a new observation belongs to by considering classes of the existing dataset with the closest distance calculated using the Minkowski distance formula as shown in equation 2.7 and assign classes based on the selected data points [116]. An example is shown in Figure 2.15 where if k=3 then the new data point would classify as class 1 since there are two existing data points from class 1 in relative closeness to the new data point as opposed to only 1 data point for class 2.

$$dist(x, y) = \left(\sum_{k=1}^{d} |x_r - y_r|^p \right)^{1/p} \tag{2.10}$$

Where (x,y) are the coordinates of the variable, $(x_r, y_r)$ is the coordinates of the closest variable, d is the distance between (x, y) and $(x_r, y_r)$, k is the number of variable used to determine class, and p is the order of the norm.

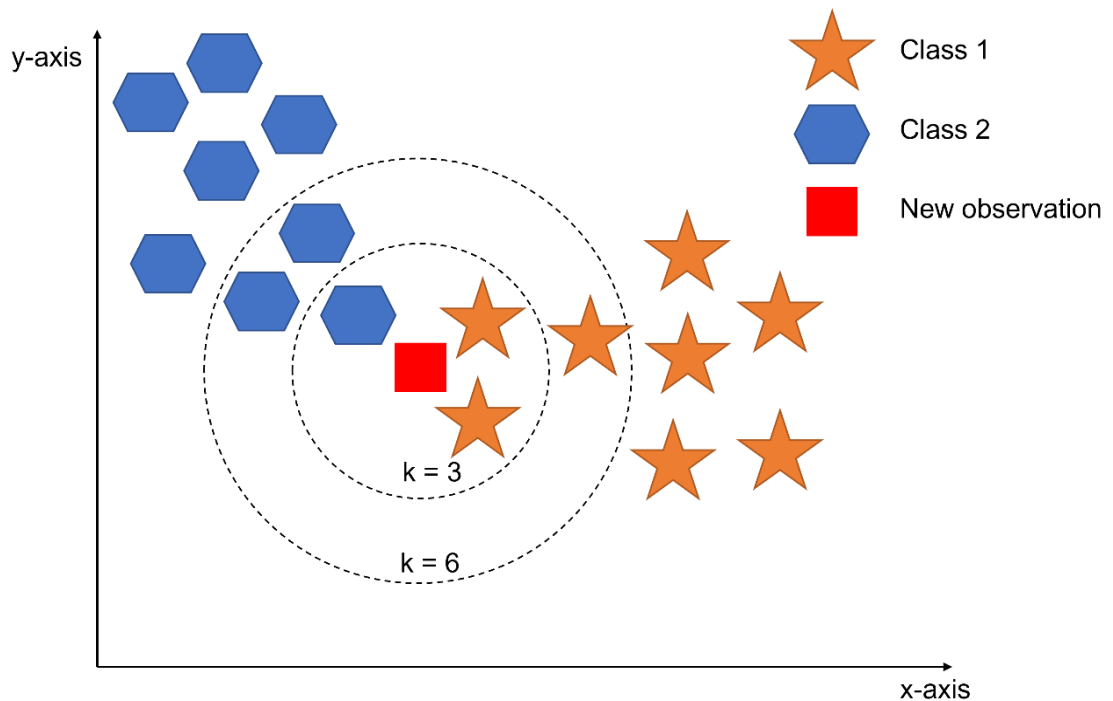

Figure 2.15 Illustration of a binary KNN classifier (reproduced from [117])

A support vector machine (SVM) is a type of classifier that can perform classification with multi-dimensionality. It uses kernel functions to reduce dimensionality and establish a hyperplane (i.e., decision boundary) that extends across feature space. This is unlike the previous two classifier (logistic regression and KNN) which are best suited for bivariate features. Another property of SVM is the small training sample needed to construct the hyperplane. The subset of training sample used is called the support vectors. The simplest kernel function is the linear function as shown in Equation 2.11 where the decision boundary is in the form of an optimal hyperplane. The goal of SVMs is to maximise the so-called margin between support vectors of different classes as shown in Figure. 2.16. This is done by calculating an optimal hyperplane in between two hyperplane generated by the support vectors. There are other kernel functions that leads to non-linear hyperplane. These includes polynomial, radial basis, and sigmoid kernel functions [118].

$$w^T x + b = 0 \tag{2.11}$$

Where x is a point on the hyperplane, w is the normal vector to the hyperplane and b is a bias term.

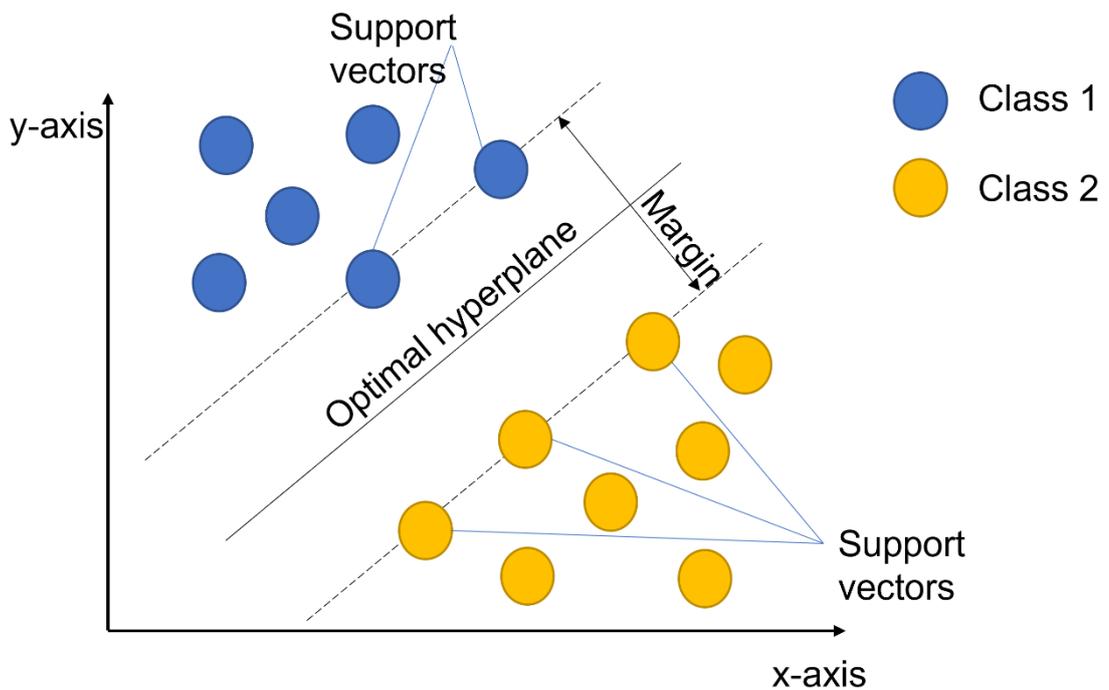

Figure 2.16 illustration of a binary class SVM classifier using linear kernel function (reproduced from [119])

Decision Trees as depicted in Figure 2.17 also known as classification and regression trees (CART for short) are a sequential, and hierarchal classifier comprised decision nodes and leaves. At the beginning node known as the root node, the whole dataset is used, and separation begins splitting the dataset into two subsets using an algorithm generated decision (e.g., is the feature above a certain value). At each sequential node, a decision is made using a randomly selected feature to further separate the dataset based on some form of purity index. Purity refers to the degree at which a data is moved into single classes. A 50/50 split between two classes is known as 100% impure while single class distribution is known as 100% pure. The decision trees always seek to maximise purity at each decision node. The Gini index as shown in Equation 2.12 is a commonly used index in CART for purity assessment of data at decision nodes. The final node is called leaf nodes where no more segregation of data can be made [118].

$$Gini\ index = \ 1 - \sum_{i=1}^{n}(P_i)^2 \qquad (2.12)$$

Where p is the probability of observations being labelled with class i and n is the number of features

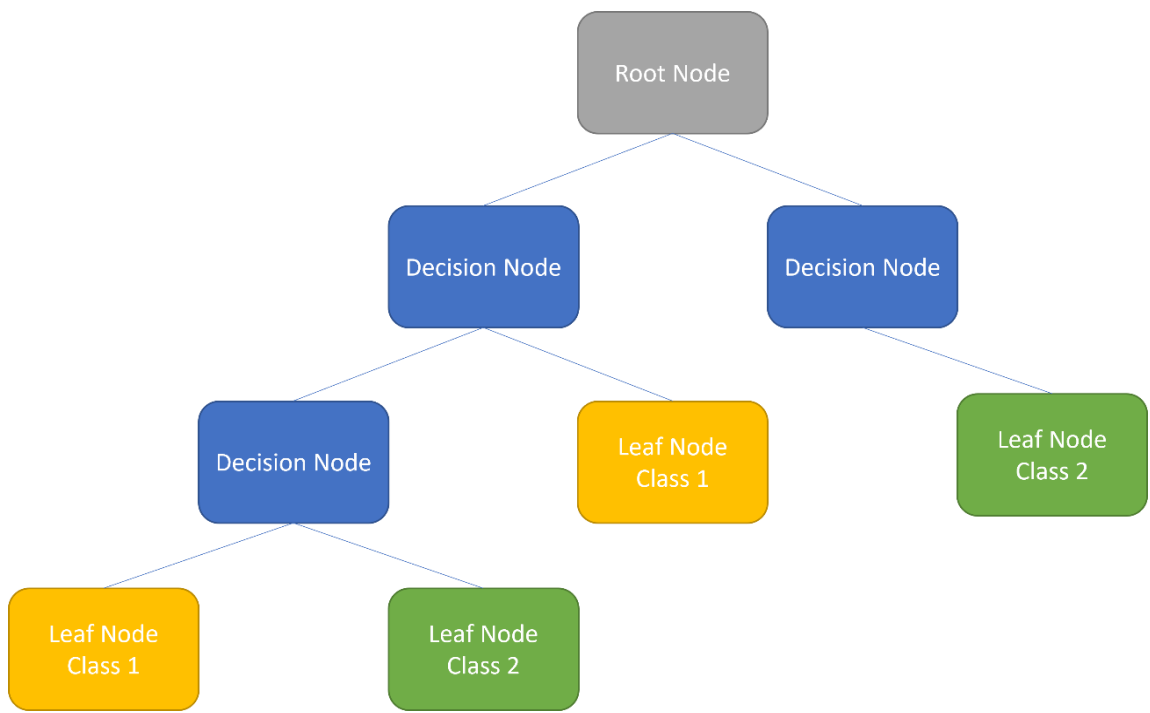

Figure 2.17 Illustration of a CART classifier for binary classification (reproduced from [120])

Ensemble learning algorithms are methods of using multiple estimators and combining the results of the individual estimators into a stronger more reliable output. There are two main types of ensemble learners for classification problem both of which utilises decision trees –boosting and bagging. The key differences of the two is highlighted in Figure 2.18. Boosting classifier align decision trees see Figure 2.18 into a workflow where the output of the individual decision trees is used as inputs for the decision tree in the next iteration. The general output of the model is determined by the last decision tree within the workflow. Bagging classifier combines the results of individual decision trees and averages out the output as a generalised model. A common boosting algorithm is the gradient boosting algorithm [121]. Random forest algorithm is widely used as a bagging algorithm for many applications [118].

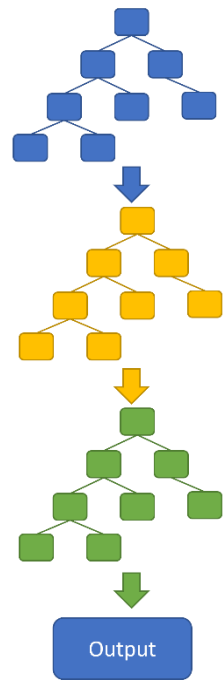
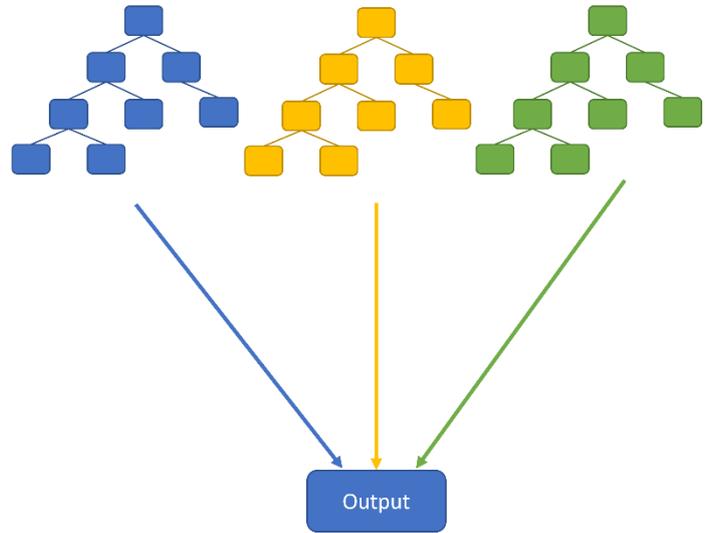

Figure 2.18 Illustrations of ensemble learning algorithms – boosting (left), bagging (right) reproduced from [120]

### 2.5.4 Evaluation metrics

Evaluation of a classifier's performance is an important element of assessing the predictability of the model. The basic metrics for evaluating the performance of a binary classification problem is the accuracy and misclassification error score. These metrics can be obtained via the construction of a confusion matrix as shown in Table 2.2. A true positive represents a match between the positive class (sometime also known as class 1) and the model prediction as positive class. True negative is vice versa with predicted negative class. As for the false positive and false negative these are incorrect prediction of data into 1 of the two classes.

Table 2.2 Typical 2x2 Confusion matrix for binary classification

|                | Predicted Positive | Predicted Negative |
|----------------|--------------------|--------------------|
| Positive class | TP                 | FN                 |
| Negative class | FP                 | TN                 |

The formulation for the accuracy and the misclassification error is as follows:

$$Accuracy = \frac{(TP+TN)}{TP+FN+TN+TP} \tag{2.13}$$

$$Error = \frac{(FP+FN)}{TP+FN+TN+TP} \tag{2.14}$$

Where TP is true positive, TN is true negative, FP is false positive, and FN is false negative.

However, in the case of class imbalance where the proportion of classes are not equal (e.g., high number of class 1 labelled data points, but very little class 2 labelled data points) these metrics are no longer representative of the performance of the classification model [122]. Hence, if one of the two classes is more important as is the case for many real-world applications the weighting needs to be skewed towards that part of the data. The F-measure provide a solution to this problem considering true positive scenarios to be more important to consider for the performance of the model [122]. The F-measure is the harmonic mean of precision and recall shown in equation 2.15 [123]. Hence, the F-measure will be used as a rigid metrics for assessing the performance of machine learning models discussed in Chapter 3.

$$`F_1 = 2 * \frac{precision * recall}{precision + recall} = \frac{TP}{TP + \frac{1}{2}(FP + FN)} \tag{2.15}$$

## 2.6 Research Objectives

To date no clear consensus has been reached for the exact mechanisms and beneficial properties of C and B microalloying additions during hot working of Ni-based superalloys. Hence, there is an interest to investigate the effects of C and B on René 41. This thesis aims to clarify the role of C and B microalloying additions in the hot working of René 41 via a multi-faceted approach. In a first section, the hot deformation behaviour of three René 41 variants with different amounts of C and B will be studied. Hot compression testing, light optical microscopy and scanning electron microscopy will be used to reveal the microstructure-processing-properties relationships. In a second section, machine learning models will be used to make informed predictions about the relationship between the bulk chemical composition of industrial ingots and their hot workability on an industrial scale. Here, an attempt will be made to correlate internal defect location via ultrasonic testing and bulk chemical composition data as provided from industrial partner WASA. The research objectives of this thesis project are as following:

- Characterisation of the initial microstructure of as received and homogenized René 41 billets. This includes grain size analysis using light optical microscopy.
- Lab-scale thermo-mechanical processing experiments in a thermomechanical processing simulator using temperatures of 1000, 1075, and 1150°C at strain rates of 0.01, 0.1, and $1s^{-1}$
- Flow stress analysis, finite element analysis and microstructural characterisation to compare hot deformation behaviour of different René 41 variants.
- Analysis of correlation between internal defects and alloy bulk chemical composition. Use machine learning models to make predictions about hot workability vs bulk chemical composition on an industrial scale.

# Chapter 3 – Experimental procedures and methodology

## 3.1 Material

The research objectives provided a rationale for investigating the influence of C and B as microalloying additions to the processability of René 41. Hot deformation behaviour is an important consideration to understand the processability of René 41 in the appropriate hot working window. Flow stress analysis, computation simulation and material characterisation methods as highlighted in section 2.4 provide ways to quantify the hot deformation behaviour of René 41. Machine learning models as highlighted in section 2.5 are used to determine linkage between chemical composition and manufacturing defects. Hence, the experimental procedures and methodology necessary to carried out the project is outlined in this section of the thesis.

Three René 41 billets with different contents of C and B microalloying additions were industrially fabricated by the industrial partner for this study. The chemical compositions collected via Inductively coupled plasma atomic emission spectroscopy (ICP-AES) are listed in wt.% in Table 3.1 with the nominal composition denoted as R41, the composition with C addition denoted as R41-C and the composition with B addition denoted as R41-B, respectively. As-received billets were produced via industrial cast & wrought processing. VIM produced ingots with a diameter of ~75 mm and weight of ~22 kg. A first homogenization step was carried out for 24 h at 1,200°C before hot rolling. Industrial hot rolling was carried out in a multi-step process. After hot rolling, billets were homogenized at 1225°C for 8 h, followed by furnace cooling to achieve comparable microstructures. Figure 3.1 depicts light optical microscopy (LOM) micrographs of the typical microstructure of the different René 41 variant billets consisting of coarse grain structure with an average grain size 200 ± 23 µm in diameter. The average grain size measurement was taken using the line intercept method as outlined in the ASTM E112 – 13. A total of 3 LOM micrographs with a dimension of 2 mm x 8 mm was used per variant. The final average grain size is the average of the three LOM micrographs. The ImageJ analysis software [124] was used to digitally process the LOM micrographs for ease of measurement. Another consideration was made to the grain size distribution of the as-homogenized microstructure. Volume fraction estimation of coarse grains was done to test the presence of bimodal grain distributions. Data was collected via a point count method using the ASTM E562 – 19. Both ASTM E930 – 18 ALA grain criteria and ASTM E1181 – 02 duplex grain criteria were used as considerations but were found to not qualify for both criteria.

Table 3.1 Chemical composition from ICP-AES of René 41 billets in wt.%

| Sample | Ni | Ti | Mo | Al | Co | Cr | C | B | Zr |
|---|---|---|---|---|---|---|---|---|---|
| R41 | Bal. | 3.21 | 9.28 | 1.60 | 10.96 | 18.10 | 0.023 | 0.008 | 0.012 |
| R41-C | Bal. | 3.23 | 9.55 | 1.57 | 11.13 | 18.33 | 0.067 | 0.008 | 0.005 |
| R41-B | Bal. | 3.24 | 9.56 | 1.57 | 11.21 | 18.13 | 0.035 | 0.014 | 0.005 |

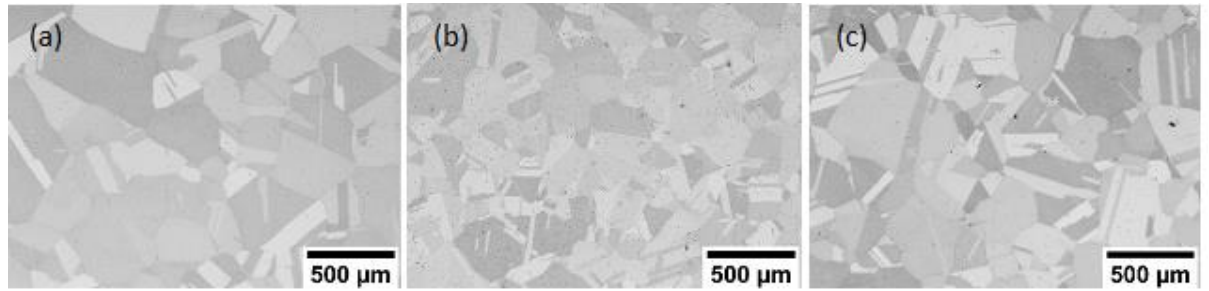

Figure 3.1 Typical optical microstructure of the as-homogenized René 41 billets (a) R41, (b) R41-C, and (c) R41-B

## 3.2 Thermomechanical processing

Sectioning of the thermomechanically processed samples was done using the Struers Labotom-3 cutting machine equipped with a SiC cutting disk. The billet sections were taken ~45 mm from the centre position then machined into cylindrical dimensions of 6 mm in diameter and 7.5 mm in height achieving a height to diameter ratio of 1.25. Single pass uniaxial compression testings were carried out using the Gleeble-3500 thermomechanical physical simulator in the set-up as shown in Figure 3.2. The Gleeble-3500 uses resistant heating coupled with a fully integrated hydraulic servo system to ensure accurate loading profile and rapid heating required by the test specification. WC anvils were used to ensure an isothermal thermomechanical schedule. A type-K (chromel-alumel) Ni-based thermocouple was spot welded onto the test specimen at the centre position for accurate temperature measurement during hot deformation. In addition, a graphite sheet was used as a high-temperature solid lubricant to reduce the friction between the contact surfaces of the hot compression test specimen and the WC anvil.

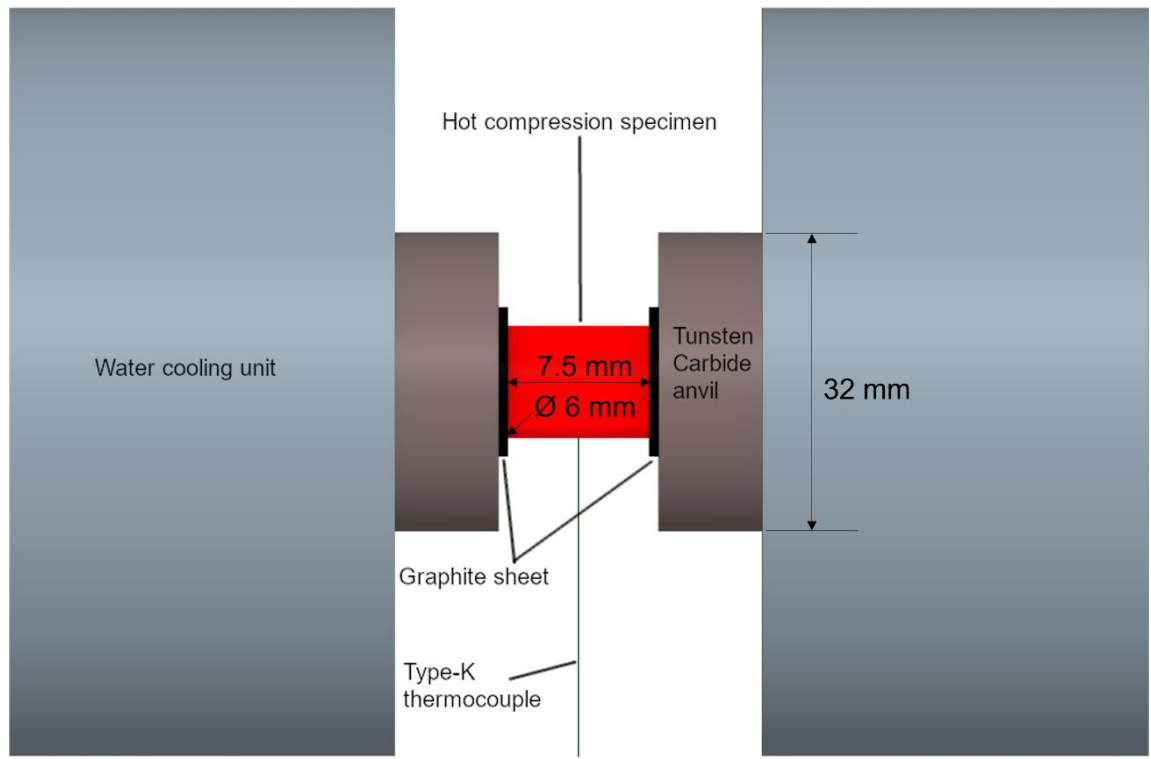

Figure 3.2 Schematics of the Gleeble-3500 thermomechanical processing set-up.

The thermomechanical schedule is depicted in Figure 3.3. The heating rate to deformation temperature $T_d$ was 15°Cs$^{-1}$ with a 180s soaking time to reach thermal equilibrium across the whole sample. The sample was then deformed to a 50% heigh reduction (corresponding to 0.69 logarithmic true strain). Deformation temperatures were 1000°C, 1075°C, and 1150°C, and strain rates were 0.01s⁻¹, 0.1s⁻¹, and 1s⁻¹. Following the deformation, samples were immediately water quenched (WQ) to retain the high temperature microstructure. For microstructural analysis, the deformed samples were sectioned mid-centre along the compression axis with the Struers Accutom-6 automatic cutting machine with a SiC cutting disk.

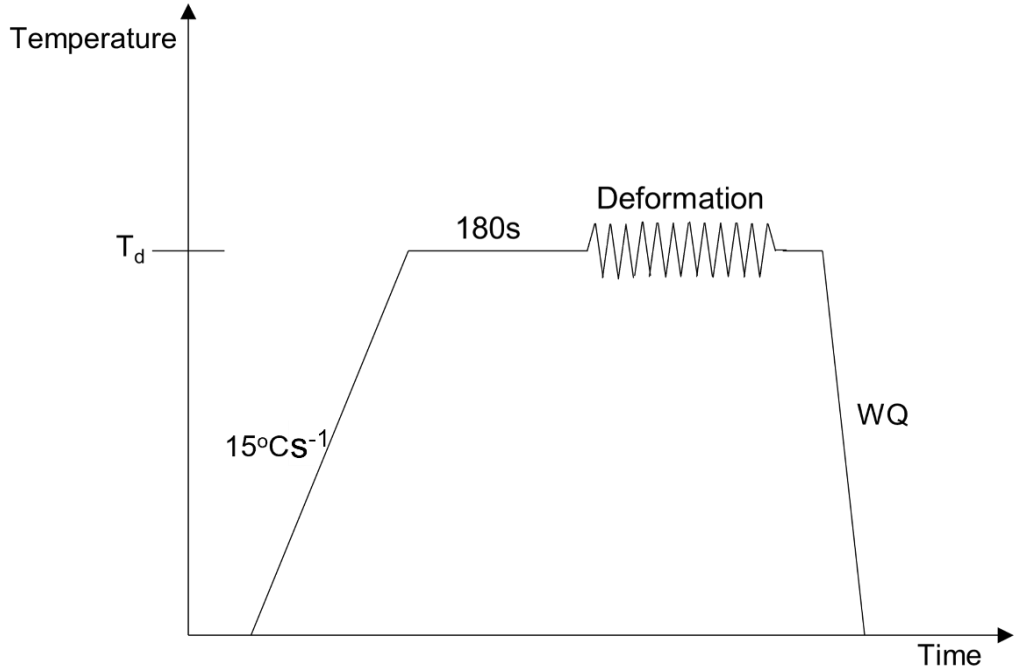

Figure 3.3 Schematics of the thermomechanical simulation schedule. $T_d$ represents deformation temperature.

The raw flow stress data was processed via the MATLAB software. The raw data was treated to a combination of locally weighted smoothing (LOESS) and Savitzky-Golay filter implementation available within MATLAB to reduce the overall noise level. The barrelling effect caused by friction was accounted for by adjusting the flow stress curves according to the constant volume solution provided by Avitzur [125] as given by,

$$\sigma_{FC} = \sigma \left(1 + \frac{2}{3\sqrt{3}} m \left(\frac{R_0}{H_0}\right) \exp\left(\frac{3\epsilon}{2}\right)\right) \tag{3.1}$$

where $\sigma_{FC}$ is the friction-corrected flow stress, $\sigma$ is the experimental flow stress, $m$ is the friction factor, $R_0$ and $H_0$ are the original radius and height of the specimen as shown in Figure 3.4, and $\epsilon$ the experimental strain. The friction factor $m$ was extracted from the experiment using the simplified friction factor $m$ equation by Ebrahimi and Najafizadeh [126] given by,

$$m = \frac{\frac{R}{H} b}{\left(\frac{4}{\sqrt{3}}\right) - \left(\frac{2b}{3\sqrt{3}}\right)} \tag{3.2}$$

where,

$$b = \frac{4\Delta R}{R}\left(\frac{H}{\Delta H}\right) \quad (3.3)$$

$$R = R_0 \sqrt{\frac{H_0}{H}} \quad (3.4)$$

$$\Delta R = R_M - R_{min} \quad (3.5)$$

$$\Delta H = H_0 - H \quad (3.6)$$

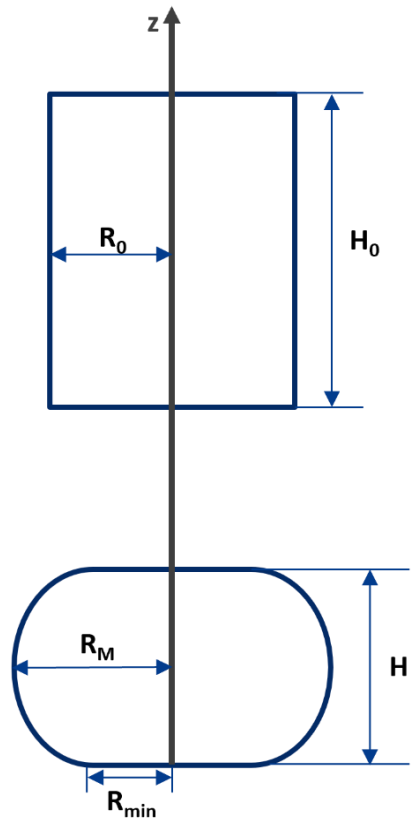

Figure 3.4 Schematics of the hot compression specimen geometry

The constitutive equation used to model the flow curves is the hyperbolic sine equation as shown in equation 2.1. The peak stress $\sigma_p$ was extracted from a given flow curve by characterising as the point where $\sigma_p = \sigma|_{\frac{d\sigma}{d\epsilon}=0}$ i.e., when the curve reaches its maxima.

The hyperbolic sine equation parameters were extracted via linear regression with regards to the following relationships as given by equation 3.8-3.12.

$$\alpha = \left(\frac{d\sigma_p}{dln\dot{\epsilon}}\right)^{-1} * \frac{dln\sigma_p}{dln\dot{\epsilon}} \qquad (3.8)$$

$$n = \frac{\partial ln\dot{\epsilon}}{\partial ln[sinh(\alpha\sigma_p)]}\bigg|_T \qquad (3.9)$$

$$Q = Rn * \frac{\partial ln[sinh(\alpha\sigma_p)]}{\partial\left(\frac{1}{T}\right)}\bigg|_{\dot{\epsilon}} \qquad (3.10)$$

$$A = \exp(lnZ - nln[sinh(\alpha\sigma_p)]) \qquad (3.11)$$

While thermal gradient variation was not assessed for the Gleeble samples, the controls in place during the Gleeble experiment ensured as close to isothermal conditions as possible. Firstly, the chamber was evacuated to low vacuum to minimise the effects of convectional thermal loss. Secondly, limited thermal conduction occurred through the graphite sheet and sample surface. Thirdly, the small geometry of the Gleeble samples allowed for uniform heating throughout the Gleeble experiment.

## 3.3 Light optical microscopy

LOM samples were hot mounted with resin. Table 3.2 listed the general procedure for LOM metallographic sample preparation by grinding and polishing. Grinding was performed using Struers Piano diamond grinding disks of the grid size 240, 600 and 1200 into a general-purpose grinding and polishing station Struers RotoPol-22 with a RotoForce-4 head. Ultrasonic cleaning was done to rid the metallographic sample surface of residuals after grinding, in between each polishing step, and on the final polished surface. Water was used as a lubricant for the grinding stages while polishing suspensions containing monocrystalline diamonds were used for 3 µm and 1 µm polishing steps along with lubrication from a mixture of ethanol and glycerol. Drying of samples was done using the Struers Drybox-2 for a duration of 5 min with a low heat setting.

Table 3.2 Grinding and polishing sequence for LOM samples

| Step | Grid Size | Force (N) | Wheel Speed (rpm) | Time (min) |
|---|---|---|---|---|
| 1 | 80 | 25 | 150 | 1.50 |
| 2 | 600 | 25 | 300 | 2.50 |
| 3 | 1200 | 25 | 300 | 3.50 |
| 4 | Ultrasonic cleaning in ethanol bath | | | |
| 5 | 3 µm | 20 | 300 | 5.00 |
| 6 | Sonication in ethanol bath | | | |
| 7 | 1 µm | 15 | 300 | 20.00 |
| 8 | Ultrasonic cleaning in ethanol bath | | | |

Etching of samples was performed using Kalling's reagent containing equal parts ethanol ($C_2H_5OH$), hydrochloric acid (HCl) and water ($H_2O$) with 1.5wt.% cupric chloride ($CuCl_2$). The etching time was between 5 to 30 seconds varying between individual samples. The samples were rinsed with water for approximately 5 minutes after etching to ensure all reagent was removed from the samples. LOM samples were observed using the Olympus BX53M light optical upright metallurgical microscope. Micrographs were captured with an attached camera and the OLYMPUS software.

## 3.4 Scanning electron microscopy

Table 3.3 listed the general procedure for scanning electron microscopy (SEM) metallographic sample preparation by grinding and polishing. SEM samples were prepared using the Struers Tegramin-20 automatic grinding and polishing station. Grinding steps were performed using the Struers MD-Piano grinding disk of the grid sizes 220, 1200 and 2000, respectively, with water as lubrication. Cleaning was done using the Struers Lavamin automatic cleaning apparatus after the grinding steps and in between polishing steps. Struers MD-Dac polishing pads were used in conjunction with 3 µm diamond suspensions. OP-U colloidal silica suspensions were used to further polish the samples after 3 µm polishing step. Vibratory polishing using the Buehler VibroMet2 was done as a final polishing step for the SEM samples. The samples were then removed from the resin and mounted onto SEM stubs with conductive silver paste to be examined.

Table 3.3 Grinding and polishing sequence for SEM samples

| Step | Grid Size | Force (N) | Wheel Speed (rpm) | Time (min) |
|---|---|---|---|---|
| 1 | 220 | 25 | 300 | 2.00 |
| 2 | 1200 | 20 | 300 | 2.00 |
| 3 | 2000 | 20 | 300 | 5.00 |
| 4 | Ultrasonic cleaning | | | |
| 5 | 3 µm | 15 | 300 | 30.00 |
| 6 | Ultrasonic cleaning | | | |
| 7 | 0.04 µm | 10 | 300 | 5.00 |
| 8 | Ultrasonic cleaning | | | |
| 9 | Vibratory polishing, Amplitude = 30%, Time = 16 h | | | |
| 10 | Ultrasonic cleaning and rinsed with iso-propanol | | | |

Electron microscopy was done using the JEOL7001f FE-SEM with an accelerating voltage of 20 kV and a probe current of size '13'. The working distance was selected to be 10 mm. Electron back-scattered diffraction (EBSD) maps were collected using the same microscope with an accelerating voltage of 20 kV and an electron probe current of '13'. Samples were tilted to 70° and an EDAX Hikari Super system with the EDAX Team data acquisition software was used. Three areas of 500 x 500 µm$^2$ each, with a step size of 0.5 µm were acquired from the sample centres. Camera parameters used were 650 in camera gain, 3.50 in camera exposure, and a range between 0.1-10 ms, 4x4 binning size, and background subtraction was also used. Additional high resolution EBSD maps were collected for the 1000°C samples to resolves features smaller than 0.5 µm. The high resolution EBSD maps have a total scan area of 30 x 30 µm$^2$ with a step size of 0.06 µm. Camera parameters for collecting the high-resolution maps were the same as above. Analysis of inverse pole figures (IPFs) maps was done via the TSL OIM analysis 8 software. The IPF maps were cleaned using grain dilation of less than 5 pixels and neighbouring confidence index correlation of less than 0.1 as removal condition for badly indexed points. The total removed pixels per scan was estimated to be around 2% which will have minimal effects on quantitative data analysis of recrystallisation characteristics. EBSD data partitions were used to separate non-recrystallised grains and recrystallised grains within each map. The grain orientation spread (GOS) was used as the criteria for separating recrystallised and non-recrystallised grain. Grains with GOS less than 2° were considered as recrystallised grains and grains with GOS more than 2° as non-recrystallised grains. Figure 3.5 displays an example of the visual representation of the filtred grains. The black portion represents the non-recrystallised region. Geometrically necessary dislocation (GND) density was also used as quantifiable information regarding the microstructure of the René 41 variants.

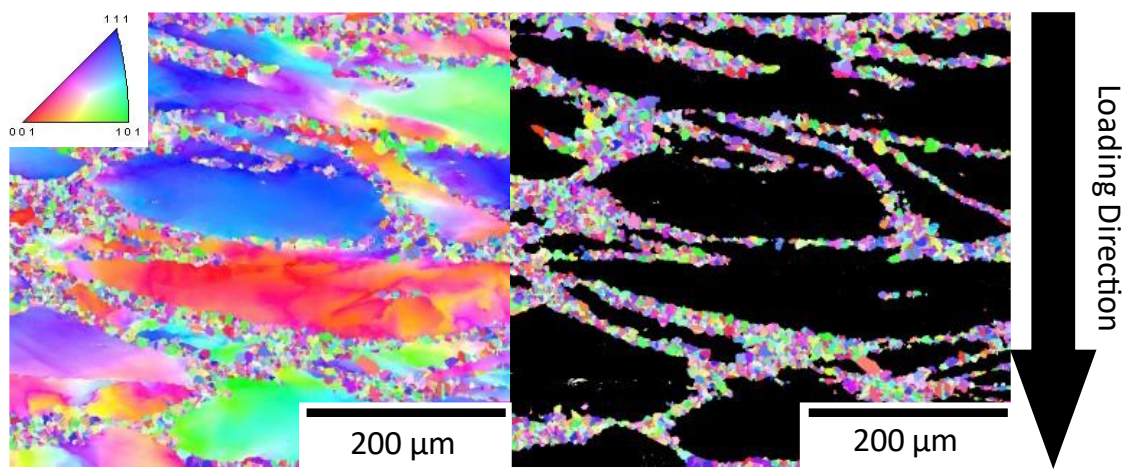

Figure 3.5 IPF maps with orientations parallel to the out-of-plane direction of the deformed René41 before (left) and after (right) excluding grains with GOS > 2°. Loading direction is shown on the left of the micrograph.

## 3.5 Finite element analysis model description

Computational simulations of the thermomechanical processing experiments were generated with general purpose finite element method (FEM) modelling software LS-DYNA® from Livermore Software Technology Corporation. LS-DYNA® uses time-explicit integration for FEM modelling. All setup for the modelling was done using LS-PrePost® version 4.8. The solid element formulation used for the model is a constant stress state 8-nodal hexahedral model to avoid volumetric locking and reduce computational power[127]. The initial state and meshing condition for the FEM simulation are shown in Figure 3.6. There are approximately a total of 30,000 nodal elements used for the simulation. Hourglass mode control was implemented in LS-DYNA [127] using the simplified Flanagan-Belytschko method [128] with an hourglass coefficient of 0.03. Critical time step sizes were set to no less than $10^{-7}$ s to reduce the computational power needed without compromising the accuracy of the model.

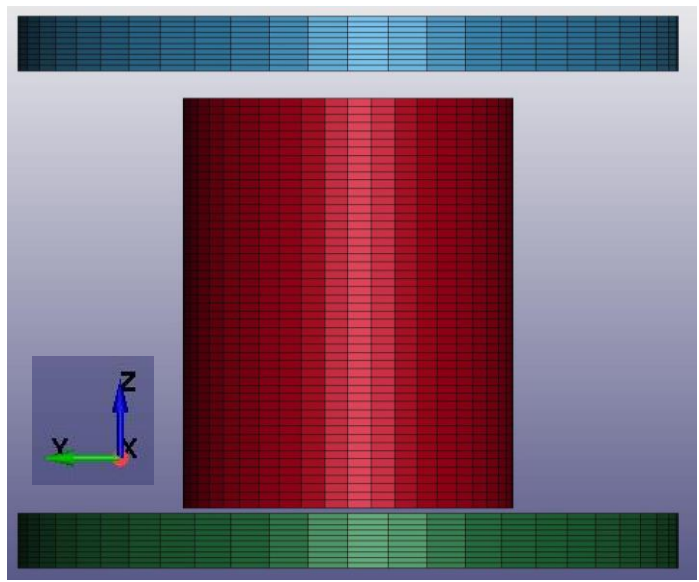

Figure 3.6 Meshing condition of FEM model.

The loading condition of the FEM model is set up to 1075°C at a true strain rate of $0.1s^{-1}$. Thermal condition is modelled to be isothermal with no heat transfer mechanisms via both thermal convection and thermal radiation. The top and bottom anvils are modelled as rigid elastic material with the built-in material model 'MAT_001' and material properties from WC as shown in Table 3.4.

The materials model chosen for the specimen was the 'MAT_102_T' model which is a close approximation of the hyperbolic sine equation used in the flow-stress analysis for the physically derived hot compression testing results. Heat generation due to adiabatic heating is estimated to be 90% of the plastic work generated during the hot deformation and is accounted for within the material model [127].

The implemented hyperbolic sine flow stress equation in LS-PrePost® [127] is given by,

$$\sigma = \frac{1}{\alpha} \sinh^{-1}\left(\left[\frac{Z}{A}\right]^{\frac{1}{n}}\right) \tag{3.13}$$

The equation parameters used in the materials model were extracted from the constitutive modelling of the flow stress data as mentioned above in equations 3.7 – 3.12. The LS-DYNA® contact algorithm AUTOMATIC_SURFACE_TO_SURFACE was used to model the dynamic responses between the specimen surface and the die. Friction implementation in LS-DYNA® [127] is a Coulomb based formulation given by,

$$F_y = \mu |f_n| \tag{3.14}$$

where $F_y$ is the yield force, $\mu$ is the coefficient of friction, and $f_n$ is the normal force.

The dynamic coefficient of friction has been assumed to equate the static coefficient of friction to reduce the complexity of the model. The coefficient of friction is determined using Bay's [129] slip line method friction factor conversion with the equation given by,

$$\mu = \frac{m}{\sqrt{27(1-m^2)}} \tag{3.15}$$

where m is the friction factor of the deformation and can be extracted from measuring the extent of barrelling effect occurred during the hot uniaxial compression testings.

Table 3.4 Material parameters chosen for the FEM model

|  | Specimen | Anvil |
|---|---|---|
|  | René41 | WC |
| Mechanical properties at 1075°C |  |  |
|    Young's modulus (MPa) | 1077 | Rigid |
|    Poisson's ratio | 0.31 | Rigid |
|    Flow stress | See Chapter 4 | Rigid |
| Thermal properties |  |  |
|    Thermal expansion coefficient ($10^{-6}$/°C) | 17 | 4.5 |
|    Thermal conductivity (W/m·K) | 29 | 28 |
|    Heat capacity (J/kg·K) | 870 | 280 |

## 3.6 Thermodynamic simulation

Thermodynamic simulation via the CALPHAD (calculation of phase diagrams) software package MatCalc. The MatCalc simulation uses the methodology as shown in Figure 3.8 based on the Scheil-Gulliver method [130] to calculate the local equilibrium of solute concentration in typical René 41 composition. The fluctuation of solute concentration in the solid phase (i.e., γ-matrix) was used qualitatively to describe the macro-segregation behaviour of different alloying elements in René 41. The elements considered in the MatCalc simulation includes major alloying elements such as Al, Co, Cr, Fe, Mo, and Ti as well as minor alloying elements such as B, C, N, Si, and Zr.

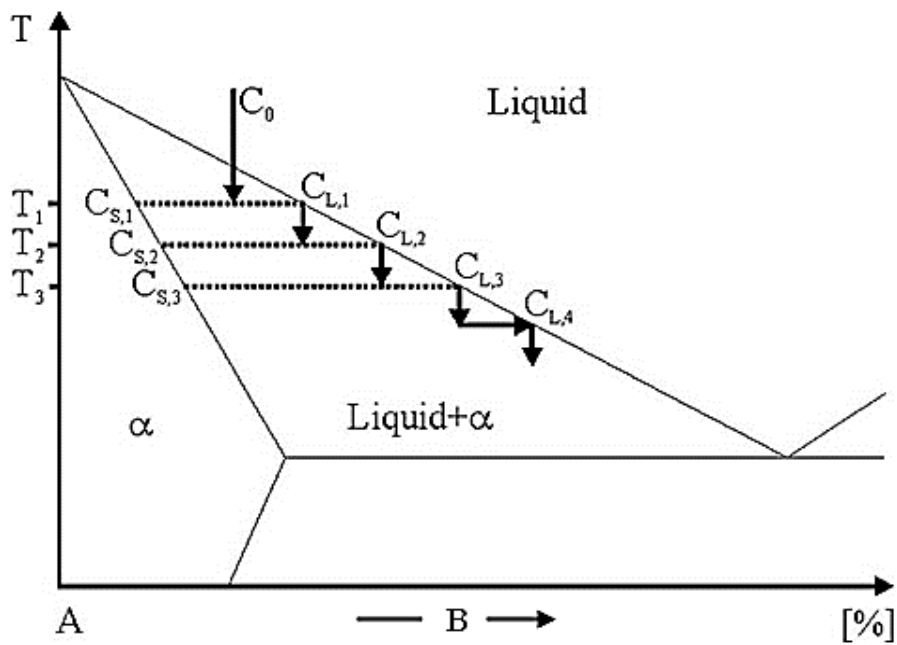

Figure 3.7 Schematic diagram of the implementation of Scheil equations via incremental changes of local solute equilibrium concentration during liquid to solid phase transformation using the lever rule [131].

## 3.7 Machine learning models

The dataset provided by WASA consists of chemical compositional data of industrially fabricated René 41 billets in accordance with industrial standard UNS N07041 as feature space. Ultrasonic testing (UT) was done in accordance with ISO 9704 level 2 personnel and encoded as binary class information (i.e., yes or no) along with the features as labels. Contact-based UT was performed using transducers in an agreement with flat bottom hole configuration. The UT defect sizes are typically around 1-5 mm in the largest UT signal orientation. A subset of the dataset consisting of ultrasonic defect locations within the billets were used to prepare data analysis on the position of UT defects within the industrially fabricated billets to give insights into the origin of the UT defects via processing route. Figure 3.8 display the indication of unacceptable defects from UT as a function of radial and longitudinal positions. These positions are compiled and stored into a dataset as fractions in the respective directions to synchronise billets with different dimensions into a uniform dataset as shown in Figure 3.9. A kernel density estimation (KDE) function was then used to construct a visual representation of the probability density of UT indication within the aggregate billets.

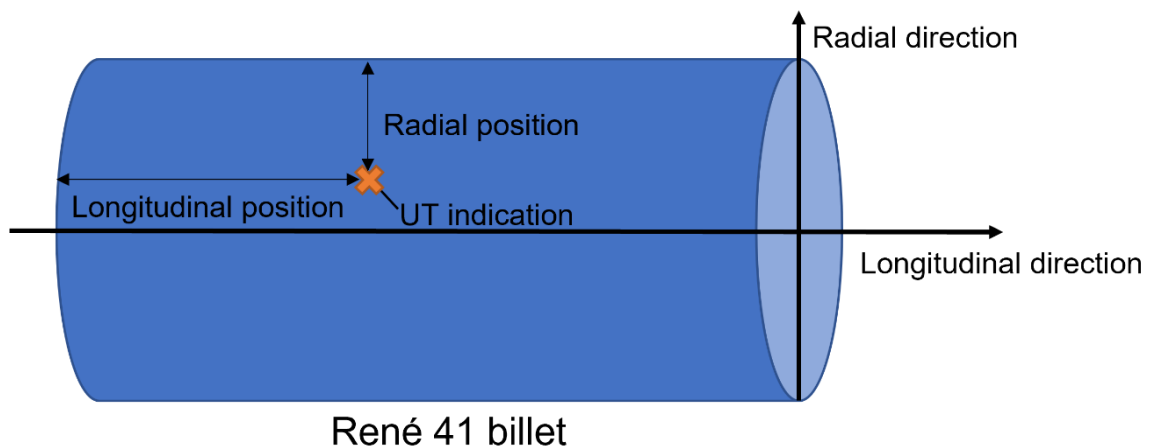

Figure 3.8 Illustration of UT indication location based on radial and longitudinal positions

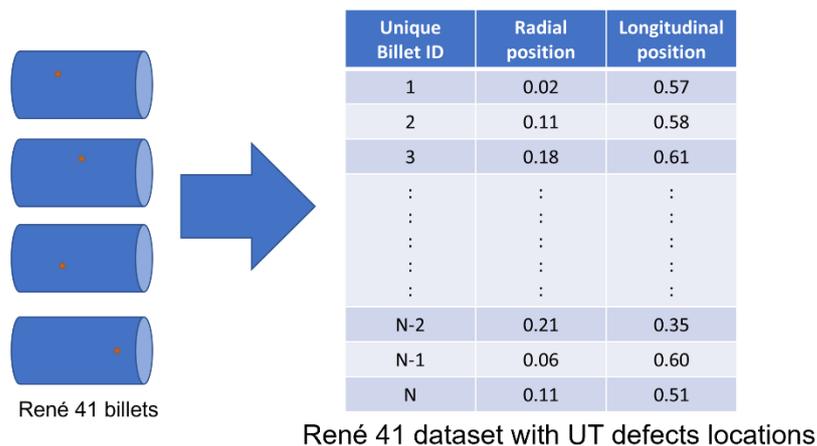

René 41 billets → René 41 dataset with UT defects locations

Figure 3.9 Workflow of data acquisition for UT defect locations as fractions in the radial and longitudinal directions

The programming environment used for the entirety of data processing, training, and testing of machine learning models was done in the Python 3.7.3 programming language. The Scikit-learning module was used for the purpose of data pre-processing, model selection and producing train-test dataset. The Pandas module was used to handle data structures. Statistical analysis was done using SciPy module within Python. The Seaborn module was used to construct the KDE plot.

A data clean-up operation was conducted onto the dataset as ML algorithms are sensitive to incomplete data entries. The data clean-up operation is to be summarised as follows: Individual features were linked together via unique chemistry identifications into a singular data structure; entries with missing data are removed; data entries with chemical composition outside of industrial standard specifications are removed; and duplicates are reduced to single entry. These operation were carried using python Pandas and NumPy modules. Approximately 75% of all entries were eliminated due to the clean-up procedures with the end results of 76 unique entries. Data imputation was not applied to the missing entries due to the sensitivity of the dataset to changes with anomaly detection.

Feature selection is used to drastically limit the number of features into the elements of the most relevance. A manual approach was taken to reduce the number of features. Chemical elements not specified in the industrial standard UNS N07041 were eliminated from the datasets to increase interpretability of the model and avoid overfitting. 16 features remained as part of the dataset. Several feature scaling techniques available by Scikit-learn have been applied and compared for the dataset namely the standardisation scaler, normalisation scaler, and robust scaler. The robust scaler was found to be the most suitable scaler for our datasets. To ensure that no information was leaked from the test split into the train dataset, the fit_transform method was applied in the training dataset and then the transform method separately in the test dataset. A 75/25 train-test split was performed onto the dataset with stratification to ensure the ratio of classes remain consistent between train and test sets.

ML algorithms used for this study as highlighted in section 2.5 are the most used classification models including the KNN, SVC, logistic regression, random forest ensemble, gradient boost ensemble and CART. All models were applied via the Scikit-learn pipeline construction. Each model variation is fitted through hyperparameter turning to allow for the optimization of models via the Scikit-learn GridSearchCV method.

Feature importance were implemented via random forest classifier and SVC classifier for further analysis of each individual feature. Evaluation metrics for the performance of ML models includes the F-measure and confusion matrices. Matplotlib module in Python and Origin Lab were primarily used for data visualisation.

# Chapter 4 – Results

## 4.1 Flow stress behaviour of hot deformed René 41 variants

Series of hot compression testings were carried out to assess the hot deformation behaviour of the three René 41 variants. The 27 raw stress-strain curves were treated to the pre-processing as outlined in section 3.2. Table 4.1-4.3 displays the peak stresses of R41, R41-C and R41-B, respectively in the different deformation conditions that were tested. Figure 4.1 depicts the friction corrected true stress-strain curves of the three René 41 variants (R41, R41-C, and R41-B) with a true strain of 0.69. Stress-strain curves with variations in temperatures are displayed in subplots with the order of increasing strain rate. The stress-strain curves follow a general trend of DRX flow curves as shown in Figure 2.6 where the flow stress increases until a peak stress $\sigma_p$ is reached. Samples with the lowest temperature (i.e., 1000°C) and highest strain rate (i.e., 1 s$^{-1}$) were observed to have the highest flow stress out of the nine given deformation parameter combinations with a peak stress of 530 MPa, 530 MPa, and 521 MPa for R41, R41-C and R41-B, respectively.

On the other hand, the lowest flow stresses recorded out of the given deformation parameters were achieved at the highest temperature (i.e., 1150°C) and the lowest strain rate (0.01s$^{-1}$) with a peak stress of 79 MPa, 81 MPa, and 78 MPa for R41, R41-C, and R41-B, respectively. This remains true for all three René 41 variants. There is a noticeable difference between the 1000°C stress-strain curves as compared to both the 1075°C and 1150°C curves for all strain rates. The difference in flow stresses between 1000°C and 1075°C samples are roughly twice the difference between 1075°C and 1150°C samples. Using R41 as an example, for a constant strain rate of 0.01 s$^{-1}$ there is a 68% increase in peak stress from 1150°C to 1075°C, whereas the peak stress from 1075°C to 1000°C saw an increase of 133%. The same trend can be observed in both 0.1 s$^{-1}$ (56% → 115%) and 1 s$^{-1}$ (47% → 81%) for the increase in peak stresses from 1150°C to 1075°C and 1075°C to 1000°C, respectively. This indicates that the impact of temperature on flow stresses are non-linear where the 75°C interval between 1000°C and 1075°C experiences a more dramatic decrease in flow stress than the 75°C interval between 1075°C and 1150°C.

The final strain level in the Gleeble experiments was insufficient to reach steady state flow stress consistently. This complicates a meaningful comparison of critical strain following the approach by Poliak and Jonas [132], and was therefore omitted.

Table 4.1 Table of peak stresses (in MPa) of R41 sample in difference deformation conditions.

| R41 | | Temperature | | |
|---|---|---|---|---|
| | | 1000°C | 1075°C | 1150°C |
| Strain Rate | 0.01s$^{-1}$ | 310 (MPa) | 132 (MPa) | 79 (MPa) |
| | 0.1s$^{-1}$ | 441 (MPa) | 205 (MPa) | 131 (MPa) |
| | 1s$^{-1}$ | 530 (MPa) | 293 (MPa) | 200 (MPa) |

Table 4.2 Table of peak stresses (in MPa) of R41-C sample in different deformation conditions.

| R41-C | | Temperature | | |
|---|---|---|---|---|
| | | 1000°C | 1075°C | 1150°C |
| Strain Rate | 0.01s$^{-1}$ | 307 (MPa) | 128 (MPa) | 81 (MPa) |
| | 0.1s$^{-1}$ | 408 (MPa) | 198 (MPa) | 133 (MPa) |
| | 1s$^{-1}$ | 530 (MPa) | 301 (MPa) | 207 (MPa) |

Table 4.3 Table of peak stresses (in MPa) of R41-B sample in different deformation conditions.

| R41-B | | Temperature | | |
|---|---|---|---|---|
| | | 1000°C | 1075°C | 1150°C |
| Strain Rate | 0.01s$^{-1}$ | 300 (MPa) | 117 (MPa) | 78 (MPa) |
| | 0.1s$^{-1}$ | 434 (MPa) | 194 (MPa) | 129 (MPa) |
| | 1s$^{-1}$ | 521 (MPa) | 300 (MPa) | 189 (MPa) |

Figure 4.2a-c depicts the peak stresses of the three René 41 variants at a given temperature as a function of log strain rates. The peak stresses of the three René 41 variants follows a general trend of decreasing for a given strain rate as a function of increasing temperatures. The overall peak stress level of René 41 variants at a given deformation condition is similar to each other. The deformation condition with the largest difference between the three René 41 variants occurs at 1000°C and 0.1 s$^{-1}$ as shown in Figure 4.4a with a value of 33 MPa. The deformation condition with the smallest difference between the peak stresses of the three René 41 variants occurs at 1150°C and 0.01 s$^{-1}$ as shown in Figure 4.4c with a value of 1 MPa.

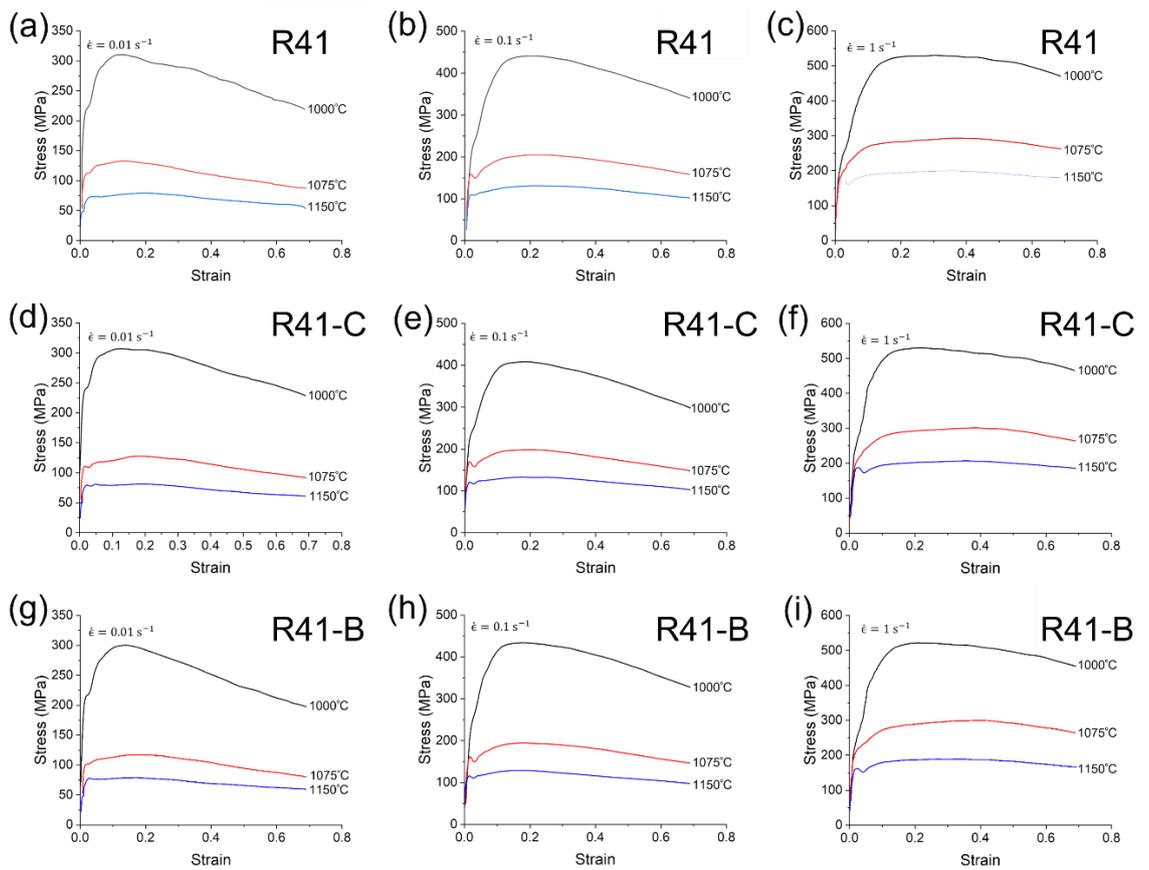

Figure 4.1 Friction corrected stress-strain curves of the three hot deformed R41 variants samples for different temperatures with (a) 0.01s$^{-1}$, (b) 0.1s$^{-1}$, and (c) 1s$^{-1}$ strain rates for R41, (d) 0.01s$^{-1}$, 0.1s$^{-1}$, and (f) 1s$^{-1}$ strain rates for R41-C, (g) 0.01s$^{-1}$, (h) 0.1s$^{-1}$, and (i) 1s$^{-1}$ strain rates for R41-B.

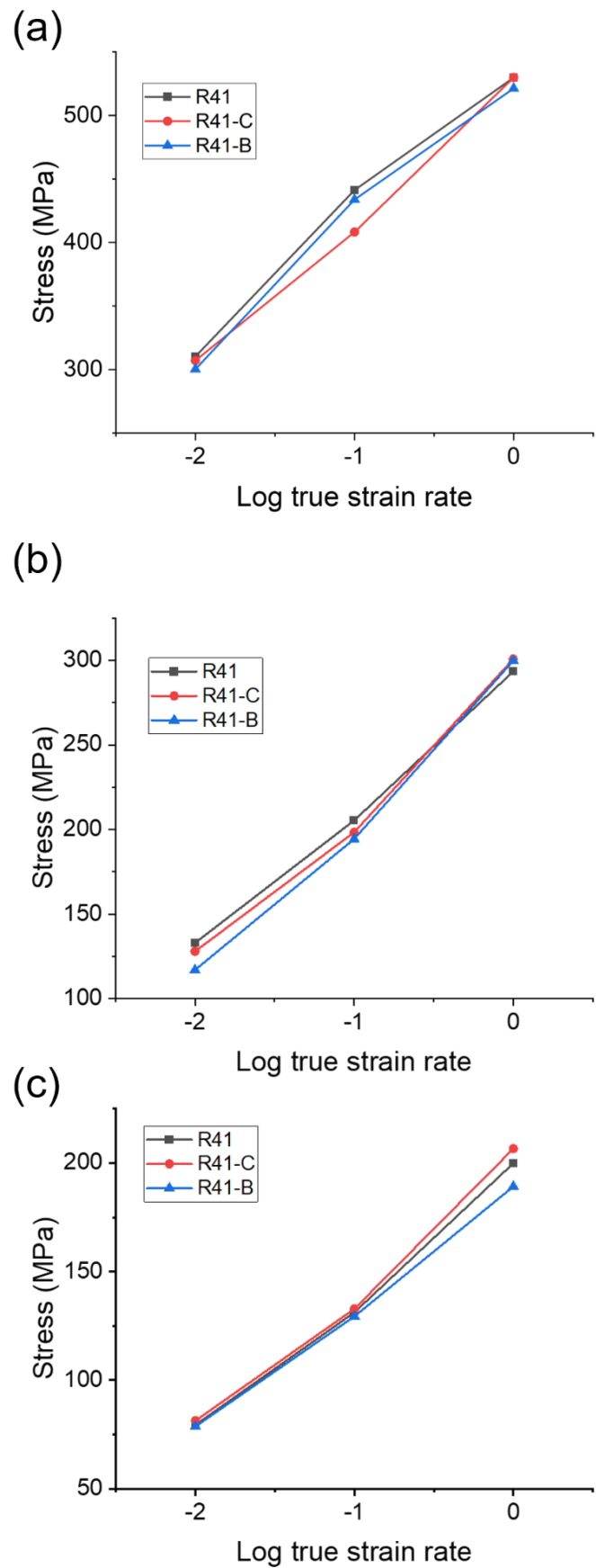

Figure 4.2 Peak stresses of the three René 41 variants with (a) 1000°C (b) 1075°C (c) 1150°C as a function of log true strain rate.

Consitutive equation constructions were carried out to provide direct comparison of hot deformation behaviour between René 41 variants and other Ni-based superalloys. Figure 4.3a-d shows with R41 as an example a series of relationships used in order to extract the hyperbolic sine equation parameters needed to construct the equation with methodology as described in previous section 3.2. Specifically, Figure 4.3a and 4.3b were used to extract the material constant $\alpha$ using the relationship in equation 3.8. Figure 4.3c and 4.3d were used to extract the stress exponent $n$ and activation energy $Q$ using the relationship in equations 3.9 and 3.10, respectively. The last material constant $A$ is extracted via Figure 4.4 using the relationship in equation 3.11. While the other René 41 variants namely R41-C and R41-B are not shown explicitly in Figure 4.3, the same procedure is used to construct the hyperbolic sine equation. Equation 4.1-4.3 are the resultant hyperbolic sine equation for R41, R41-C and R41-B, respectively. It can be noted that the stress exponent $n$ ranges from 4.25 for R41-B to 4.46 for R41 with R41-C in between the two.

Another general comparison between the René 41 variants is that the R41-C sample possesses the lowest activation energy for deformation $Q$ of 697 kJmol$^{-1}$ amongst the three variants. Equations 4.1-4.3 represent the hyperbolic sine equation for R41, R41-C and R41-B respectively.

$$\dot{\epsilon}_{R41} = (1.37 * 10^{28})[\sinh(0.0048\sigma_p)]^{4.46} * \exp\left(-\frac{757*10^3}{RT}\right) \quad (4.1)$$

$$\dot{\epsilon}_{R41-C} = (4.36 * 10^{25})[\sinh(0.0045\sigma_p)]^{4.33} * \exp(-\frac{697*10^3}{RT}) \quad (4.2)$$

$$\dot{\epsilon}_{R41-B} = (6.39 * 10^{26})[\sinh(0.0046\sigma_p)]^{4.25} * \exp(-\frac{728*10^3}{RT}) \quad (4.3)$$

FEM simulation models using experimentally derived constitutive models were established via LS-DYNA to gauge the plastic strain distribution within the René 41 samples. Figure 4.5a, b, and c show the different FEM simulations of the effective plastic strain distributions at the cross-sections of the three René variants using the method highlighted in previous section 3.4. The strain distribution is concentrated at the centre region with plastic strain values of 1.25, 1.30 and 1.245 for the R41, R41-C, and R41-B samples, respectively. The top and bottom of the sample cross-sections experienced the least amount of plastic strain with a value of 0.022 strain for all three René 41 samples. Another region with higher plastic strain is located at the four corners of the samples in an oval shape extending into the cross-section in a 'X' pattern. No noticeable difference can be discerned for the three René 41 samples, other than that R41 have slightly higher overall plastic strain level than compared with both R41-C and R41-B.

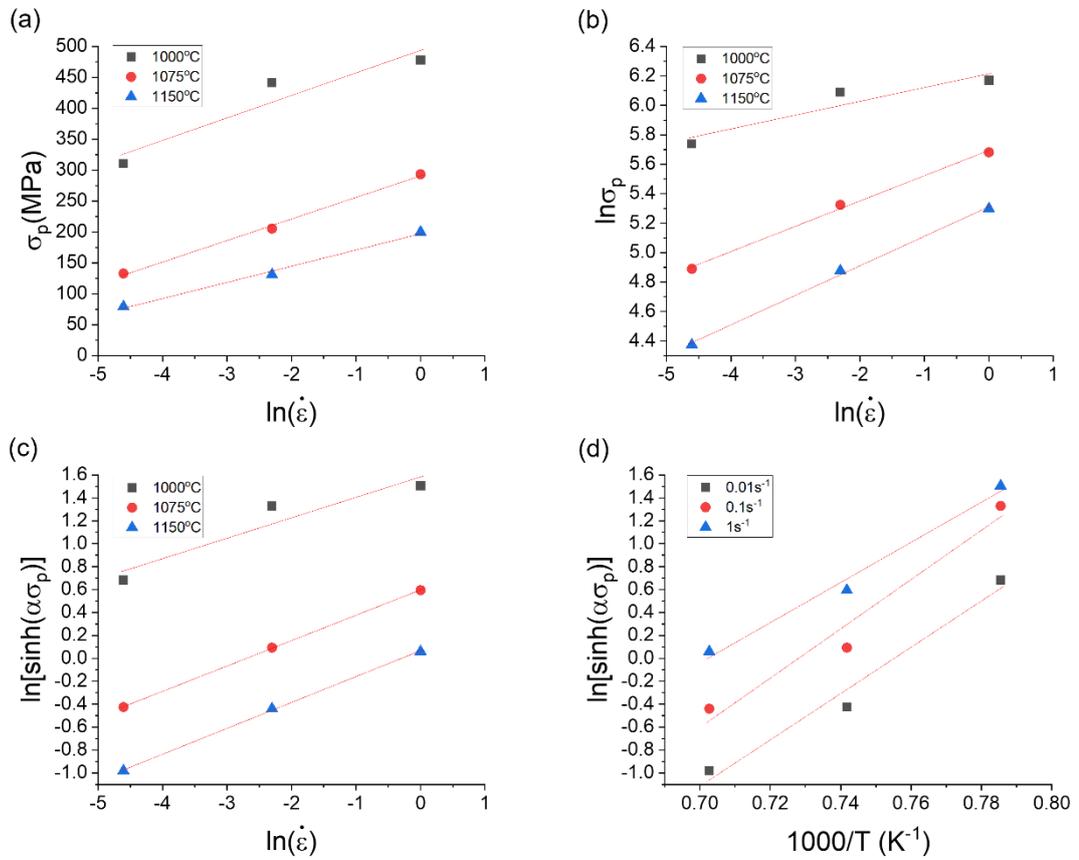

Figure 4.3 Relationships between peak stress derived parameters used for construction of hyperbolic sine equation using R41 samples (a) $\sigma_p - ln\dot{\varepsilon}$ ,(b) $ln\sigma_p - ln\dot{\varepsilon}$ ,(c) $ln[sinh\,(\alpha\sigma_p)] - ln\dot{\varepsilon}$ ,(d) $ln[sinh\,(\alpha\sigma_p)] - 1000T^{-1}$

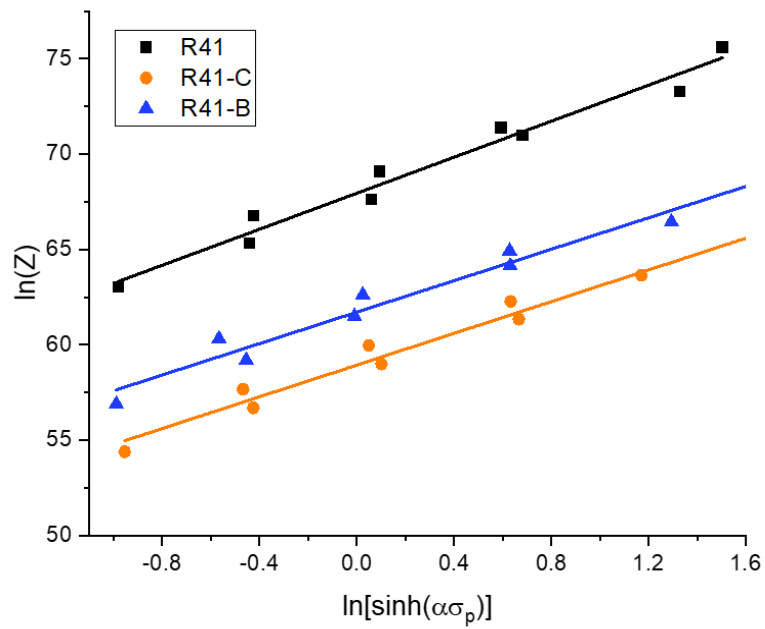

Figure 4.4 $lnZ - ln\,[sinh\,(\alpha\sigma_p)]$ relationship of the three René 41 variants

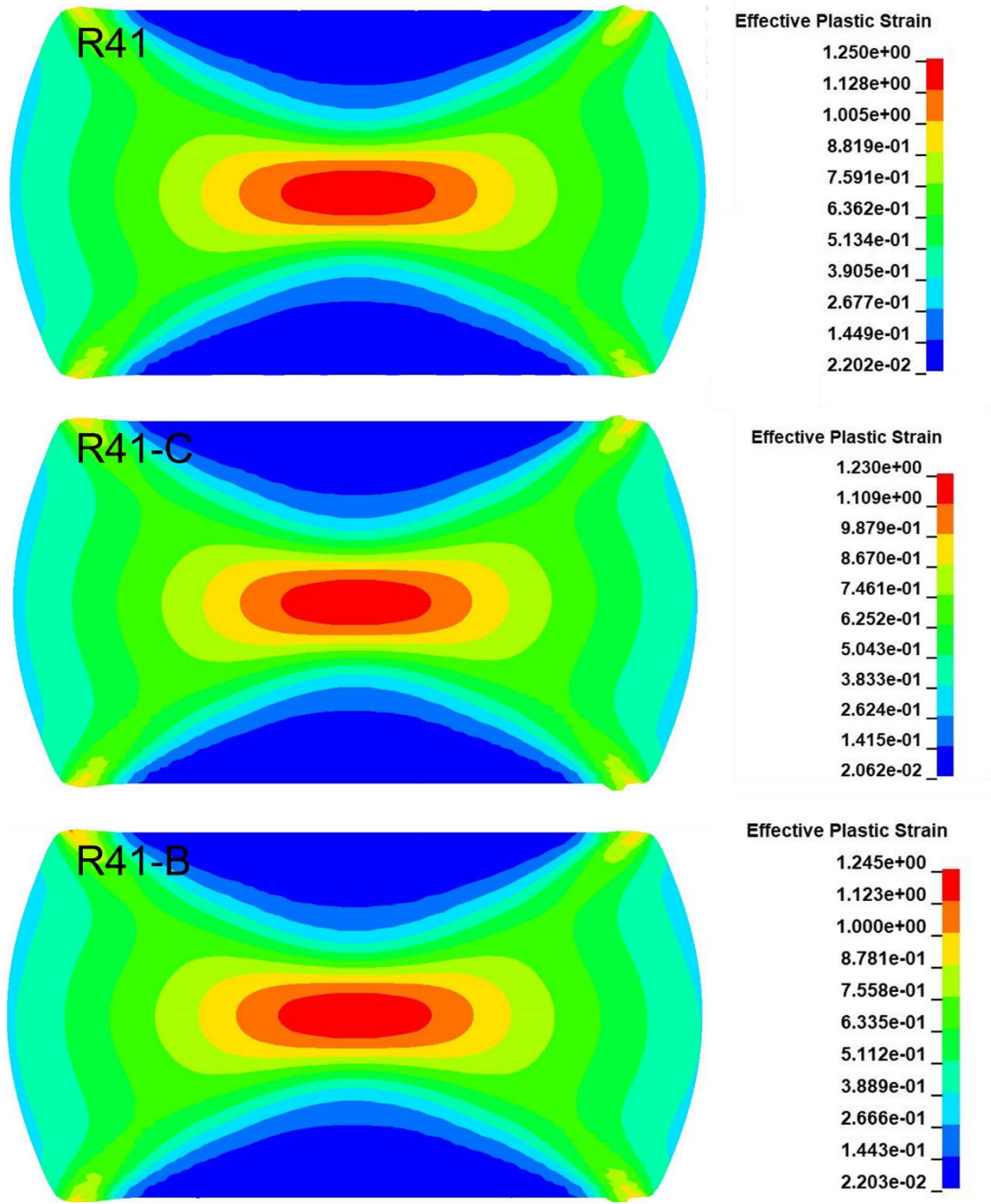

Figure 4.5 FEM simulations of effective plastic strain distribution for (a) R41 (b) R41-C (c) R41-B sample at 1075°C and 0.1s⁻¹.

## 4.2 Material Characterization of hot deformed René 41 variants

As part of material characterisation methods, qualitative analysis of LOM micrographs were conducted. This helps narrow down important deformation parameters to be used further in quantitative analysis via EBSD data. The figures as shown in Figure 4.8, 4.9 and 4.10 are etched light optical microstructures of the three René 41 variants under different deformation conditions, respectively. The 1000°C samples appear to have elongated grain structures with little to no visible recrystallised grains regardless of strain rates for all three René 41 variants. In addition to the elongated grains, shear bands can be observed in the 1000°C sample series as highlighted with blue arrows, with increasing intensity for increasing in strain rates. For the 1075°C samples, dynamically recrystallised grains can be observed nucleating from the prior grain boundaries of the deformed grains, also known as a necklace grain structure. Increasing strain rates reduce the size and number of recrystallised grains observed in the micrographs. This is true for all three René 41 variants in both 1075°C and 1150°C samples. Larger recrystallised grains and higher volume fractions of recrystallised grains can be observed qualitatively for the 1150°C samples when compared to the 1075°C at the same strain rate.

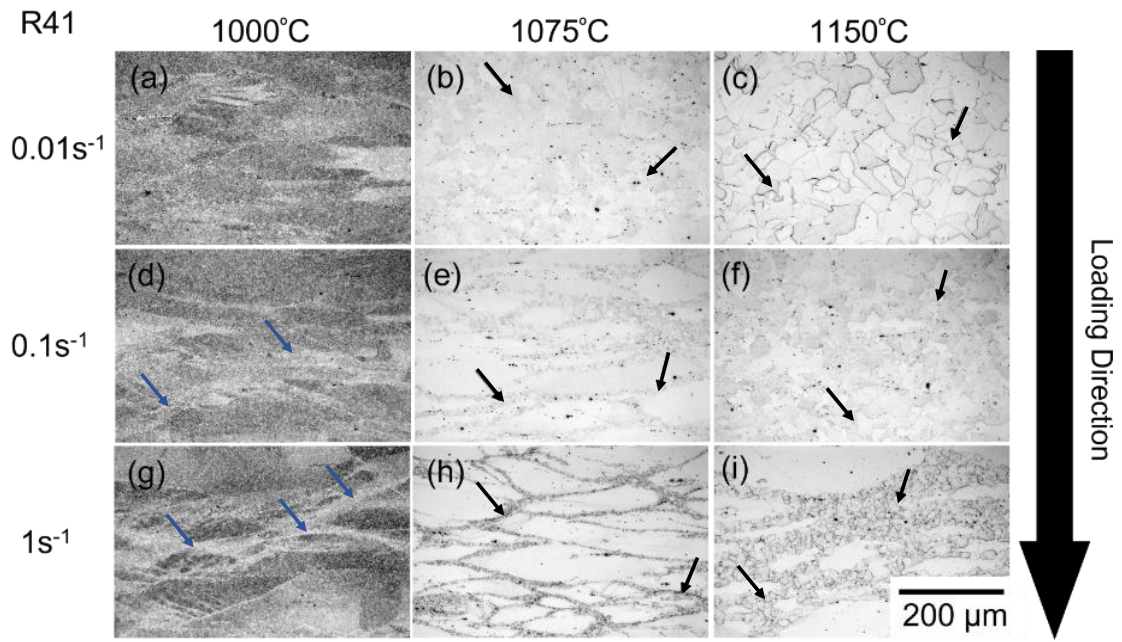

Figure 4.6 LOM micrographs of R41 samples under different deformation conditions Black arrows indicate evidence of recrystallised grains, blue arrow indications shear bands. Loading direction is shown on the left of the micrograph.

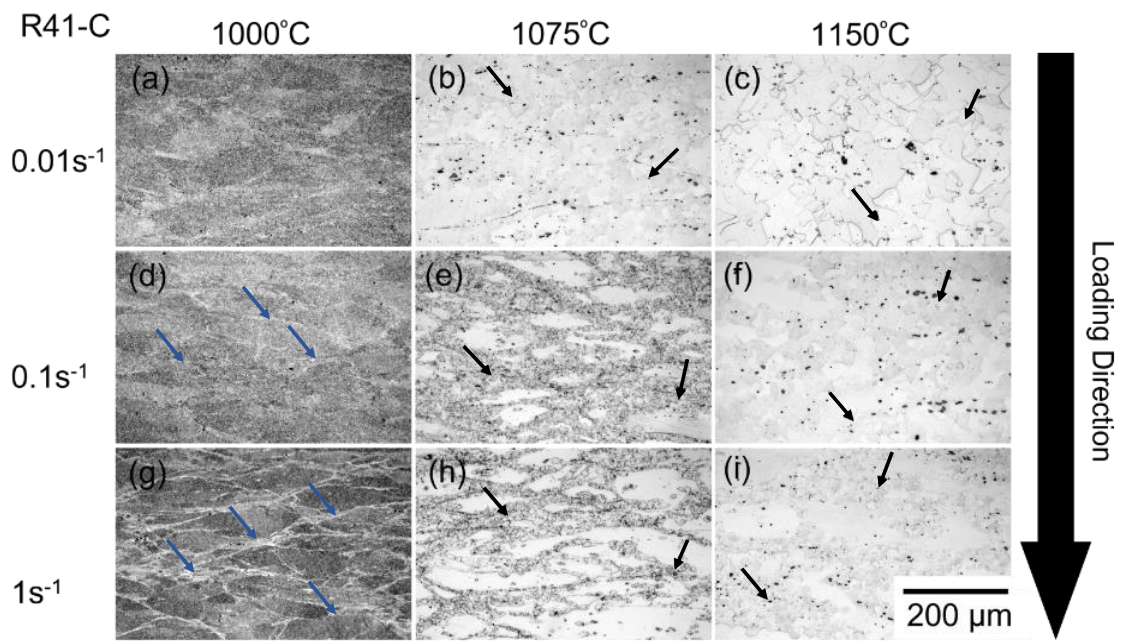

Figure 4.7 LOM micrographs of R41-C samples under different deformation conditions. Black arrows indicate evidence of recrystallised grains, blue arrow indications shear bands. Loading direction is shown on the left of the micrograph.

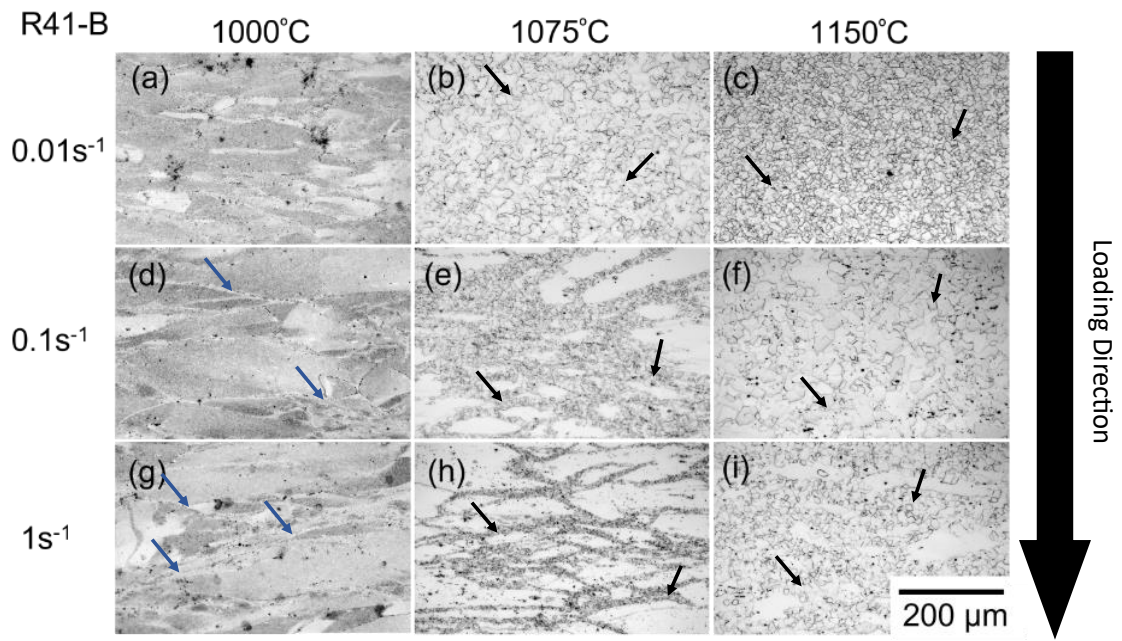

Figure 4.8 LOM micrographs of R41-B samples under different deformation conditions. Black arrows indicate evidence of recrystallised grains, blue arrow indications shear bands. Loading direction is shown on the left of the micrograph.

Figure 4.11 are the higher magnification LOM micrographs of secondary precipitate populations in René 41 variants at two distinct regions – the centre region for 4.9a, 4.9c and 4.9e and the low strain region around top/bottom of the samples for 4.9b, 4.9d, and 4.9f. In Figure 4.11a, the secondary precipiates observed have a blocky morphology located sparsely in the microstructure. As shown in Figure 4.9c the secondary precipitate population appears to be of the same nature. However, Figure 4.9 exhibits more secondary precipitates stringers. These secondary precipitates appear to be intragranular with no preference for grain boundaries. However, when compared with LOM micrographs taken in the low strain region around the top and the bottom of the samples, the locations of secondary precipitate populations seem to suggest a preference for grain boundaries in the case of R41 and R41-B as shown in Figure 4.9b and 4.9f. However, the microstructure, as R41-C secondary precipitate population remains scattered across the microstructure as Figure 4.11d displayed.

The inverse pole figure maps (IPFs) displayed in Figure 4.10 were collected at the centre region of the $0.1s^{-1}$ deformation condition of the three René 41 variants. The IPF maps show the orientation of individual grains within the microstructures. At 1150°C, there are mostly smaller equiaxed grains with the occassional larger grain structure that can be observed. At 1075°C, a necklace structure similar to the LOM micrographs can be observed with smaller equiaxed grains located around the grain boundaries of deformed grains. At 1000°C, a large portion of the IPF maps have noisy features within the microstructure. At higher magnification as shown in Figure 4.11 using both BSE imaging and EBSD, this is confirmed to be sub micron scale recrystallised grain nuclei. In Figure 4.11, γ' precipitates can also be observed to be present in all three René 41 variants.

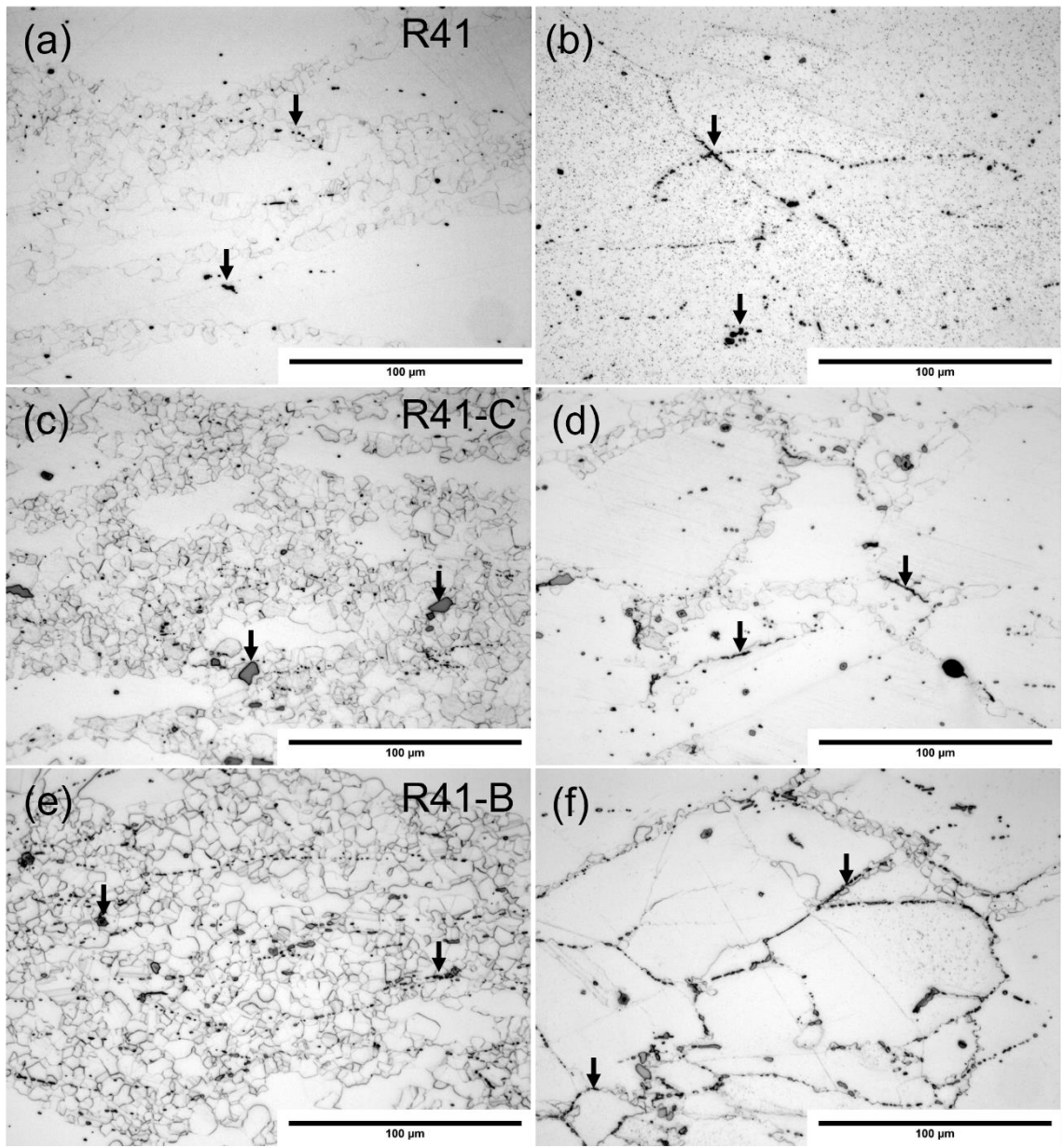

Figure 4.9 LOM micrographs of secondary precipitate populations in centre region (a) R41 (c) R41-C (e) R41-B and top/bottom region (b) R41 (d) R41-C (f) R41-B at 1075°C and 0.1 s$^{-1}$. Arrows indicates presence of secondary precipitates.

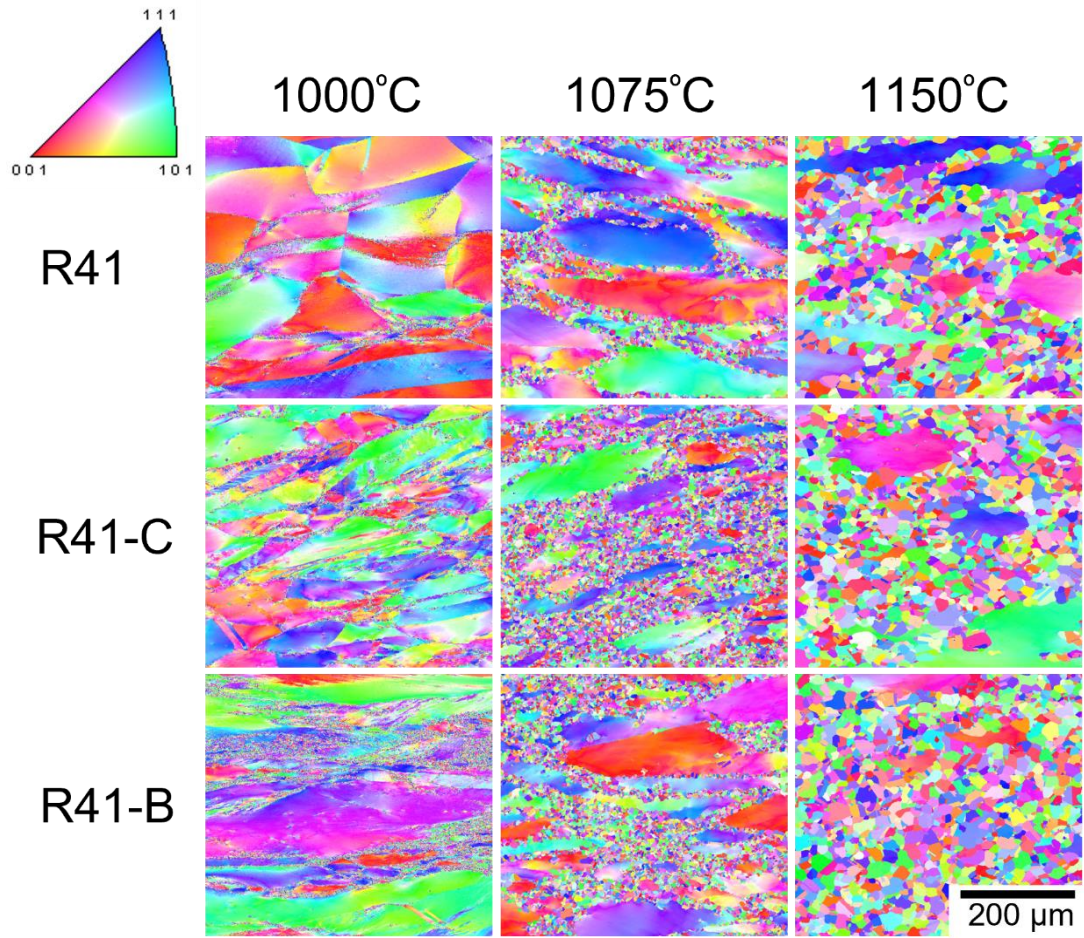

Figure 4.10 IPF maps parallel to the out-of-plane direction of the deformed René 41 variants at 0.1s$^{-1}$ and three different temperatures.

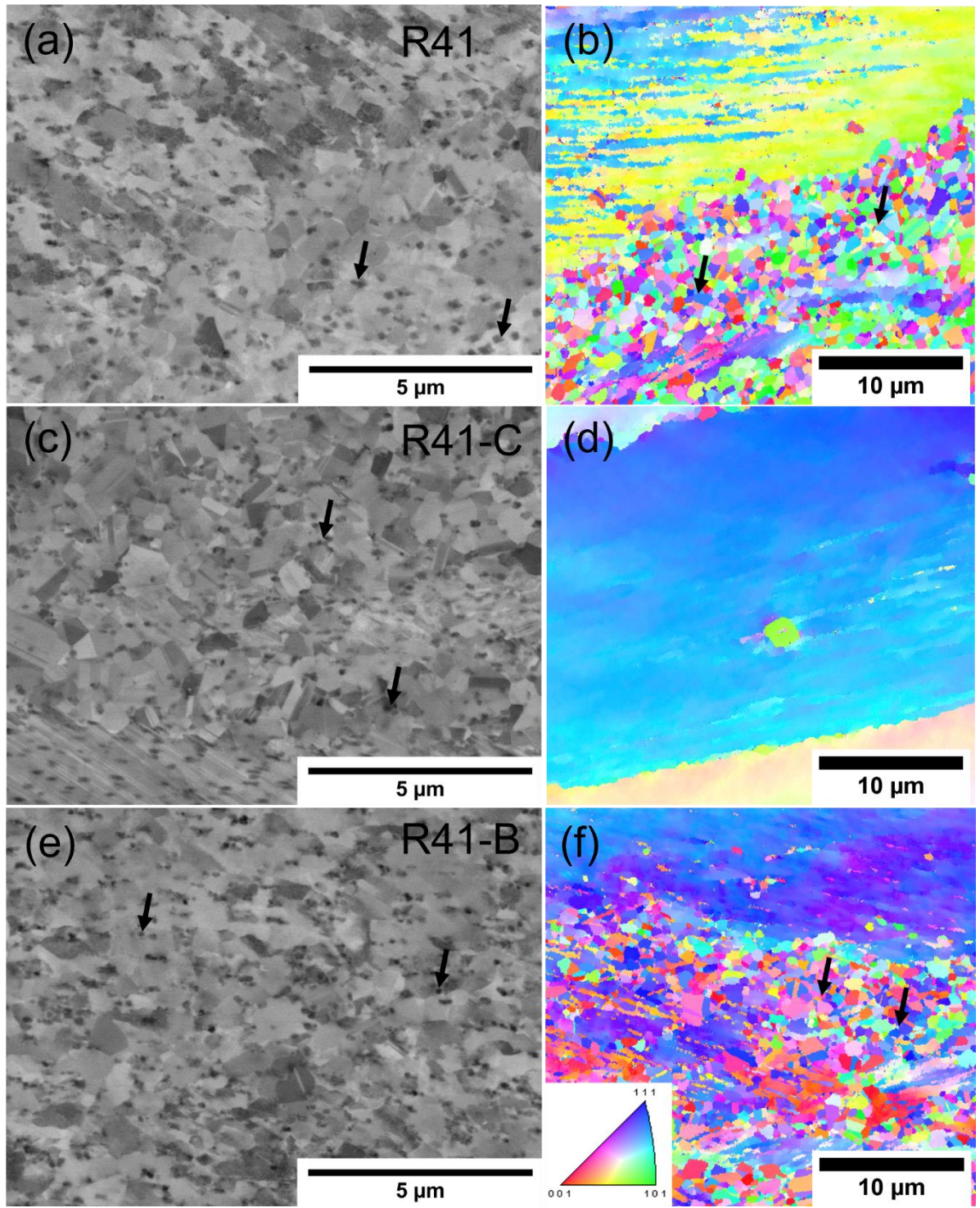

Figure 4.11 High resolution micrographs of the three René41 variants at 1000°C and 0.1s$^{-1}$ deformation condition. BSE imaging from sample (a) R41, (b) R41-C, and (c) R41-B. Arrow indicates evidence of intragranular γ' precipitates. IPF maps are parallel to the out-of-plane direction to the loading direction for (b) R41, (d) R41-C, and (f) R41-B. Arrow indicates evidence of sub-micron recrystallised nuclei

Data collected from EBSD was also used to quantify the recrystallisation behaviour of the three René 41 variants. Specifically, the recrystallised grain fraction, average grain size and geometrically necessary dislocation (GND) densities as shown in Figure 4.12, 4.13 and 4.14, respectively were evaluated. A core assumption of the analysed data is that both the sectioned view of the sample and the selected area of interest are assumed to be representative of the bulk material. This conveys that the grain structure as viewed in Figure 4.10-4.11 are assumed to be equiaxed in all directions inside the material. Hence, the grain size was calculated based on equiaxed grain structure. Recrystallised grain fraction and GND densities are collected as spatial data and therefore only require the selection of a representative selected area. In Figure 4.12, the recrystallised grain fractions are 14, 15, and 22 vol.% for 1000°C sample series of R41, R41-C, and R41-B, respectively. The recrystallised fractions increases as temperature increases with 77, 88, and 91 vol.% at 1150°C for R41, R41-C and R41-B, respectively. This is consistent with observations from both the LOMs and IPF maps. While all three René 41 variants have similar levels of recrystallised grain fractions at 1000°C, R41 is significantly lower than in both the R41-C and R41-B at both 1075°C and 1150°C. The largest difference between variants can be observed at 1075°C, where R41-C has twice the recrystallised grain fraction at a value of 65 vol.% when compared to R41 which only contains 31 vol.%. In Figure 4.13, the average recrystallised grain size is similar across all temperatures. R41-B has a slightly smaller average grain size of 11.5 μm at 1150°C when compared to R41 and R41-C which both have 12.5 μm. The trend in Figure 4.14 is inversely proportional to Figure 4.12, where the increase in temperature results in a decrease in GND densities across all three René 41 variants. For example, R41 has a GND density of $96.3*10^{12}/m^2$ at 1000°C, but decrease to $31.78*10^{12}/m^2$ at 1150°C.

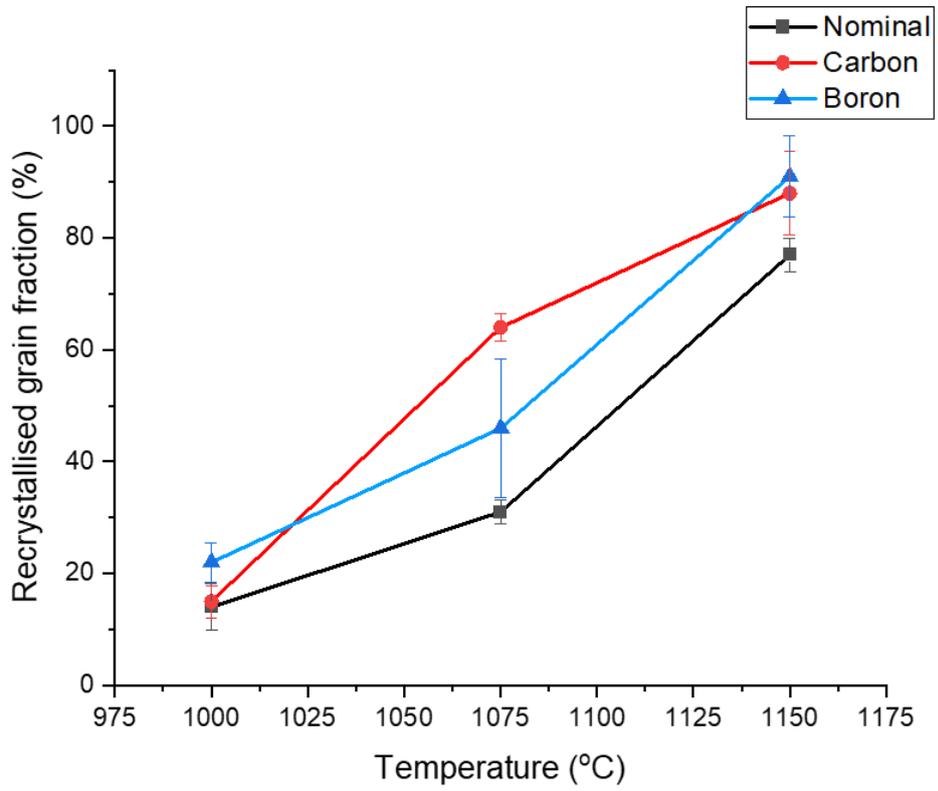

Figure 4.12 Recrystallised fraction of the three René 41 variant

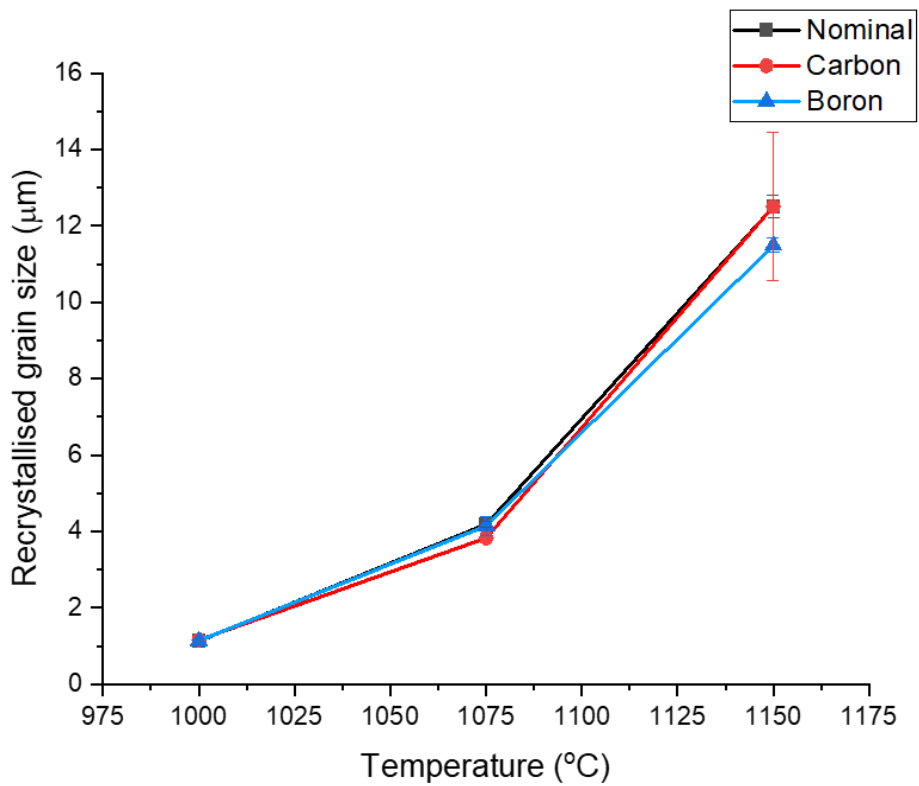

Figure 4.13 Average recrystallised grain size of the three René 41 variants

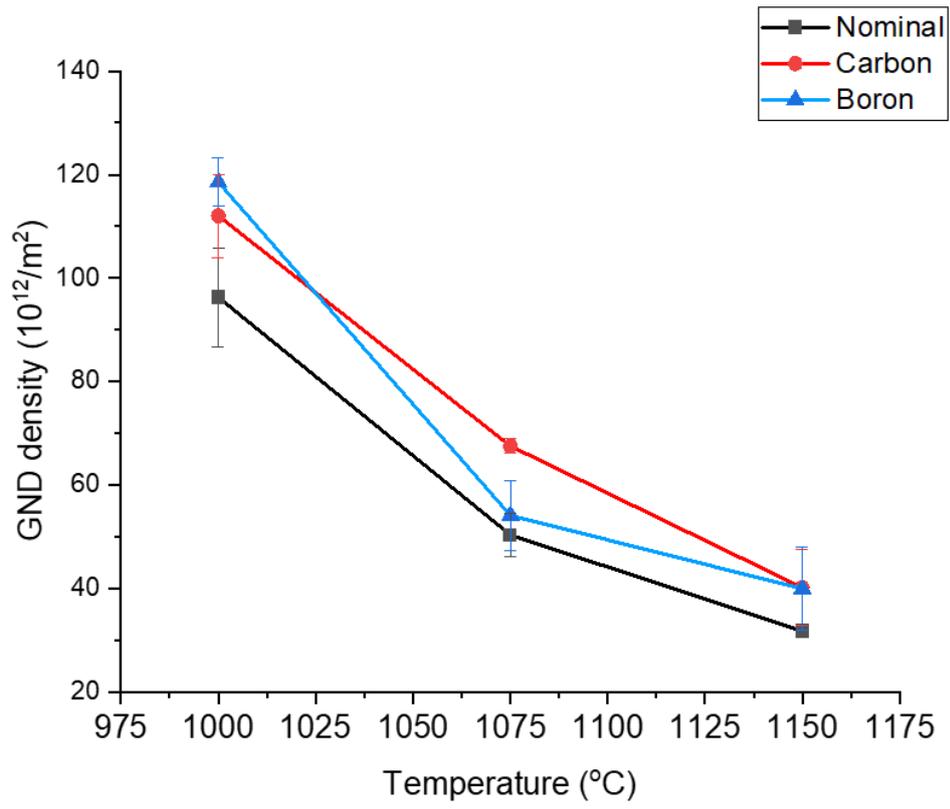

Figure 4.14 Average GND densities of the three René 41 variants

## 4.4 Supervised classification learning of René 41 industrial dataset

UT indications with respect to the aggregate billet position were visualised to better understand the origin of UT defects in René 41 billets. Figure 4.15 is a KDE plot of ultrasonic defect locations in the cross-sectional view of industrial René 41 billets. The typical dimensions of the René 41 billets are 1-3 m in length and diameter of 20-30 cm. This plot represents the probability density of UT defects within the billet at a specific region from the bottom to the top of the billet, as well as from the centre position to the surface of the billet. The KDE plot shows the contour with the maximum probability density of UT defects occur at locations within the centre region along the middle length of the billet at a magnitude of 5.5 UT defects per unit area. There is also another minor intensity region at a magnitude of 3.5 UT defects per unit area located at the 0.75 fraction point of the billet in the longitudinal direction. On the other hand, the observation can be made that there are no UT defects detected above the mid-centre region to the surface region for the radial direction.

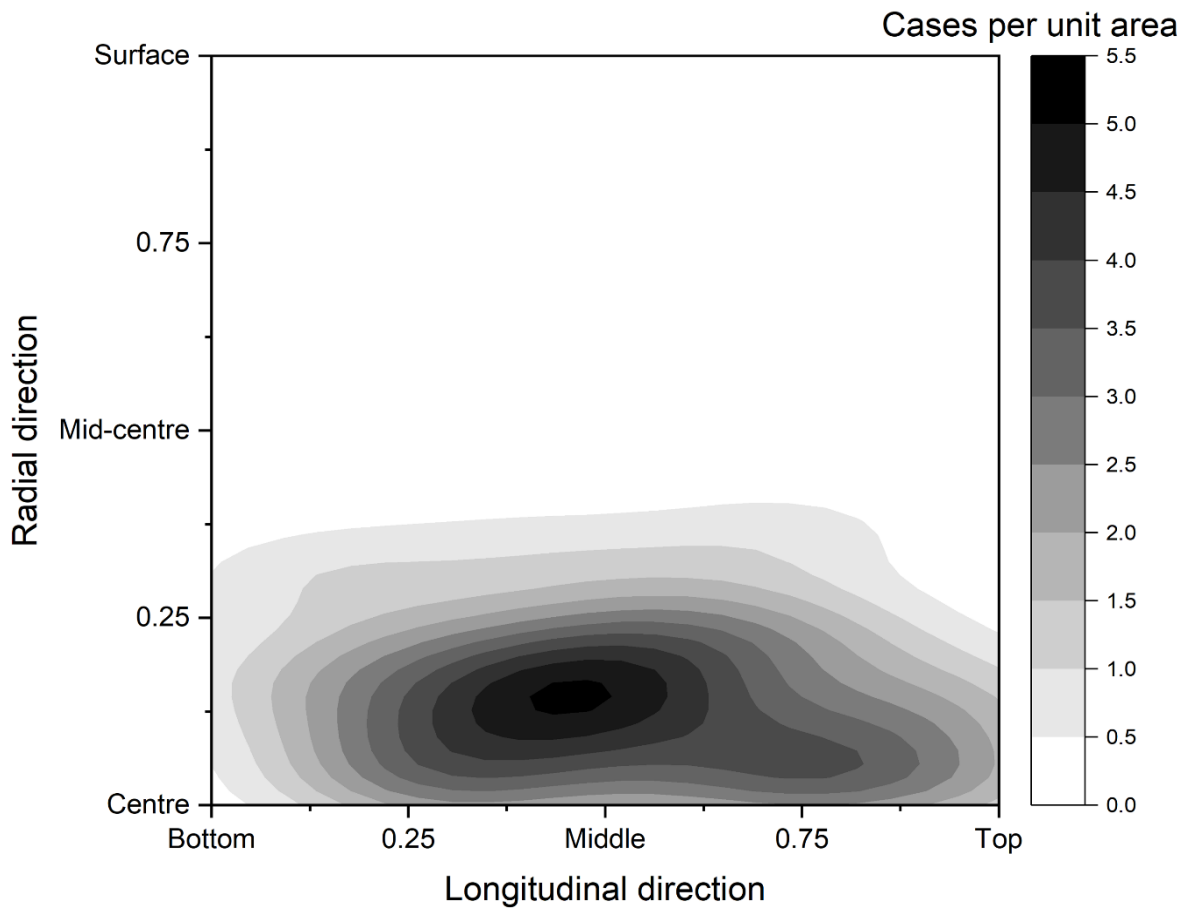

Figure 4.15 Kernel density estimation contour of UT defects of aggregate billet cross-sections.

Data analysis of René 41 chemical composition distributions were done as shown in Figure 4.16-4.17 to provide data interpretability for machine learning model results. Figures 4.16 and 4.17 display the chemical element distributions of the C content and B content, respectively in industrial René 41 datasets as a function of wt.% differences of sample population mean value $\bar{x}$. There are two overlapping distributions for each figure indicating the differences between the C distribution with indications of UT defects ($UT_C$) and C distribution without indications of UT defects (non-$UT_C$). KDE fitted curves are also implemented in both distributions for easier interpretation of the shape of the distributions. In Figure 4.16, it can be observed that the overall shape of $UT_C$ and non-$UT_C$ resemble the general Gaussian-like distribution. However, there is a small increase at 0.2 wt.% C difference for non-$UT_C$, while $UT_C$ does not have the same increase. Non-$UT_C$ has a trailing end of outliers extending from -0.02 wt.% C difference to -0.06 wt.% C difference, while in $UT_C$ the outliers are only up to -0.04 wt.% C difference. This is reflected in the skewness of both distributions measured at -2.29 while the skewness of non-$UT_C$ is -2.60. Hence, the non-$UT_C$ is more skewed towards lower C content than the $UT_C$ as a result. The boron content as shown in Figure 4.17 is also of similar nature, where single peak Gaussian-like distributions can be observed to be the case for both the B distribution with UT defects ($UT_B$) and the B distribution without UT defects (non-$UT_B$). However, key differences can be observed in the $UT_B$, where symmetry is not preserved when compared to $UT_C$. Non-$UT_B$ have a skewness of 0.5, while $UT_B$ have a skewness of 0.08. The low skewness of $UT_B$ can be attributed to the lack of outliers, as seen as other distributions. No noticeable differences can be discerned by manual evaluation between the chemical element distributions with/without UT defects.

Another way to approach data analysis of distributions is to rearrange the distributions into ratios which highlights its variance when compared against one another. Figures 4.18 and 4.19 depict the relative non-UT/UT ratios of the C and B distributions after exclusion of outliers with bins containing less than 2 samples. In Figure 4.18, the relative concentration of UT increases from 20% at lower C content to 40% at higher C content. Note that it is easier to observe in Figure 4.19 that the peak of $UT_C$ is shifted away from the sample population mean than in Figure 4.17. In similar trend, Figure 4.19, the relative concentration of UT peaked with 27% at lower B content before dropping gradually to 20% at B content closer to the sample population mean. Hence, C distribution with UT defects have a bias towards higher C content, while the opposite can be said for B content.

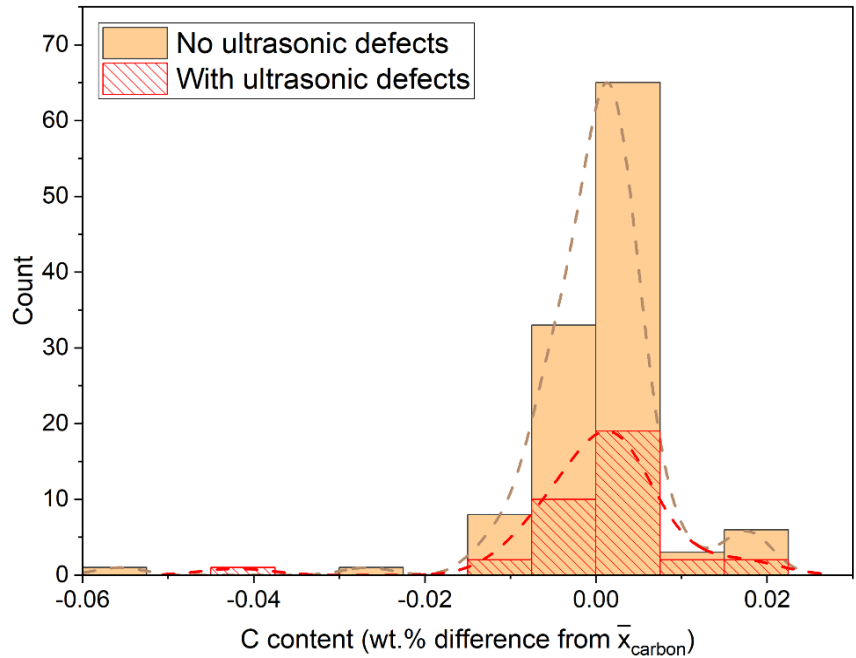

Figure 4.16 Carbon content distributions of René 41 billets in wt.% differences from mean carbon content both with ultrasonic defects and without ultrasonic defects. Dash lines represents KDE fitted curves of the distributions.

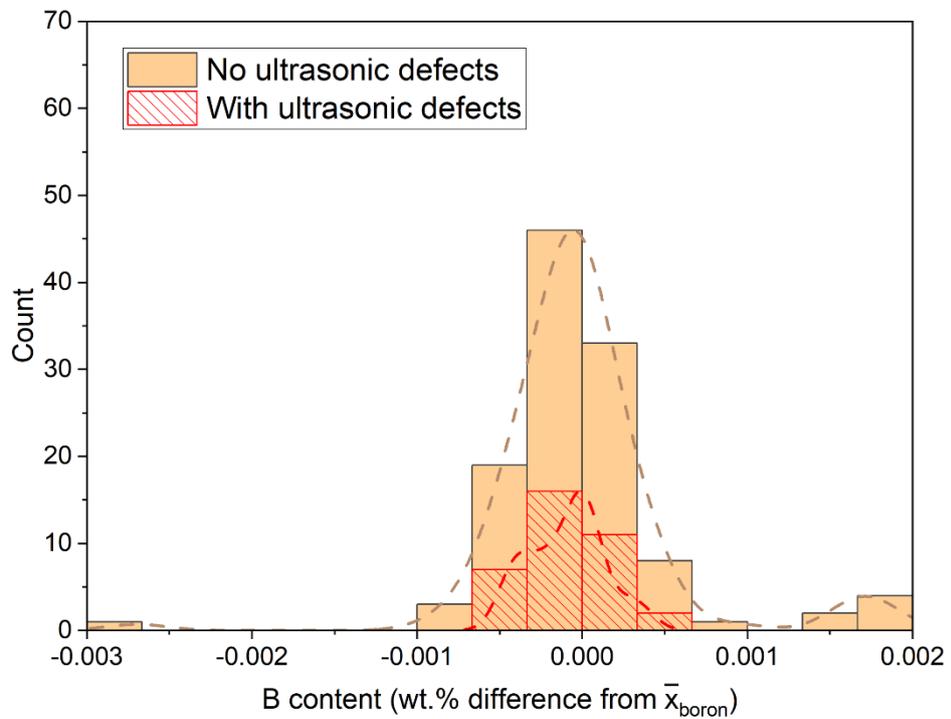

Figure 4.17 Boron content distribution of René 41 billets in wt. % differences from mean boron content both with ultrasonic defects and without ultrasonic defects. Dash lines represents KDE fitted curves of the distributions.

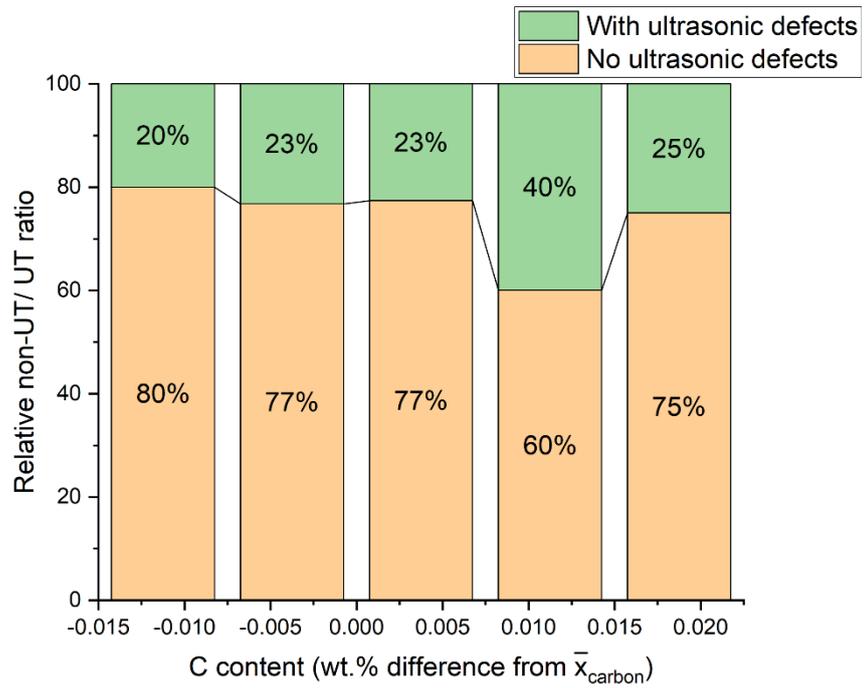

Figure 4.18 Relative non-UT/UT ratios of René 41 billets in % differences from mean carbon content both with ultrasonic defects and without ultrasonic defects

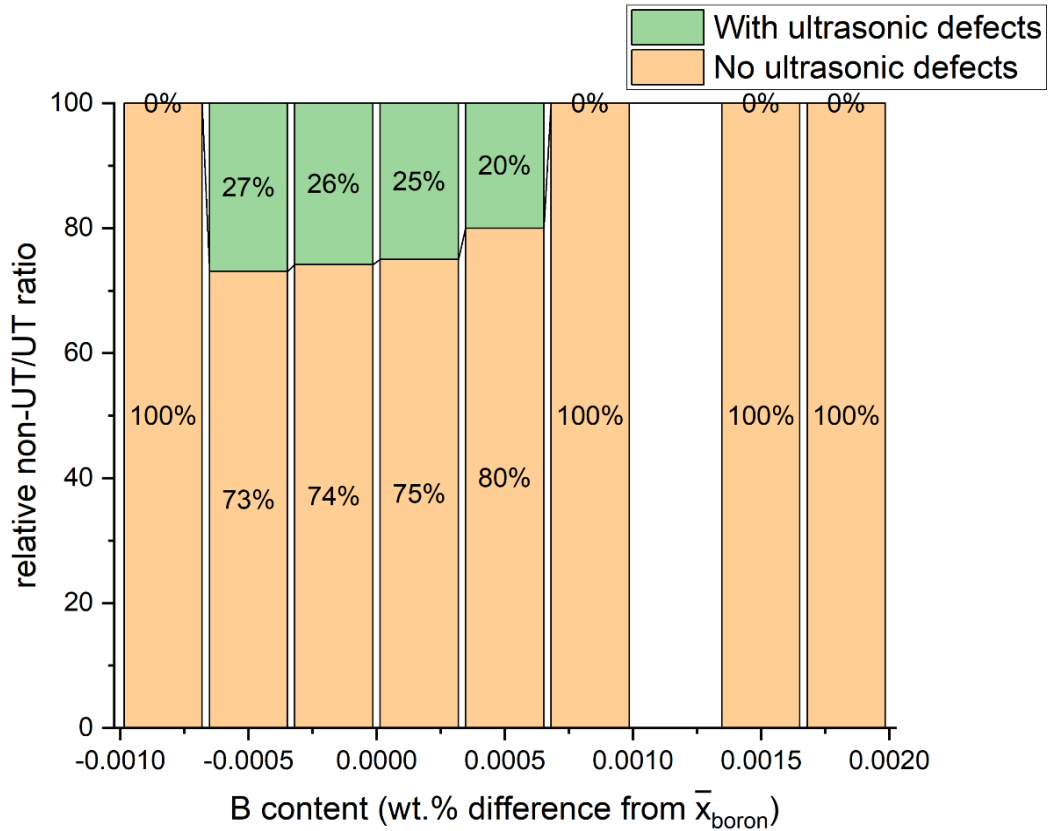

Figure 4.19 Relative non-UT/UT ratio of René 41 billets in wt.% differences from mean boron content both with ultrasonic defects and without ultrasonic defects

Figure 4.20 displays the performance of different classification models on the René 41 datasets based on the F-measure scoring metric. In general, the models have similar performance at around 60-80% accuracy, varying depending on the individual classification model. It can be observed that SVC model achieved the highest f1-score out of all the competing models and is the only model that is greater than 0.72. The relative closeness of the f1-score can indicate that the outcome predicted by the models are independent of model selection, hence, interpretation using various models can be done. In the case of a CART classifier, a visual representation of the outcome predicted by the model is shown in Figure 4.21. The blue leaf node represents the UT classification while the orange leaf node represents the non-UT classification. The decision nodes are built upon separating features into the two classes using the Gini impurity index as mentioned in previous section 2.5. The feature importance diagram as shown in Figure 4.22 lists and ranks the features according to their relative importance in computing the Random Forest classifier. It can be observed that C, N and Ti are the three highest-ranking chemical elements out of the 16 chosen features. Fe, B, Zr ranked the lowest amongst the 16 chosen features.

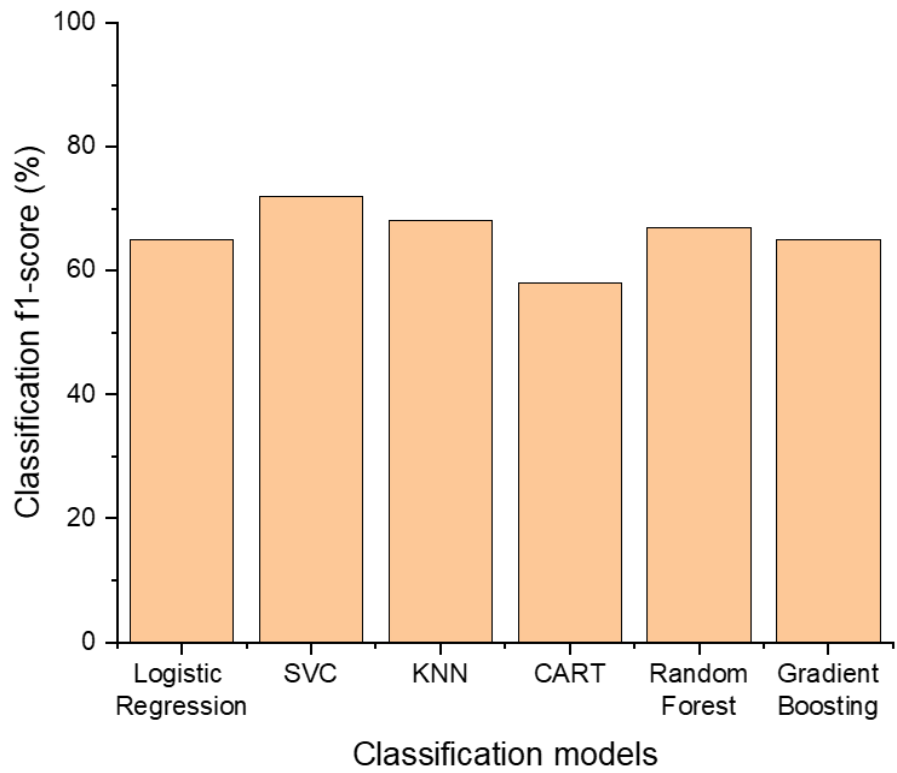

Figure 4.20 performance of supervised learning models based on f1-score

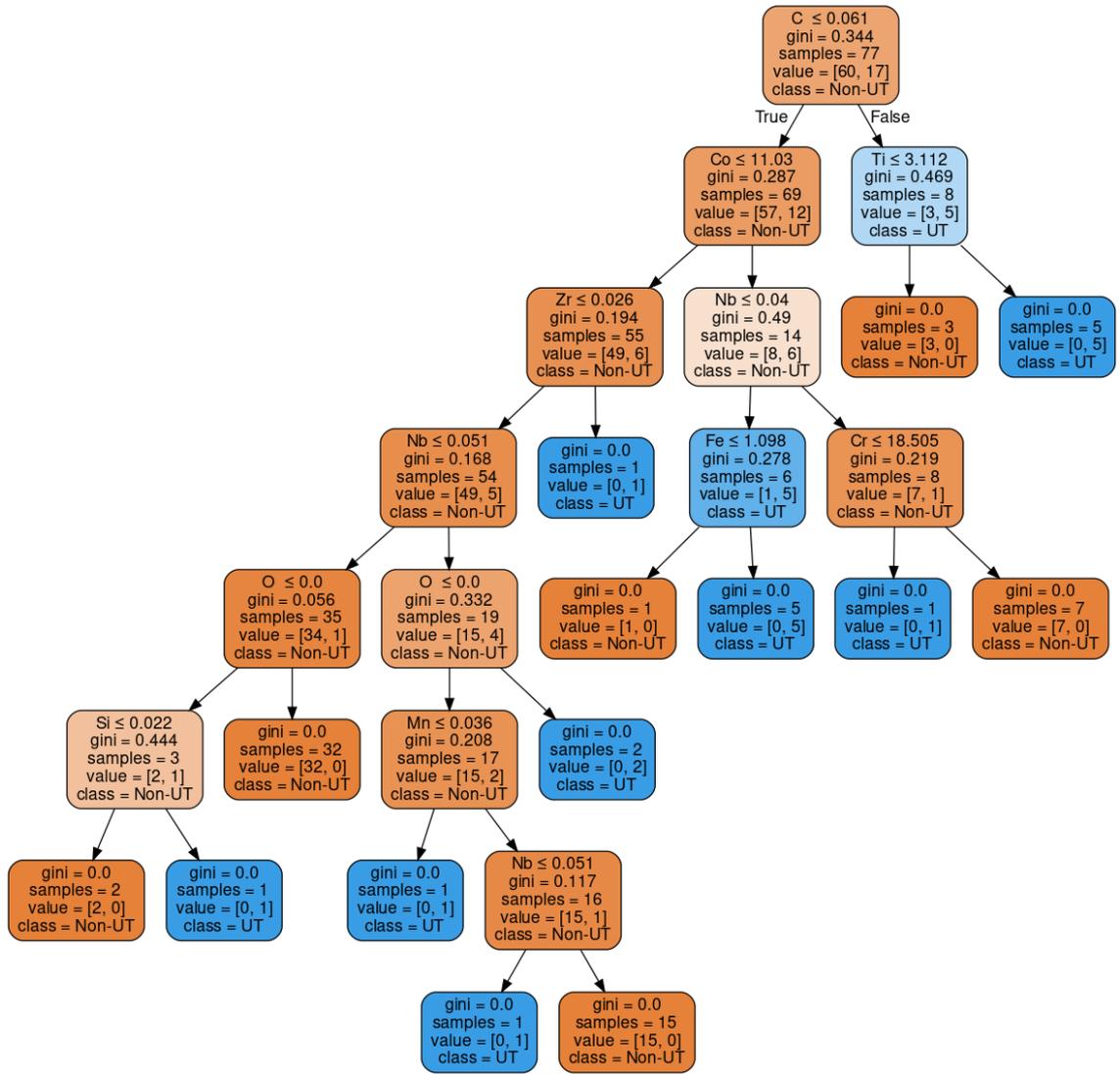

Figure 4.21 Visual representation of results from CART classification model

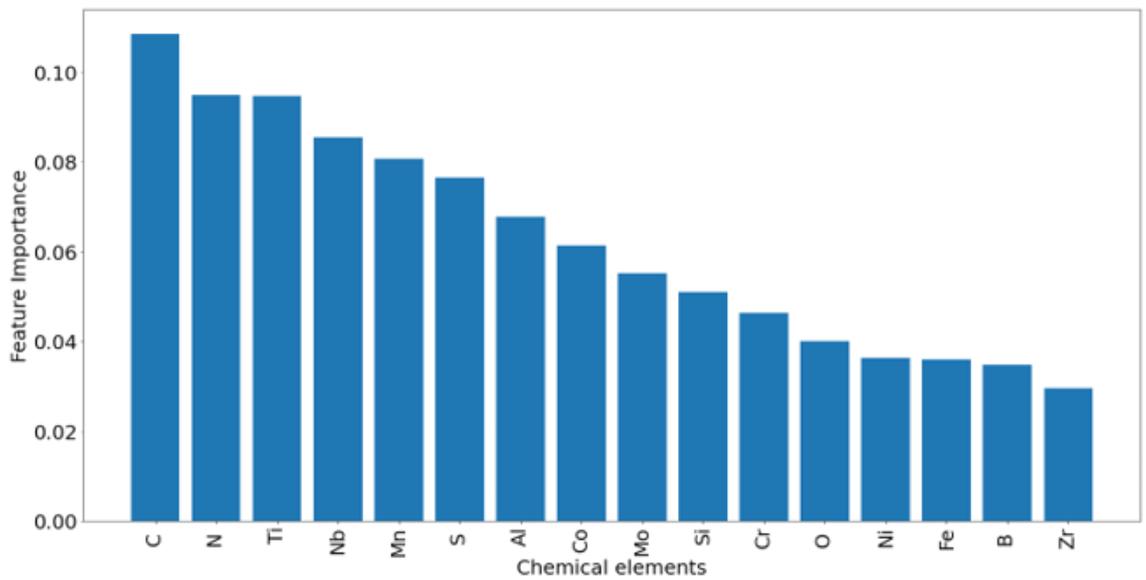

Figure 4.22 Feature importance diagram of chemical elements using Random Forest classifier

# Chapter 5 – Discussion

## 5.1 Hot deformation behaviours of René 41 variants

Understanding the underlying mechanisms governing the hot deformation behaviour of René 41 pave the way for improvements in processability during processing route. First, the influence of forming parameters on the hot deformation behaviour of René 41 has been researched and is discussed in section 5.1 to understand René 41's baseline response to hot deformation. Then key differences between the hot deformation behaviours of the three René 41 variants are examined with context to the influence of microstructural features. Lastly, an attempt is made to provide a generalisation of the observed behaviours of René 41 during hot deformation with reference to other existing Ni-based superalloys.

### 5.1.1 Influence of hot deformation parameters on the behaviours of René 41

Section 4.1 and 4.2 presented results regarding the different aspects of hot deformation behaviour of the three René 41 variants. The stress-strain curves as shown in Figure 4.1 demonstrated the rapid increases in flow stress during the initial hot compression exceeding 500 MPa in 1000°C sample series due to work hardening processes. As the strain increases, the work hardening rate decreases due to competing flow softening mechanisms such as DRX and DRV. The peak stress observed in the stress-strain curves indicates that DRX is the dominant mechanism of the two [83]. This is reinforced by the observations of necklace structure in Figure 4.6-4.8 with temperatures above 1000°C. DDRX characterised by the grain boundary bulging of DRX grains often results in necklace grain structure due to the preference for nucleation events to occur at the prior deformed grain boundaries [83]. This is consistent with hot deformation studies of wrought superalloys with similar bulk chemical composition such as Waspaloy [133] and Haynes 282 [134].

The impact of temperature on René 41 can be seen from the sharp increase in flow stresses for the 1000°C samples at any given strain rates, as can be seen in Figure 4.1. This can be attributed to the presence of γ' precipitates as evidenced in Figures 4.11a, 4.11c, and 4.11e, where nanoscale precipitates fitting the description of γ' precipitates can be observed in the microstructure via BSE imaging. Sahithya et al. [91] found that the flow stresses of DMR-740 Ni-based superalloy behaved differently between super and sub γ' solvus and used hyperbolic sine equations to establish two constitutive models for quantifying the differences [91]. The sharp change in in flow stresses as observed in Figure 4.1 is, thus, the result of crossing the sub/super γ' solvus line. For René 41, this temperature window agrees with existing literature where the γ' solvus is reported to be around 1060°C [7]. Another contributing factor to the increases in flow stress can be attributed to the coarse grain starting microstructure of the René 41 samples. The starting microstructure is an important consideration for evaluating hot deformation behaviour of any metals since variation in the grain structure can have influence onto the flow stresses based on changes in the microstructural evolution during hot deformation [135]. Sarkar et al. [136] demonstrated that in Cu-Cr-Zr-Ti alloy hot deformation of coarse-grained samples experienced increased flow stresses as compared to fine grained samples. This is attributed by the authors to be a mixed influence of increased barriers for DRX nucleation and grain boundary sliding (GBS) which causes strain to accumulate within the microstructure. Another study by Liu et al. [90] shows that the Udimet 720Li wrought Ni-based superalloy exhibits the same trend where flow stresses in coarse grain samples were increased by ~100 MPa as compared to fine grained samples under the same deformation conditions. The authors also made efforts to distinguish between the influence of coarse grain structures below the γ' solvus and above the γ' solvus. Above the γ' solvus with only γ-matrix present in the microstructure, the authors reported that DDRX was the dominant flow softening mechanism and this was dependent on the prior deformation grain boundary area in which the new DRX grains nucleated from. Hence, recrystallisation kinetics was slower in coarse grain structure samples due to the limited grain boundary area as compared to fine grain structure samples. Thereby, the work hardening rate is outweighed by the flow softening rate of DDRX in the fine grain structure sample much earlier than the coarse grain structure sample and the flow stress is comparatively lower. For the sub γ' solvus region, it is known that γ' precipitates provide a strong pinning effect as can be seen in the DDRX kinetics [89,92]. In the case of a coarse grain structure, dislocations accumulated at the deformed grain boundaries increase the stored energy to induce DRX nucleation. However, these DRX nuclei are pinned by the γ' population and inhibit growth. For the René 41 samples, the sub-micron scale recrystallised grain nuclei observed in the 1000°C samples as shown in the IFP maps in Figure 4.11 matches the coarse grain DDRX nucleation description from Liu et al [90]. This can be attributed to the same recrystallisation pinning mechanism by γ' population. This contrasts with the fine grain structure sample observed in the same [90] study where DDRX nucleation were less likely to occur resulting in dynamic recovery as the dominant softening mechanism instead.

The influence of strain rate is also apparent in the hot deformation behaviour of René 41. Increases in flow stresses can be observed in Figures 4.1-4.3 for higher strain rates. This result agrees with other wrought Ni-based superalloy studies for the strain rates up to 1 s$^{-1}$ [88,104,137]. It is reported by Nicolaÿ et al. [138] that in this strain rate range, DRX is governed by the deformation time. Increasing strain rate causes deformation time to shorten, hence, a lower volume fraction of DRX grains is found in the microstructure. A deeper analysis has been made by Jiang et al. [139] on the influence of strain rates to the microstructural evolution of γ' strengthened Ni-based superalloy both below and above the γ' solvus. Under 1s$^{-1}$, the DRX kinetics was impeded by increasing strain rates due to changes in the critical dislocation density required for DRX nucleation [140] and the time-dependence of grain boundary migration necessary for DRX grain growth [141]. Roberts and Ahlblom [140] formulated an expression as shown in equation 5.1 for the critical dislocation density for DRX nucleation.

$$\frac{\rho_m^3}{\dot{\epsilon}} > \frac{2\gamma_b}{KMLGb^5} \tag{5.1}$$

Where $\gamma_b$ is the grain boundary surface energy, $K$ is a material constant, $M$ is the grain boundary mobility, $L$ is the dislocation mean slip distance, $G$ is the shear modulus, and $b$ is the Burger's vector. It can be seen in equation 5.1 that increasing the strain rate $\dot{\epsilon}$ will increase the critical dislocation density $\rho_m^3$. The consequence of requiring more stored energy for DRX leads to an increase in the dynamic recovery rate as an alternative flow softening process. This can be observed by the increasing plateau of flow stresses towards the higher strain when comparing the higher strain rate sample conditions in Figures 4.1c, 4.1f, 4.11i to slower strain rate sample in Figures 4.1a, 4.1d, and 4.1g. This effect is most visible in the 1000°C stress-strain curve as the DRX inhibition is compounded upon with the effect of γ' precipitates.

Another effect of increasing strain rate can be observed in Figure 4.6a, 4.6d, and 4.6g where the shear band intensity increases as strain rate increases. Past studies have shown that shear band formation in hot working is caused by frictions between die and sample surfaces creating a dead zone at the top and bottom of the cross-section [77,142]. This is confirmed by the FEM simulations shown in Figure 4.7 where there is a clear strain gradient formed within the cross-section of the hot compressed samples in the compression direction. The FEM simulation results display a 'X' profile portraying a medium strain localization profile according to similar FEM results from Tang et al. [143]. Adiabatic heating is the conversion of work into thermal energy during plastic deformation. This may result in localised heating due to insufficient heat transfer via conduction. The higher the strain rate, the more pronounced the effects of adiabatic heating becomes [144]. It has been reported from Qiang and Bassim [145] that the strain rate is closely associated with the occurrence of shear band formation. From Armstrong's [146] work, a mechanism has been proposed for the formation of shear bands during isothermal hot compression. Dislocations piled up at grain boundaries are suddenly released into the neighbouring grain in an avalanche manner in terms of mobile dislocation density. This release of dislocations causes a high density of mobile dislocations to simultaneously move through the same maximum shear plane within the grain interior and is associated with adiabatic heating in the microstructure [145]. Hence, the formation of shear bands occurs as plastic deformation is localised within these flow instability region. This mechanism can be used to explain the observations from Figure 4.9 where the occurrence of shear bands is intensified for increasing strain rate as adiabatic heating within the microstructure also increases.

5.1.2 Influence of C and B additions on hot deformation behaviours of René 41

The major alloying elements of the R41, R41-C and R41-B samples are in close wt.% of each other as shown from Table 3.1, hence, bulk chemical composition are not affecting the differences seen in the René 41 variants. R41 as baseline sample has the lowest C and B additions compared to the R41-C which have C additions and R41-B which have with B additions. The hot deformation behaviour of R41, R41-C and R41-B as highlighted in section 4.1 shows relative similarity in terms of flow stresses across all deformation parameters. The respective constitutive models have also been presented in equation 4.1-4.3 to describe each of the René 41 samples. The comparison is better highlighted in Figure 4.8 and equation 4.1-3., R41-C has the lowest activation energy for hot deformation at 697kJ mol$^{-1}$. This gives indication to the R41-C as the material with the least flow resistance in the hot working window tested. Furthermore, Figure 4.12 indicates that R41-C have 64 vol.% recrystallised fraction at 1075°C with twice as much compared to the 31 vol.% of R41. However, in Figure 4.13 the recrystallised grain size remains close to the other samples at all deformation conditions. This indicates that the contributing factor to the increase in recrystallised fraction observed in R41-C is due to the number of DRX grains or nucleation events rather than the growth of pre-existing recrystallised grains.

The lack of growth in the recrystallised grains observed can be explained by the short deformation time and lack of post dynamic recrystallisation mechanisms due to immediate water quenching of the samples [134]. This drastically interrupts the growth kinetics of the DRX grains while nucleation events which occurs at a much faster rate will not be as severely influenced. The difference in DRX grain nucleation for the René 41 variants can be explained by the difference in microstructures. This is best highlighted in Figure 4.11a, 4.11c, and 4.11e where the secondary precipitate populations differ from each other. As mentioned previously, the coarse grain structure of the samples of the three René 41 variants provides limited grain boundary area for DDRX to occur. Therefore, additional microstructural features will have beneficial effects on both nucleation of DRX grains [147] and the generation of dislocations.

Particle stimulated nucleation (PSN) is a mechanism observed in low strain deformation of many superalloys with secondary precipitates [148–152]. The strain level of the current study exceeds any strain level where direct evidence of PSN could be observed, however, the compounding effects of extra nucleation sites during low strain level can lead to increases in the overall recrystallised fraction [147]. This is best explained by considering the microstructural evolution during the hot deformation of the samples. At initial stage of loading, the coarse grain structures within R41-C and R41-B possess additional DRX nucleation sites compared to R41 due to the presence of secondary precipitates. The overall recrystallisation is now faster due to more DRX nuclei simultaneously growing within the microstructure. The growth of these DRX nuclei will in turn lead to more grain boundary areas, thereby, as discussed earlier, will convert to a higher density of DDRX nucleation sites. Hence, a higher recrystallised volume fraction is observed. However, in the Eriksson et al. [153] study, the authors found no apparent changes between the hot deformation behaviour of Haynes 282 superalloy with/without grain boundary carbides. One key difference is that the starting microstructure is a fine grained as-forged condition whereas the superalloy studies with PSN tend to have an either as-cast or coarse grain structure as initial microstructure. Therefore, it is postulated that secondary precipitates might be more effective in a coarser grain structure than in a finer grain structure. Another mechanism that secondary precipitates can influence the final recrystallised fraction is the increase in stored energy within the microstructure. Sahithya et al.'s [151] work on as-cast Ni-based superalloys observed Orowan looping mechanisms occurring at coarse MC carbides. Dislocations by-passing a secondary precipitate via Orowan looping can lead to increases in the dislocation density. This is evident in the increase in GND density observed for the R41-C sample at 1075°C in Figure 4.16. This correlates to the increased stored energy within the microstructure. It should be noted that difference in the population of secondary precipitates also has an influence on the increase in dislocation density. While at 1150°C both R41-C and R41-B have similar behaviours in terms of increases in GND density in Figure 4.16, the 1075°C R41-C sample has around 10% more recrystallised fraction than R41-B. This can be attributed to the formation of $M_6C$ carbides in the R41-C sample. While direct phase identification was not done for this study, previous literature [154] has highlighted the solvus ranges for $M_6C$ in René 41 would be below 1150°C but above 1075°C. For the R41-B, borides are to be expected instead of carbides. Here, borides tend to have a higher solvus than secondary carbides as reported in LSHR Ni-based superalloy [155].

Another consequence of secondary precipitates can be observed in Figure 4.10 where the IPF maps for the sub-γ' solvus condition of the R41-C and R41-B samples have a more homogeneous deformation as compared to the R41 sample. From Figure 4.14, it can be observed that the GND densities at 1000°C for both R41-C and R41-B are higher than R41. As highlighted before, higher dislocation density leads to stronger tendency to form shear band due to the dislocation pile-up mechanism [146]. However, comparing the shear band formation within the LOMs in Figure 4.8g, 4.9g, and 4.10g, the R41 sample have the strongest shear band intensity out of the three René 41 variants. One explanation is that while the secondary precipitate population increases the overall dislocation density within the microstructure, it promotes dispersing dislocations throughout the grain interior. Thus, dislocations in the R41 sample will accumulate intensely at the deformed grain boundaries due to the lack of precipitate population and promote shear band formations [147].

On the other hand, grain boundary migration during recrystallisation can influence the distribution and morphology of secondary precipitates. From Figure 4.11 it can be observed that secondary precipitates heavy decorated the grain boundaries for low strain regions located at the top and bottom of the cross-section. Whereas the secondary precipitate population are more intergranular and dispersed in the higher strain region located at the centre of the cross-section. When recrystallisation occurs, the movement of grain boundaries can drag grain boundary precipitates across grains to collide with other grain boundary precipitates. A merging of smaller carbides has been observed in Qin et al.'s [156] study on the hot deformation behaviour of Alloy 602 CA Ni-based superalloy. This merging mechanism induced by the compressive force of hot deformation causes smaller carbides to merge with larger carbides. Once merging of precipitates occurs, the grain boundary precipitates will reach enough size to overcome the drag force by grain boundary migration while spacing between these precipitates also increase due to the merging of nearby precipitates. This can explain the appearance of disperse secondary precipitate population at the higher strain region in the René 41 samples.

### 5.1.3 Comparison of hot deformation behaviour of René 41 with other superalloys

Table 5.1 summarises the literature regarding the constitutive modelling results of cast and wrought Ni-based superalloys of similar applications to René 41 [103,157–160]. All deformation temperatures listed are within the 900-1210°C range as it is the primary hot working window for wrought Ni-based superalloys. The same consideration is applied to strain rates as strain rates below 20 s$^{-1}$ are common in hot working of wrought Ni-based superalloys. While the starting microstructure is designed to be consistent between the three René 41 samples, other studies have varying processing routes producing different starting microstructures to René 41 samples. A study with a coarse starting grain structure will have noticeably higher flow stresses than the exact same Ni-based superalloy with a finer grain structure. This can have an impact the hyperbolic sine equations which are based on the flow stresses. Nonetheless, insights can be obtained by conducting qualitative analysis and comparison using the literature in Table 5.1. The activation energies represent qualitatively the amount of work needed to deform the material under hot working condition. Similar values of activation energies convey a similar level of working needed for the thermomechanical processing of the material. This can be influenced by bulk chemical composition as is the case for Waspaloy and René 41 as they are the closest comparison in terms of bulk chemical compositions. However, the same could be said with regards to GH4282 which also has a similar bulk chemical composition to René 41, but due to the differences in γ' forming elements (i.e., Ti and Al), its activation energy is much lower. On the other hand, Udimet 720 which has a much higher (Ti + Al) content also has a significantly higher γ' volume fraction and γ' solvus temperature, allowing for strength to be retained at higher temperatures. Hence, Udimet720 has a much higher activation energy for hot deformation than René 41. Also, for γ'' strengthened superalloys such as Alloy 718 and Alloy 625, the γ'' solvus temperature is much lower than γ' [158] at around 925°C [161]. Therefore, the activation energy of hot deformation is lower due to the loss of primary strengthening mechanism as result of dissolution of γ'' precipitates in the hot working window.

Table 5.1 Comparison of reported stress exponents and activation energies values for different Ni-based superalloy studies.

| Materials | Processing Route | Temperatures (°C) | Strain rates ($s^{-1}$) | Stress exponent | Activation energy ($kJmol^{-1}$) | Reference |
| --- | --- | --- | --- | --- | --- | --- |
| R41 | Hot-rolled + homogenised | 1000-1150 | 0.01-1 | 4.46 | 757 | Equation 4.1 |
| R41-C | Hot-rolled + homogenised | 1000-1150 | 0.01-1 | 4.33 | 697 | Equation 4.2 |
| R41-B | Hot-rolled + homogenised | 1000-1150 | 0.01-1 | 4.25 | 728 | Equation 4.3 |
| Waspaloy | As forged | 1040-1120 | 0.01-10 | 4.35 | 670 | [157] |
| GH4282 | As forged | 950-1210 | 0.01-10 | 4.30 | 498 | [158] |
| Alloy 740H | Wrought + solution treated | 1050-1200 | 0.1-20 | 3.48 | 357 | [159] |
| Udimet 720 | Forged + solution treated | 1000-1175 | 0.001-1 | 4.34 | 1552 | [162] |
| Nimonic 80A | Wrought + solution treated | 900-1200 | 0.02-20 | 4.40 | 424 | [163] |
| Alloy 718 | Wrought + solution treated | 950-1100 | 0.001-10 | 3.56 | 429 | [160] |
| Alloy 625 | As-cast + homogenised | 1050-1200 | 0.01-10 | 4.17 | 520 | [103] |

### 5.1.4 Discontinuous behaviours of flow stresses in René41

The discontinuous yielding phenomena is sometime observed in the hot deformation of metals as shown in Figure 4.1 where during the initial loading, the flow stresses of René 41 experienced a disrupted yield drop followed by resuming normal flow stress behaviour. A similar behaviour has been observed by Guimaraes and Jonas [163] in Waspaloy and Inconel 718 where yield drops were seen in the temperature range of 1050-1150°C and 870-1090°C, respectively [163]. The authors proposed that the observed phenomena are due to the locking of dislocations via short range ordering of dissolved γ' former elements such as Al, Ti, and Co for Waspaloy and solid solution elements such as Cr, Co, and Fe for Inconel 718 [163]. This mechanism is rooted in the formation of a Cottrell atmosphere first proposed by Cottrell and Bilby in 1949 as observed on the effect of C solute atoms in mild steel [164]. This mechanism was later generalised as the static theory on discontinuous yielding where a yield drop occurs via the locking of dislocations by solute atoms and upon reaching a critical level of dislocation density, the dislocations are released all at once on a rapid level [165]. For another theory on discontinuous yielding mechanism, recrystallisation mechanism plays an important role rather than the solute atoms. This mechanism was first proposed by Johnston and Gilman in 1959 [166] where during loading rapid flow softening occurs due to a sudden influx of dislocations, thus, leads to discontinuous yielding [165]. Zhao et al. observed this behaviour in GH4049 wrought Ni-based superalloy for high-speed deformation ranges between 10-50s$^{-1}$ [167]. It is likely that dynamic yielding is the driving force for the discontinuous yielding as DDRX is the main dynamic recrystallisation mechanism for René 41 which relies on the total grain boundary area for nucleation sites. As the initial grain size for the René 41 Gleeble samples are large at around 200 μm relative to the total volume of around 212 mm$^3$, the total grain boundary area is limited during initial loading. Hence, rapid work hardening occurs. Once the critical dislocation density is reached, rapid flow softening due to DDRX occurs at once and temporarily lower the flow stress shortly before increasing again with the newly formed recrystallised grains.

## 5.2 Correlation analysis of industrial René 41 dataset

### 5.2.1 Analysis of ultrasonic defects based on manufacturing history

The KDE plot shown in Figure 4.17 represents the location of UT defects detected along the billet radial and longitudinal direction. The most noticeable feature is the concentration of UT defects within the centre to the mid-centre position in the radial direction for all UT defects. This is an important observation as it conveys the preference for UT defects to reside close to the centre of the billet. A possible explanation for the origin of the UT defects is the occurrence of white spots, specifically dendritic white spots. UT inspection have been reported to pick up white spots as defects in Ni-based superalloys [169]. As mentioned in section 2.3.1 dendritic white spots are the remnants of fall-in material arising from the arc instability around solidification defects in the VIM electrode such as solidification pipes and porosity. These dendritic white spots have a solute-lean composition, hence, have a higher solidus temperature than the melt pool composition during VAR and remain in the final solidification structure. However, Jackman et al. [71] reported that UT defects are more likely to occur in 'dirty' discrete white spots than dendritic white spots as inclusion clusters have a strong tendency to form cracks. The same authors also reported discrete white spots with inclusion clusters to have much more severe consequences on mechanical properties than dendritic white spots, hence, the likelihood of dendritic white spots as the root cause for UT defects is low.

Another mechanism arising from the observation of UT defect location is the possibility of macro-segregation of solute elements. Cui et al. [170] reported that in a numerically simulated VAR processed Ni-based superalloy, macro-segregation of solute elements such as Al was found in both the radial and longitudinal direction. The solute segregation ratio as displayed in Figure 5.1 increases linearly from the bottom of the ingot to the mid-length position where a sharp drop can be observed. This pattern can be observed to overlap in Figure. 4.17 where the middle section has the most concentration of UT defects. This also explains why the bottom of the René 41 billets were the least impacted region in the longitudinal direction. However, it should be noted that the chemical composition of the studied Ni-based superalloy by Cui et al. [170] is different to the René 41 chemical composition, thereby, the segregation behaviour may change for different elements. The following MatCalc simulations are calculated using the Scheil solidification equation as shown in Figure 5.2-5.6. These MatCalc simulations represent the changes in equilibrium concentration of different phases and solute elements during solidification. This can be used as qualitative analysis to investigate the segregation behaviour in René 41 during VAR ingot processing. The solidus as shown in Figure 5.2 is estimated to be around 1340°C with phase fractions during the solidification stage mainly consisting of the liquid and γ-matrix with a small fraction of less than 0.01 at.% total phase fraction of primary MCN present in the microstructure. For major elements in Figure 4.3, Mo, and Cr is observed to have a tendency for positive segregation (enrichment of solutes) while Ni is observed to have a tendency for negative segregation (depletion of solutes). As for the minor alloying elements, Si has a positive segregation while C and Mn tend to segregate erratically. The chemical elements mentioned display changes in equilibrium composition during solidification, hence, will contribute to macro-segregation behaviour. On the other hand, elements such as Al, Ti, Co, and Fe for the major alloying elements as well as B and Zr for the minor alloying elements reported much lower differences between transitioning from the 2-phase region to the solidus composition. This leads to lower segregation in the ingot.

It should also be noted that the composition of MCN favours N at temperatures above the solidus but as favours C at temperatures below the solidus. As the phase fraction of MCN is significant mostly below the liquidus, the majority of primary MCN precipitates are expected to be in the form of MC carbides in the solidified microstructure. The segregation of Cr, and Mo can lead to changes in the distribution of secondary carbide populations inside the René 41 billets. As mentioned in section 2.1, secondary carbides such as $M_6C$ and $M_{23}C_6$ have been observed in René 41 literature. Therefore, regions with enrichment in Cr and Mo are expected to have increased populations of $M_6C$ and $M_{23}C_6$ carbides at their respective sub-solvus temperatures [39,60]. Another consideration is the distribution of MC carbides as the decomposition of MC carbides contribute for the formation of secondary carbides as highlighted before in equation 2.1 and 2.2. Hence, there is a possibility that segregation behaviour of carbide formers leads to clustering of carbides which can lead to UT defects [71].

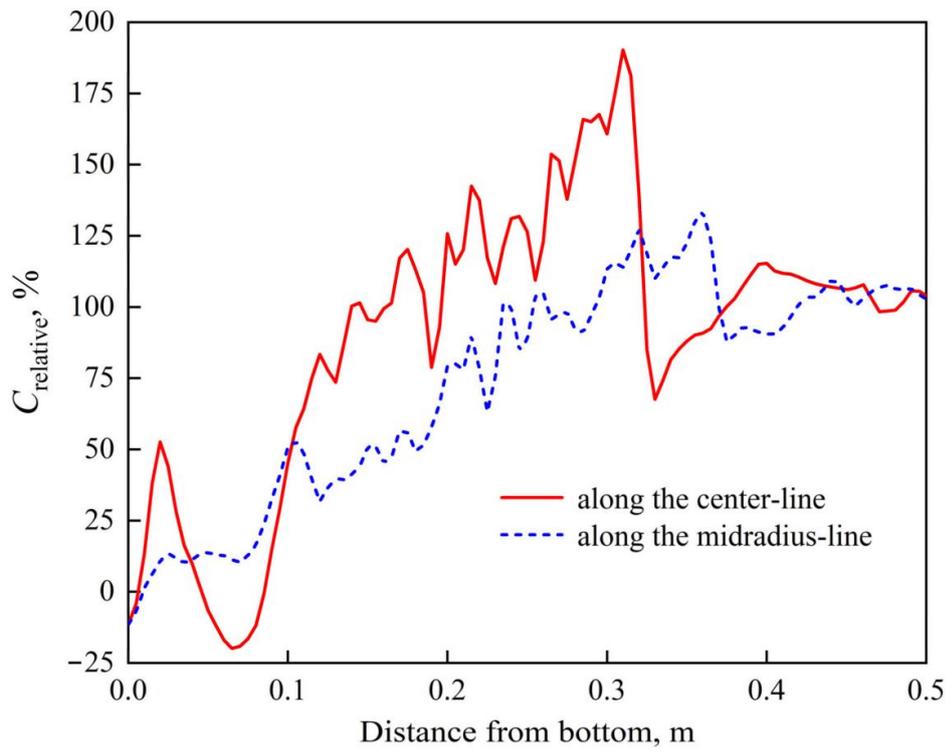

Figure 5.1 Relative solute segregation ratio in the longitudinal direction of simulated VAR ingot [162]

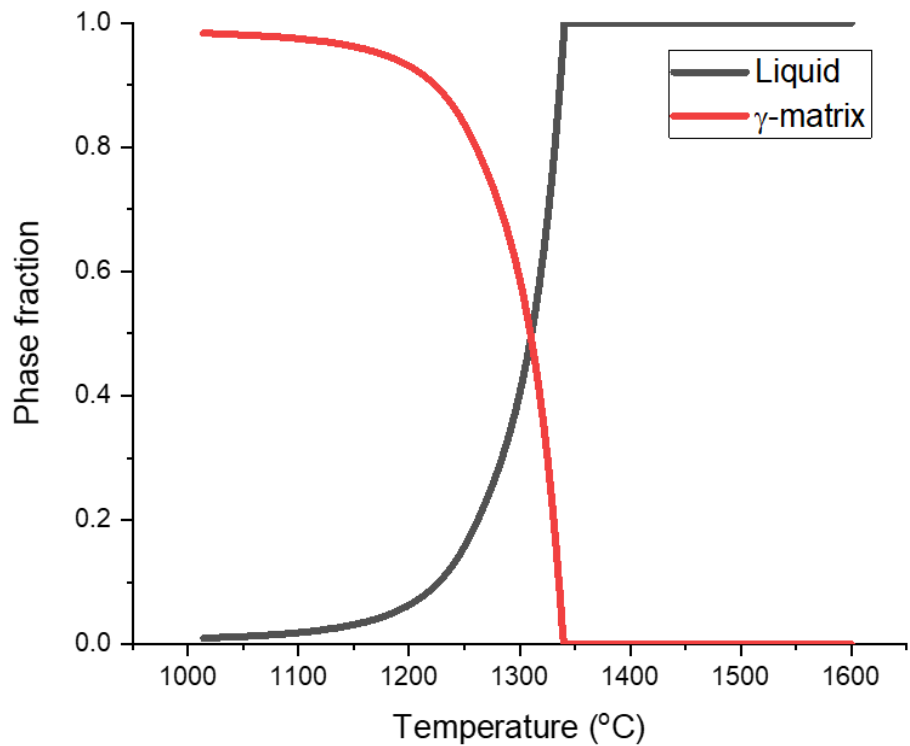

Figure 5.2 MatCalc simulation of liquid and matrix phase fractions during solidification

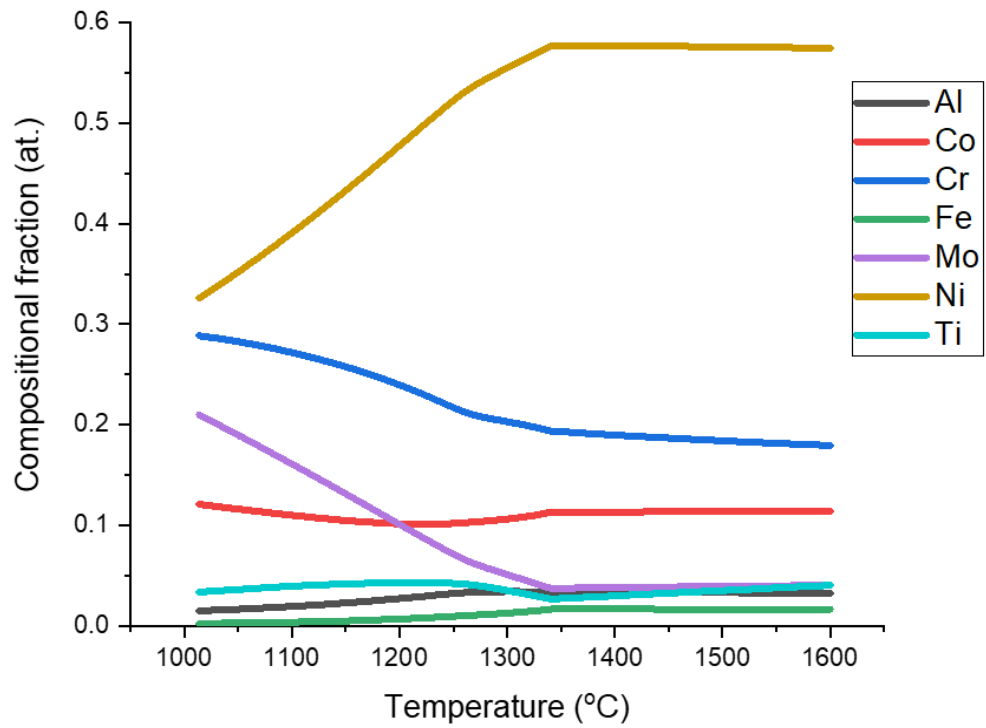

Figure 5.3 MatCalc simulation of γ-phase major alloying element compositions during solidification.

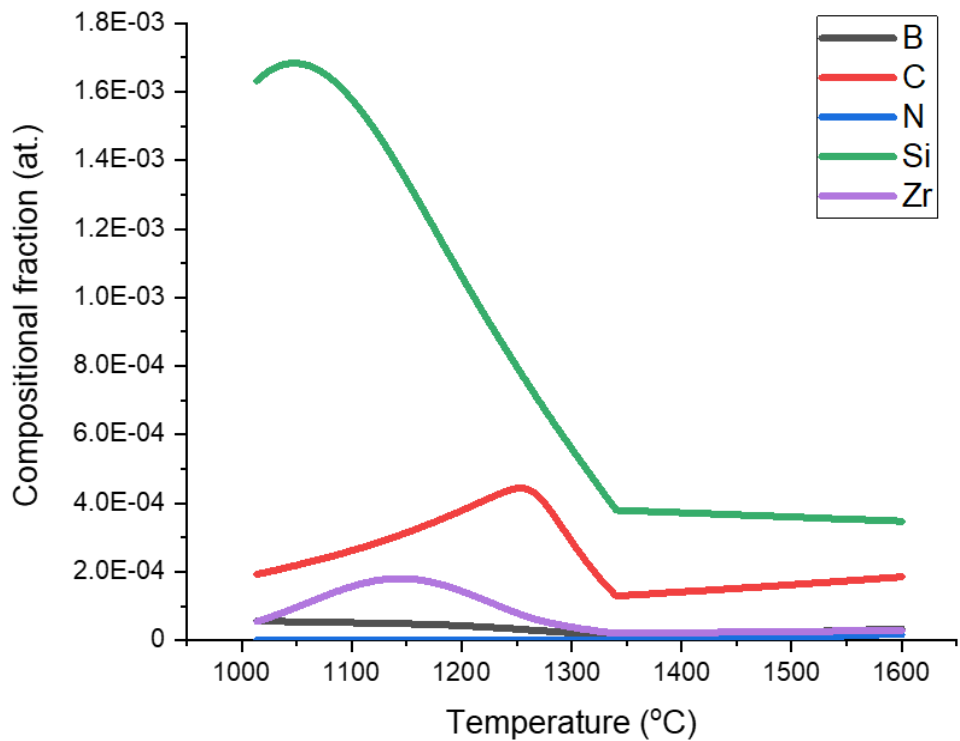

Figure 5.4 MatCalc simulation of γ-phase minor alloying element compositions in the during solidification

### 5.2.2 Analysis of ultrasonic defects based on Machine learning model evaluation

Based on literature review of carbides and borides in section 2.1.3, both carbides and borides are expected to have very similar forming elements such as Cr and Mo. Analysis from section 5.2.1 shows C and B are expected to have a different segregation behaviour as B content remains relatively constant during solidification but C content display fluctuation during solidification. This could have a strong influence on UT defects but as shown in Figures 4.16 and 4.17, no clear patterns could be observed between distribution with/without UT defects, considering the differences in samples size and relative closeness of the distributions for both C and B content. However, rearrangement of the data representation in Figures 4.18 and 4.19 provided better clarify in differentiate between distributions with/without UT defects. This conveys that there are subtle differences within the C and B distributions that can be hard to detect by conventional evaluation methodologies.

The results of the different classification models for predicting UT defects based on bulk chemical composition is shown in Figure 4.22. The highest scoring classification model was the SVC with a score of 72. The other classification models are within 10% of SVC which indicates that the model performance is relatively insensitive to the classification model selected. Since each classification model has been pre-processed identically, the effects of pre-processing on the comparability of model performances should be minimal. SVC have been reported to have a better performance than other classifications in terms of handling problems with class imbalances [171]. Additionally, studies with SVMs reported better generalisation capabilities than other conventional classification algorithms [110]. Therefore, SVC have a competitive advantage against all the classification models chosen. However, there needs to be further improvement into the model performance if the classification model were to be used in industrial application as rejection criteria. One major issue of training the model is that there is severe class imbalance present in the industrial René 41 datasets. This is susceptible to increased error in training classification models [122]. Another problem is the relatively small dataset size of 76 unique data points for developing the classification models. Huang and Li [111] reported similar issues in model performances due to small sample sizes for their study. Data imputation is an alternative to artificially increase the data size by interpolating addition data points based on available data [172]. Jakobsen et al. [173] highlighted the benefits of data imputation in clinical trial statistical analysis. In addition to increasing sample size, increasing the number of relevant features can also been of use. Incorporating various forming parameters such as forging strain rates, furnace temperatures, and strain level could establish more reliable models [112].

Figure 4.23 shows the visual presentation of the CART classification model based on the chemical compositions used. It can be observed that the UT which represents chemical compositions with UT defects and non-UT which represents chemical composition without UT defects is separated via the decision nodes which perform inequality criteria based on the model calculation. The random forest classifier is of the same concept but pools the outputs of many CART trees together as the main decision criteria. In the random forest classifier used in this study, a total of 1000 CART trees were used. Feature importance diagram collects data on the number of times a feature (i.e., chemical elements) is used in CART. These are then ranked amongst the features to project the most important feature to the least important feature for the Random Forest classifier. It is a useful metric to gain insights on the relevancy of specific features to the UT class. In Figure 4.24, the feature importance diagram shows that C, N and Ti are amongst the highest-ranking features for the random forest classifier. This conveys that these elements are associated by the classifier to be the most relevant for detecting UT defects in each René 41 chemical composition. However, Fe, B and Zr are the lowest ranking features for the random forest classifier. Interpretation of the high-ranking features requires human intervention as machine learning algorithms does not have the capability to interpret its own output. C, N and Ti consequentially is also the main forming element for primary MCN precipitates. The prediction of the model gives insights as to the mechanisms behind the UT defect detections. However, unless a more detailed investigation is conducted into the UT defects the mechanism can only be speculated in this study.

In section 5.1 the presence of secondary phases is reported to be important for hot deformation behaviour of René 41, specifically, the lack of secondary phases will cause restoration mechanisms to be impeded especially in material with a coarse grain structure. Another factor to be considered is the segregation behaviour of solute elements during solidification stage. Solute segregation in VAR processes can have large difference across both the longitudinal and radial directions. These differences allow for agglomeration of primary MC type carbides located across the billet length. This in addition to the presence of γ' strengthening precipitates during sub γ' solvus deformation which increases tendency to form of shear bands. Tang et al. [173] reported that plane-strain compression which is typically experienced cogging causes shear banding to be more intense than uniaxial compression loading. Ultimately, these shear band formations will give rise to cracks within the microstructure that can be detected during ultrasonic inspection. Hence, it is likely that controlling the MCN precipitate distribution along the billet is the key to controlling the UT defects population.

# Conclusions

In this thesis, two approaches were combined to investigate the influence of C and B additions on the processability of René 41. First, three experimental compositions of René 41 (nominal, high C, and high B) were fabricated to assess their hot deformation behaviour. Second, data analysis and machine learning models were developed to investigate chemical compositions linkage to ultrasonic defects found in industrially fabricated René 41 billets.

The major conclusions are outlined as follows:

- Coarse grain structure and γ' precipitates have a large influence on the hot deformation behaviour and the underlying dynamic recrystallisation mechanism of René 41
- High C content René 41 exhibits the best hot workability with an activation energy of 697 kJ/mol
- The presence of carbides and borides in the high B variant and high C variant are postulated to be the main factor in the additional recrystallised fraction observed in both the variants as compared to the nominal René 41
- Kernel density estimation plot shows all the ultrasonic defects found within René 41 billets were concentrated at the centre region along the billet length and around the centre to mid-centre region in the radial direction.
- Feature importance shows that controlling C, N, and Ti elements is key to predicting ultrasonic defects. Combined with thermodynamic simulation results where Cr, Mo, C are found to be the elements with the highest segregation behaviours during solidification process, it is postulated that agglomeration of MCN type carbides are the probable causes for ultrasonic defects.

Based on findings in this thesis, the microalloying of B and C should be adjusted in René 41 so that a mixed population of carbides and borides can co-exist to promote better hot workability without the risk of forming internal defects. Going forward, a more thorough experimental and theoretical analysis involving hot deformation under different strain to present direct evidence of the recrystallisation mechanisms occurring in René 41 would be extremely valuable in improving the processability of René 41. Another area to improve in future research is the phase identification of borides and carbides present within the René 41 microstructure. Lastly, material characterisation of the UT defect and its surrounding microstructure can be used as an experimental validation for the claim as outlined by the machine learning model on the linkage between UT defect and chemical composition. All of this would be invaluable in unravelling the interrelationship between C and B microalloying additions and the processability of René 41.

# References

[1]     W. Betteridge, S.W.K. Shaw, Development of superalloys, Materials Science and Technology (United Kingdom). 3 (1987) 682–694. https://doi.org/10.1179/mst.1987.3.9.682.

[2]     C.T. Sims, A History of Superalloy Metallurgy for Superalloy Metallurgists, Superalloys 1984 (Fifth International Symposium). (1984) 399–419. https://doi.org/10.7449/1984/Superalloys_1984_399_419.

[3]     R.C. Reed, SUPERALLOYS II, Cambridge University Press, Cambridge, UK, 2006.

[4]     M.J. Donachie, S.J. Donachie, SUPERALLOYS Second Edition, 2002. https://doi.org/10.1089/jwh.1.1999.8.637.

[5]     R.E. Schafrik, R. Sprague, Gas turbine materials, Advanced Materials and Processes. 162 (2004) 41–46. https://doi.org/10.1016/0016-0032(58)90740-3.

[6]     M.C. Hardy, M. Detrois, E.T. McDevitt, C. Argyrakis, V. Saraf, P.D. Jablonski, J.A. Hawk, R.C. Buckingham, H.S. Kitaguchi, S. Tin, Solving Recent Challenges for Wrought Ni-Base Superalloys, Metall Mater Trans A Phys Metall Mater Sci. 51 (2020) 2626–2650. https://doi.org/10.1007/s11661-020-05773-6.

[7]     A.M. Sabroff, F.W. Boulger, H.J. Henning, Forging Materials and Practices, Reinhold Book Corp., New York, 1968.

[8]     L. Weisenberg, R. Morris, How to fabricate Rene' 41, Metal Progress. (1960) 70–74.

[9]     superalloys-market12, (n.d.). https://www.alliedmarketresearch.com/superalloys-market (accessed April 25, 2019).

[10]    P. Kontis, E. Chauvet, Z. Peng, J. He, A.K. da Silva, D. Raabe, C. Tassin, J.J. Blandin, S. Abed, R. Dendievel, B. Gault, G. Martin, Atomic-scale grain boundary engineering to overcome hot-cracking in additively-manufactured superalloys, Acta Mater. 177 (2019) 209–221. https://doi.org/10.1016/j.actamat.2019.07.041.

[11]    Z. Asqary, S.M. Abbasi, M. Seifollahi, M. Morakabati, The effect of boron and zirconium on the microstructure and tensile properties of Nimonic 105 superalloy, Mater Res Express. 6 (2019). https://doi.org/10.1088/2053-1591/ab4676.

[12]    J. Grodzki, N. Hartmann, R. Rettig, E. Affeldt, R.F. Singer, Effect of B, Zr, and C on Hot Tearing of a Directionally Solidified Nickel-Based Superalloy, Metall Mater Trans A Phys Metall Mater Sci. 47 (2016) 2914–2926. https://doi.org/10.1007/s11661-016-3416-8.

[13] S. Floreen, J.M. Davidson, Effects of B and Zr On The Creep and Fatigue Crack Growth Behavior of a Ni-Base Superalloy, Metallurgical Transactions. A, Physical Metallurgy and Materials Science. 14 A (1983) 895–901. https://doi.org/10.1007/bf02644294.

[14] K. Zhao, Y.H. Ma, L.H. Lou, Improvement of creep rupture strength of a liquid metal cooling directionally solidified nickel-base superalloy by carbides, J Alloys Compd. 475 (2009) 648–651. https://doi.org/10.1016/j.jallcom.2008.07.127.

[15] J.W. Freeman, R.F. Decker, The mechanism of beneficial effects of boron and zirconium on creep properties of a complex heat-resistant alloy, Trans. Metall. Soc. AIME. (1960).

[16] Y.C. Lin, X.M. Chen, A critical review of experimental results and constitutive descriptions for metals and alloys in hot working, Mater Des. 32 (2011) 1733–1759. https://doi.org/10.1016/j.matdes.2010.11.048.

[17] Y.S. Na, J.T. Yeom, N.K. Park, J.Y. Lee, Simulation of microstructures for Alloy 718 blade forging using 3D FEM simulator, J Mater Process Technol. 141 (2003) 337–342. https://doi.org/10.1016/S0924-0136(03)00285-1.

[18] B. Roebuck, J.D. Lord, M. Brooks, M.S. Loveday, C.M. Sellars, R.W. Evans, Measurement of flow stress in hot axisymmetric compression tests, Materials at High Temperatures. 23 (2006) 59–83. https://doi.org/10.1179/mht.2006.005.

[19] A.J. Schwartz, B.L. Adams, D.P. Field, M. Kumar, Electron Backscatter Diffraction in Materials Science, 2nd ed., Springer US, 2009.

[20] H.A. Roth, C.L. Davis, R.C. Thomson, Modeling solid solution strengthening in nickel alloys, Metall Mater Trans A Phys Metall Mater Sci. 28 (1997) 1329–1335. https://doi.org/10.1007/s11661-997-0268-2.

[21] Y. Xiong, A. Yang, Y. Guo, W. Liu, L. Liu, Effect of fine-grained structure on the mechanical properties of superalloys K3 and K4169, Sci Technol Adv Mater. 2 (2001) 7–11. https://doi.org/10.1016/S1468-6996(01)00018-3.

[22] M.A. Meyers, L.E. Murr, A model for the formation of annealing twins in F.C.C. metals and alloys, Acta Metallurgica. 26 (1978) 951–962. https://doi.org/10.1016/0001-6160(78)90046-9.

[23] M. Qian, J.C. Lippold, The effect of annealing twin-generated special grain boundaries on HAZ liquation cracking of nickel-base superalloys, Acta Mater. 51 (2003) 3351–3361. https://doi.org/10.1016/S1359-6454(03)00090-9.

[24] B.G. Choi, Eta phase formation and its effect on mechanical properties in ni-base superalloy, Superalloys. (2004) 163–173.

[25] F. Long, Y.S. Yoo, C.Y. Jo, S.M. Seo, Y.S. Song, T. Jin, Z.Q. Hu, Formation of η and σ phase in three polycrystalline superalloys and their impact on tensile properties, Materials Science and Engineering A. 527 (2009) 361–369. https://doi.org/10.1016/j.msea.2009.09.016.

[26] H.M. Lalvani, M.A. Rist, J.W. Brooks, Effect of delta phase on the hot deformation behaviour and microstructural evolution of inconel 718, Adv Mat Res. 89–91 (2010) 313–318. https://doi.org/10.4028/www.scientific.net/AMR.89-91.313.

[27] R.O. Williams, Origin of strengthening on precipitation: Ordered particles, Acta Metallurgica. 5 (1957) 241–244. https://doi.org/10.1016/0001-6160(57)90096-2.

[28] A.J. Goodfellow, Strengthening mechanisms in polycrystalline nickel-based superalloys, Materials Science and Technology (United Kingdom). 34 (2018) 1793–1808. https://doi.org/10.1080/02670836.2018.1461594.

[29] D.P. Pope, Mechanical properties of ni3ai and nickel-base alloys with high volume fraction of y', International Metals Reviews. 29 (1984) 136–167. https://doi.org/10.1179/imtr.1984.29.1.136.

[30] M.R. Ahmadi, E. Povoden-Karadeniz, L. Whitmore, M. Stockinger, A. Falahati, E. Kozeschnik, Yield strength prediction in Ni-base alloy 718Plus based on thermo-kinetic precipitation simulation, Materials Science and Engineering A. 608 (2014) 114–122. https://doi.org/10.1016/j.msea.2014.04.054.

[31] R.W. Kozar, A. Suzuki, W.W. Milligan, J.J. Schirra, M.F. Savage, T.M. Pollock, Strengthening mechanisms in polycrystalline multimodal nickel-base superalloys, Metall Mater Trans A Phys Metall Mater Sci. 40 (2009) 1588–1603. https://doi.org/10.1007/s11661-009-9858-5.

[32] M. Mehl, D. Hicks, C. Toher, O. Levy, R. Hanson, G. Hart, S. Curtarolo, The AFLOW Library of Crystallographic Prototypes: Part 1, Comput Mater Sci. 136 (2017) 1–828.

[33] A. Jain, S. Ong, G. Hautier, W. Chen, W. Richards, S. Dacek, S. Cholia, D. Gunter, D. Skinner, G. Ceder, K. Persson, Commentary: The Materials Project: A materials genome approach to accelerating materials innovation, APL Mater. 1 (2013).

[34] P.B. Hirsch, Kear-wilsdorf locks, jogs and the formation of antiphase-boundary tubes in Ni3Al, Philosophical Magazine A: Physics of Condensed Matter, Structure, Defects and Mechanical Properties. 74 (1996) 1019–1040. https://doi.org/10.1080/01418619608242174.

[35] R.J. Mitchell, Polycrystalline Nickel-Based Superalloys: Processing, Performance, and Application, Encyclopedia of Aerospace Engineering. (2010) 1–12. https://doi.org/10.1002/9780470686652.eae217.

[36] T.M. Pollock, S. Tin, Nickel-Based Superalloys for Advanced Turbine Engines: Chemistry, Microstructure and Properties, J Propuls Power. 22 (2006) 361–374. https://doi.org/10.2514/1.18239.

[37] M.P. Jackson, R.C. Reed, Heat treatment of UDIMET 720Li: the effect of microstructure on properties, 1999.

[38] G.P. Sabol, R. Stickler, Microstructure of Nickel-Based Superalloys, Physica Status Solidi (B). 35 (1969) 11–52. https://doi.org/10.1002/pssb.19690350102.

[39] B. Geddes, H. Leon, X. Huang, Superalloys Alloying and Performance, Materials Park, Ohio, 2010. www.asminternational.orgwww.asminternational.org.

[40] K. Kuo, The formation of η carbides, Acta Metallurgica. 1 (1953) 301–304. https://doi.org/10.1016/0001-6160(53)90103-5.

[41] L. Jiang, X.X. Ye, Z.Q. Wang, C. Yu, J.S. Dong, R.B. Xie, Z.J. Li, L. Yan, T.K. Sham, X.T. Zhou, A.M. Li, The critical role of Si doping in enhancing the stability of M6C carbides, J Alloys Compd. 728 (2017) 917–926. https://doi.org/10.1016/j.jallcom.2017.09.042.

[42] X. Dong, X. Zhang, K. Du, Y. Zhou, T. Jin, H. Ye, Microstructure of Carbides at Grain Boundaries in Nickel Based Superalloys, J Mater Sci Technol. 28 (2012) 1031–1038. https://doi.org/10.1016/S1005-0302(12)60169-8.

[43] J. Yang, Q. Zheng, X. Sun, H. Guan, Z. Hu, Relative stability of carbides and their effects on the properties of K465 superalloy, Materials Science and Engineering A. 429 (2006) 341–347. https://doi.org/10.1016/j.msea.2006.05.091.

[44] X.B. Hu, Y.L. Zhu, N.C. Sheng, X.L. Ma, The Wyckoff positional order and polyhedral intergrowth in the $M_3B_2$ - and $M_5B_3$ -type boride precipitated in the Ni-based superalloys, Sci Rep. 4 (2014) 2–10. https://doi.org/10.1038/srep07367.

[45] X.B. Hu, Y.L. Zhu, X.L. Ma, Crystallographic account of nano-scaled intergrowth of $M_2B$-type borides in nickel-based superalloys, Acta Mater. 68 (2014) 70–81. https://doi.org/10.1016/j.actamat.2014.01.002.

[46] X.B. Hu, H.Y. Niu, X.L. Ma, A.R. Oganov, C.A.J. Fisher, N.C. Sheng, J.D. Liu, T. Jin, X.F. Sun, J.F. Liu, Y. Ikuhara, Atomic-scale observation and analysis of chemical ordering in $M_3B_2$ and $M_5B_3$ borides, Acta Mater. 149 (2018) 274–284. https://doi.org/10.1016/j.actamat.2018.02.055.

[47] N. Sheng, X. Hu, J. Liu, T. Jin, X. Sun, Z. Hu, $M_3B_2$ and $M_5B_3$ Formation in Diffusion-Affected Zone During Transient Liquid Phase Bonding Single-Crystal Superalloys, Metall Mater Trans A Phys Metall Mater Sci. 46 (2015) 1670–1677. https://doi.org/10.1007/s11661-014-2733-z.

[48] M.J. Kaufman, V.I. Levit, Characterization of chromium boride precipitates in the commercial superalloy GTD 111 after long-term exposure, Philos Mag Lett. 88 (2008) 259–267. https://doi.org/10.1080/09500830801905445.

[49] D. Wang, C. Liu, J. Zhang, L. Lou, Evolution of Grain Boundary Precipitates in a Directionally Solidified Ni-Base Superalloy during High Temperature Creep, Superalloys 2012. (2012) 363–368. https://doi.org/10.1002/9781118516430.ch39.

[50] P. Kontis, H.A.M. Yusof, S. Pedrazzini, M. Danaie, K.L. Moore, P.A.J. Bagot, M.P. Moody, C.R.M. Grovenor, R.C. Reed, On the effect of boron on grain boundary character in a new polycrystalline superalloy, Acta Mater. 103 (2016) 688–699. https://doi.org/10.1016/j.actamat.2015.10.006.

[51] H.E. Collins, Relative Stability of Carbide and Intermetallic Phases in Nickel-Base Superalloys, in: TMS Superalloys 1968, 1968: pp. 171–198. https://doi.org/10.7449/1968/superalloys_1968_171_198.

[52] R.T. Holt, W. Wallace, R.T. Holt, W. Wallace, Impurities and trace elements in nickel-base superalloys Impurities and trace elements in nickel-base superalloys, 4590 (2013). https://doi.org/10.1179/imtr.1976.21.1.1.

[53] R.T. Holt, W. Wallace, Impurities and trace elements in nickel-base superalloys, International Metals Reviews. 21 (1976) 1–24. https://doi.org/10.1179/imtr.1976.21.1.1.

[54] J.E. Doherty, A.F. Giamei, B.H. Kear, The importance of grain boundary morphology and cohesion on intergranular strength, Canadian Metallurgical Quarterly. 13 (1974) 229–236. https://doi.org/10.1179/cmq.1974.13.1.229.

[55] L.R. Liu, T. Jin, N.R. Zhao, Z.H. Wang, X.F. Sun, H.R. Guan, Z.Q. Hu, Effect of carbon additions on the microstructure in a Ni-base single crystal superalloy, Mater Lett. 58 (2004) 2290–2294. https://doi.org/10.1016/j.matlet.2004.01.038.

[56] K.A. Al-Jarba, G.E. Fuchs, Effect of carbon additions on the as-cast microstructure and defectformation of a single crystal Ni-based superalloy, Materials Science and Engineering A. 373 (2004) 255–267. https://doi.org/10.1016/j.msea.2004.01.030.

[57] A.H. Cottrell, Boron and carbon in nickel, iron, and Ni3Al, Materials Science and Technology (United Kingdom). 7 (1991) 585–586. https://doi.org/10.1179/mst.1991.7.7.585.

[58] S. Sanyal, U. v. Waghmare, P.R. Subramanian, M.F.X. Gigliotti, Effect of dopants on grain boundary decohesion of Ni: A first-principles study, Appl Phys Lett. 93 (2008) 10–13. https://doi.org/10.1063/1.3042211.

[59] J.S. Crompton, J.W. Martin, Crack growth in a single crystal superalloy at elevated temperature, Metallurgical Transactions A. 15 (1984) 1711–1719. https://doi.org/10.1007/BF02666354.

[60] A.K. Jena, M.C. Chaturvedi, The role of alloying elements in the design of nickel-base superalloys, J Mater Sci. 19 (1984) 3121–3139. https://doi.org/10.1007/BF00549796.

[61] S. GAO, J. shan HOU, Y. an GUO, L. zhang ZHOU, Phase precipitation behavior and tensile properties of as-cast Ni-based superalloy during heat treatment, Transactions of Nonferrous Metals Society of China (English Edition). 28 (2018) 1735–1744. https://doi.org/10.1016/S1003-6326(18)64817-4.

[62] L.Z. He, Q. Zheng, X.F. Sun, H.R. Guan, Z.Q. Hu, A.K. Tieu, C. Lu, H.T. Zhu, Effect of carbides on the creep properties of a Ni-base superalloy M963, Materials Science and Engineering A. 397 (2005) 297–304. https://doi.org/10.1016/j.msea.2005.02.038.

[63] K. Song, M. Aindow, Grain growth and particle pinning in a model Ni-based superalloy, Materials Science and Engineering A. 479 (2008) 365–372. https://doi.org/10.1016/j.msea.2007.09.055.

[64] J.M. Walsh, B.H. Rear, Errata: Direct evidence for boron segregation to grain boundaries in a nickel-base alloy by secondary ion mass spectrometry, Metallurgical Transactions A. 6 (1975) 950. https://doi.org/10.1007/BF02672328.

[65] Z. Jie, J. Zhang, T. Huang, H. Su, Y. Zhang, L. Liu, H. Fu, Effects of boron and zirconium additions on the fluidity, microstructure and mechanical properties of IN718C superalloy, J Mater Res. 31 (2016) 3557–3566. https://doi.org/10.1557/jmr.2016.383.

[66] K.K. Sharma, P.S. Misra, N.C. Birla, S.N. Tewari, Effect of boron on the structure and properties of a PM nickel base superalloy, Key Eng Mater. 29–31 (1989) 429–442. https://doi.org/10.4028/www.scientific.net/kem.29-31.429.

[67] R.F. Decker, J.P. Rowe, J.W. Freeman, BORON AND ZIRCONIUM FROM CRUCIBLE REFRACTORIES IN A COMPLEX HEAT·RESISTANT ALLOY, 1958. https://ntrs.nasa.gov/search.jsp?R=19930092370.

[68] R.F. Decker, J.W. Freeman, The Mechanism Of Beneficial Effect Of Boron And Zirconium On Creep Properties Of A Complex Heat-Resistant Alloy, Ann Arbor, Michigan USA, 1961.

[69] P. Kontis, A. Kostka, D. Raabe, B. Gault, Influence of composition and precipitation evolution on damage at grain boundaries in a crept polycrystalline Ni-based superalloy, Acta Mater. 166 (2019) 158–167. https://doi.org/10.1016/j.actamat.2018.12.039.

[70] S. Sanyal, U. v. Waghmare, T. Hanlon, E.L. Hall, P.R. Subramanian, M.F.X. Gigliotti, Interfaces in Ni-Based Superalloys and Implications for Mechanical Behavior and Environmental Embrittlement: A First-Principles Study, Superalloys 2012. (2012) 531–536. https://doi.org/10.1002/9781118516430.ch58.

[71] L.A. Jackman, G.E. Maurer, S. Widge, WHITE SPOTS IN SUPERALLOYS, in: E.A. Loria (Ed.), Superalloys 718, 625, 706 and Various Derivatives, The Minerals, Metals & Materials Society, 1994: pp. 153–166.

[72] E. Samuels, J.A. Domingue, G.E. Maurer, Characterizing Solute-Lean Defects in Superalloys, JOM. 42 (1990) 27–30.

[73] S. v Thamboo, MELT RELATED DEFECTS IN ALLOY 706 AND THEIR EFFECTS ON MECHANICAL PROPERTIES, in: E.A. Loria (Ed.), Superalloys 718, 725, 706 and Various Derivatives, The Minerals, Metals & Materials Society, 1994: pp. 137–152.

[74] K. 0 Yu, J.A. Domingue, CONTROL OF SOLIDIFICATION STRUCTURE IN VAR AND ESR PROCESSED ALLOY 718 INGOTS, in: E.A. Loria (Ed.), Superalloys 718 - Metallurgy and Application, The Minerals, Metals & Materials Society, 1989: pp. 33–48.

[75] K. Takachio, T. Nonomura, improvement in the Quality of Superalloy VAR Ingots, ISIJ International. 36 (1996) 85–88.

[76] J.W. Brooks, Forging of superalloys, Mater Des. 21 (2000) 297–303.

[77] G.E. Dieter, H.A. Kuhn, S.L. Semiatin, Handbook of Workability and Process Design, ASM International, 2003.

[78] H.K. Zhang, H. Xiao, X.W. Fang, Q. Zhang, R.E. Logé, K. Huang, A critical assessment of experimental investigation of dynamic recrystallization of metallic materials, Mater Des. 193 (2020) 108873. https://doi.org/10.1016/j.matdes.2020.108873.

[79] A. Gill, A. Telang, S.R. Mannava, D. Qian, Y.-S. Pyoun, H. Soyama, V.K. Vasudevan, Comparison of mechanisms of advanced mechanical surface treatments in nickel-based superalloy, (2013). https://doi.org/10.1016/j.msea.2013.04.021.

[80] F.J. Humphreys, M. Hatherly, Recrystallization and Related Annealing Phenomena, Second Edi, 2004. https://doi.org/10.1017/CBO9781107415324.004.

[81] R.D. Doherty, D.A. Hughes, F.J. Humphreys, J.J. Jonas, D. Juul Jensen, M.E. Kassner, W.E. King, T.R. McNelley, H.J. McQueen, A.D. Rollett, Current issues in recrystallization: A review, Materials Science and Engineering A. 238 (1997) 219–274. https://doi.org/10.1016/S0921-5093(97)00424-3.

[82] K. Sahithya, I. Balasundar, P. Pant, T. Raghu, H.K. Nandi, V. Singh, P. Ghosal, M. Ramakrishna, Deformation behaviour of an as-cast nickel base superalloy during primary hot working above and below the gamma prime solvus, Materials Science and Engineering A. 754 (2019) 521–534. https://doi.org/10.1016/j.msea.2019.03.083.

[83] K. Huang, R.E. Logé, A review of dynamic recrystallization phenomena in metallic materials, Mater Des. 111 (2016) 548–574. https://doi.org/10.1016/j.matdes.2016.09.012.

[84] T. Sakai, A. Belyakov, R. Kaibyshev, H. Miura, J.J. Jonas, Dynamic and post-dynamic recrystallization under hot, cold and severe plastic deformation conditions, Prog Mater Sci. 60 (2014) 130–207. https://doi.org/10.1016/j.pmatsci.2013.09.002.

[85] D. Jia, W. Sun, D. Xu, F. Liu, Dynamic recrystallization behavior of GH4169G alloy during hot compressive deformation, J Mater Sci Technol. 35 (2019) 1851–1859. https://doi.org/10.1016/j.jmst.2019.04.018.

[86] H. Jiang, L. Yang, J. Dong, M. Zhang, Z. Yao, The recrystallization model and microstructure prediction of alloy 690 during hot deformation, Mater Des. 104 (2016) 162–173. https://doi.org/10.1016/j.matdes.2016.05.033.

[87] X. Qin, D. Huang, X. Yan, X. Zhang, M. Qi, S. Yue, Hot deformation behaviors and optimization of processing parameters for Alloy 602 CA, J Alloys Compd. 770 (2019) 507–516. https://doi.org/10.1016/j.jallcom.2018.08.144.

[88] Y.C. Lin, X.Y. Wu, X.M. Chen, J. Chen, D.X. Wen, J.L. Zhang, L.T. Li, EBSD study of a hot deformed nickel-based superalloy, J Alloys Compd. 640 (2015) 101–113. https://doi.org/10.1016/j.jallcom.2015.04.008.

[89] B. Xie, H. Yu, T. Sheng, Y. Xiong, Y. Ning, M.W. Fu, DDRX and CDRX of an as-cast nickel-based superalloy during hot compression at γ′ sub-/super-solvus temperatures, J Alloys Compd. 803 (2019) 16–29. https://doi.org/10.1016/j.jallcom.2019.06.202.

[90] F. Liu, J. Chen, J. Dong, M. Zhang, Z. Yao, The hot deformation behaviors of coarse, fine and mixed grain for Udimet 720Li superalloy, Materials Science and Engineering A. 651 (2016) 102–115. https://doi.org/10.1016/j.msea.2015.10.099.

[91] K. Sahithya, I. Balasundar, P. Pant, T. Raghu, Primary hot working characteristics of an as-cast and homogenized nickel base superalloy DMR-742 in the sub and super-solvus temperature regime, J Alloys Compd. 821 (2020) 153455. https://doi.org/10.1016/j.jallcom.2019.153455.

[92] Z. Wan, L. Hu, Y. Sun, T. Wang, Z. Li, Microstructure evolution and dynamic softening mechanisms during high-temperature deformation of a precipitate hardening Ni-based superalloy, Vacuum. 155 (2018) 585–593. https://doi.org/10.1016/j.vacuum.2018.06.068.

[93] M.A. Charpagne, J.M. Franchet, N. Bozzolo, Overgrown grains appearing during sub-solvus heat treatment in a polycrystalline γ-γ' Nickel-based superalloy, Mater Des. 144 (2018) 353–360. https://doi.org/10.1016/j.matdes.2018.02.048.

[94] V.M. Miller, A.E. Johnson, C.J. Torbet, T.M. Pollock, Recrystallization and the Development of Abnormally Large Grains After Small Strain Deformation in a Polycrystalline Nickel-Based Superalloy, Metallurgical and Materials Transactions A. 47 (2016) 1566–1574. https://doi.org/10.1007/s11661-016-3329-6.

[95] L. Wang, G. Xie, J. Zhang, L.H. Lou, On the role of carbides during the recrystallization of a directionally solidified nickel-base superalloy, Scr Mater. 55 (2006) 457–460. https://doi.org/10.1016/j.scriptamat.2006.05.013.

[96] S.A. Hosseini, S.M. Abbasi, K.Z. Madar, H.M.K. Yazdi, The effect of boron and zirconium on wrought structure and γ-γ′ lattice misfit characterization in nickel-based superalloy ATI 718Plus, Mater Chem Phys. 211 (2018) 302–311. https://doi.org/10.1016/j.matchemphys.2018.01.076.

[97] R.A. Ricks, A.J. Porter, R.C. Ecob, The growth of γ′ precipitates in nickel-base superalloys, Acta Metallurgica. (1983). https://doi.org/10.1016/0001-6160(83)90062-7.

[98] A. Chamanfar, H.S. Valberg, B. Templin, J.E. Plumeri, W.Z. Misiolek, Development and validation of a finite-element model for isothermal forging of a nickel-base superalloy, Materialia (Oxf). 6 (2019). https://doi.org/10.1016/j.mtla.2019.100319.

[99]  Y.C. Lin, M.S. Chen, J. Zhong, Constitutive modeling for elevated temperature flow behavior of 42CrMo steel, Comput Mater Sci. 42 (2008) 470–477. https://doi.org/10.1016/j.commatsci.2007.08.011.

[100] C.M. Sellars, J.M. Mc.Tegart, On the mechanism of hot deformation, Acta Matallurgica. 14 (1966) 1136–1138.

[101] A. Amiri, S. Bruschi, M.H. Sadeghi, P. Bariani, Investigation on hot deformation behavior of Waspaloy, Materials Science and Engineering A. 562 (2013) 77–82. https://doi.org/10.1016/j.msea.2012.11.024.

[102] Y. Wang, J. Wang, J. Dong, A. Li, Z. Li, G. Xie, L. Lou, Hot deformation characteristics and hot working window of as-cast large-tonnage GH3535 superalloy ingot, J Mater Sci Technol. 34 (2018) 2439–2446. https://doi.org/10.1016/j.jmst.2018.04.001.

[103] A.K. Godasu, U. Prakash, S. Mula, Flow stress characteristics and microstructural evolution of cast superalloy 625 during hot deformation, J Alloys Compd. 844 (2020) 156200. https://doi.org/10.1016/j.jallcom.2020.156200.

[104] D.X. Wen, Y.C. Lin, H. bin Li, X.M. Chen, J. Deng, L.T. Li, Hot deformation behavior and processing map of a typical Ni-based superalloy, Materials Science and Engineering A. 591 (2014) 183–192. https://doi.org/10.1016/j.msea.2013.09.049.

[105] F. Chen, J. Liu, H. Ou, B. Lu, Z. Cui, H. Long, Flow characteristics and intrinsic workability of IN718 superalloy, Materials Science and Engineering A. 642 (2015) 279–287. https://doi.org/10.1016/j.msea.2015.06.093.

[106] J. Schmidt, M.R.G. Marques, S. Botti, M.A.L. Marques, Recent advances and applications of machine learning in solid-state materials science, NPJ Comput Mater. 5 (2019). https://doi.org/10.1038/s41524-019-0221-0.

[107] L. Meng, B. McWilliams, W. Jarosinski, H.Y. Park, Y.G. Jung, J. Lee, J. Zhang, Machine Learning in Additive Manufacturing: A Review, Jom. 72 (2020) 2363–2377. https://doi.org/10.1007/s11837-020-04155-y.

[108] E. Ruiz, D. Ferreño, M. Cuartas, A. López, V. Arroyo, F. Gutiérrez-Solana, Machine learning algorithms for the prediction of the strength of steel rods: an example of data-driven manufacturing in steelmaking, Int J Comput Integr Manuf. 33 (2020) 880–894. https://doi.org/10.1080/0951192X.2020.1803505.

[109] Y.C. Lin, F.Q. Nong, X.M. Chen, D.D. Chen, M.S. Chen, Microstructural evolution and constitutive models to predict hot deformation behaviors of a nickel-based superalloy, Vacuum. 137 (2017) 104–114. https://doi.org/10.1016/j.vacuum.2016.12.022.

[110] C. del Vecchio, G. Fenu, F.A. Pellegrino, D.F. Michele, M. Quatrale, L. Benincasa, S. Iannuzzi, A. Acernese, P. Correra, L. Glielmo, Support Vector Representation Machine for superalloy investment casting optimization, Appl Math Model. 72 (2019) 324–336. https://doi.org/10.1016/j.apm.2019.02.033.

[111] D.J. Huang, H. Li, A machine learning guided investigation of quality repeatability in metal laser powder bed fusion additive manufacturing, Mater Des. 203 (2021) 109606. https://doi.org/10.1016/j.matdes.2021.109606.

[112] E. Hamouche, E.G. Loukaides, Classification and selection of sheet forming processes with machine learning, Int J Comput Integr Manuf. 31 (2018) 921–932. https://doi.org/10.1080/0951192X.2018.1429668.

[113] G. Bonaccorso, Mastering Machine Learning Algorithms (2nd Edition), Second Editions, 2020.

[114] A.C. Muller, S. Guido, Introduction to Machine Learning with Python : A Guide for Data Scientists, O'Reily Media, Incorporated, 2016.

[115] M. Galarnyk, Understanding Train Test Split (Scikit-Learn + Python), Towards Data Science. (2022). https://towardsdatascience.com/understanding-train-test-split-scikit-learn-python-ea676d5e3d1 (accessed May 18, 2022).

[116] P. Gupta, N.K. Sehgal, Introduction to Machine Learning in the Cloud with Python, Springer International Publishing, 2021. https://doi.org/10.1007/978-3-030-71270-9.

[117] R. Shah, Introduction to k-Nearest Neighbors (kNN) Algorithm, Artifical Intelligence in Plain English. (2021). https://ai.plainenglish.io/introduction-to-k-nearest-neighbors-knn-algorithm-e8617a448fa8 (accessed May 18, 2022).

[118] P. Mather, B. Tso, Classification Methods for Remotely Sensed Data, 2nd ed., Taylor & Francis Group, 2009.

[119] Renesh Bedre, Support Vector Machine (SVM) basics and implementation in Python, Reneshbedre.Com. (2021). https://www.reneshbedre.com/blog/support-vector-machine.html (accessed May 18, 2022).

[120] R. Silipo, K. Melcher, From a Single Decision Tree to a Random Forest, Towards Data Science. (2019). https://towardsdatascience.com/from-a-single-decision-tree-to-a-random-forest-b9523be65147 (accessed May 18, 2022).

[121] A. Subasi, Machine learning techniques, in: Practical Machine Learning for Data Analysis Using Python, Elsevier, 2020: pp. 91–202. https://doi.org/10.1016/b978-0-12-821379-7.00003-5.

[122] H. He, Y. Ma, IMBALANCED LEARNING, IEEE Press, Piscataway, NJ USA, 2013.

[123] A. Subasi, Introduction, in: Practical Machine Learning for Data Analysis Using Python, Elsevier, 2020: pp. 1–26. https://doi.org/10.1016/b978-0-12-821379-7.00001-1.

[124] C.A. Schneider, W.S. Rasband, K.W. Eliceiri, NIH Image to ImageJ: 25 years of image analysis, Nat Methods. 9 (2012) 671–675.

[125] B. Avitzur, Metal forming : processes and analysis, McGraw-Hill, New York ; London, 1968.

[126] R. Ebrahimi, A. Najafizadeh, A new method for evaluation of friction in bulk metal forming, J Mater Process Technol. 152 (2004) 136–143. https://doi.org/10.1016/j.jmatprotec.2004.03.029.

[127] J.O. Hallquist, LS-DYNA ® THEORY MANUAL, 2006. www.lstc.com.

[128] D.P. Flanagan~, T. Belytschko~, A UNIFORM STRAIN HEXAHEDRON AND QUADRILATERAL WITH ORTHOGONAL HOURGLASS CONTROL, 1981.

[129] N. Bay, FRICTION STRESS AND NORMAL STRESS IN BULK METAL-FORMING PROCESSES, Journal of Mechanical Working Technology. 14 (1987) 203–223.

[130] E. Kozeschnik, W. Rindler, B. Buchmayr, Scheil-Gulliver simulation with partial redistribution of fast diffusers and simultaneous solid-solid phase transformations, International Journal of Materials Research. 98 (2013) 826–831.

[131] MatCalc Engineering, T11: Simulation of solidification of 0.7C 3Mn steel, (n.d.). https://www.matcalc.at/wiki/doku.php?id=tutorials:t11 (accessed May 26, 2022).

[132] E.I. Poliak, J.J. Jonas, Critical strain for dynamic recrystallization in variable strain rate hot deformation, ISIJ International. 43 (2003) 692–700. https://doi.org/10.2355/isijinternational.43.692.

[133] G. Shen, S.L. Semiatin, R. Shivpuri, Modeling Microstructural Waspaloy Development during the Forging of Waspaloy, Metallurgical and Materials Transactions A. 26 (1995) 1795–1803.

[134] E. Eriksson, M. Hörnqvist Colliander, Dynamic and post-dynamic recrystallization of haynes 282 below the secondary carbide solvus, Metals (Basel). 11 (2021) 1–24. https://doi.org/10.3390/met11010122.

[135] C. Rehrl, S. Kleber, O. Renk, R. Pippan, Effect of forming conditions on the softening behavior in coarse grained structures, Materials Science and Engineering A. 528 (2011) 6163–6172. https://doi.org/10.1016/j.msea.2011.04.043.

[136] A. Sarkar, M.J.N.V. Prasad, S.V.S.N. Murty, Effect of initial grain size on hot deformation behaviour of Cu-Cr-Zr-Ti alloy, Mater Charact. 160 (2020) 110112. https://doi.org/10.1016/j.matchar.2019.110112.

[137] X.M. Chen, Y.C. Lin, D.X. Wen, J.L. Zhang, M. He, Dynamic recrystallization behavior of a typical nickel-based superalloy during hot deformation, Mater Des. 57 (2014) 568–577. https://doi.org/10.1016/j.matdes.2013.12.072.

[138] A. Nicolaÿ, G. Fiorucci, J.M. Franchet, J. Cormier, N. Bozzolo, Influence of strain rate on subsolvus dynamic and post-dynamic recrystallization kinetics of Inconel 718, (2019). https://doi.org/10.1016/j.actamat.2019.05.061.

[139] H. Jiang, J. Dong, M. Zhang, Z. Yao, A Study on the Effect of Strain Rate on the Dynamic Recrystallization Mechanism of Alloy 617B, Metall Mater Trans A Phys Metall Mater Sci. 47 (2016) 5071–5087. https://doi.org/10.1007/s11661-016-3664-7.

[140] W. Roberts, B. Ahlblom, A NUCLEATION CRITERION FOR DYNAMIC RECRYSTALLIZATION DURING HOT WORKING, Acta Metallurgica. 26 (1978) 801–813.

[141] G. Gottstein, L.S. Shvindlerman, Grain Boundary Migration in Metals, Taylor & Francis Group, 2010.

[142] S.L. Semiatin, G.D. Lahoti, The Occurrence of Shear Bands in Isothermal, Hot Forging, Metallurgical Transaction A. 13 (1982) 275–287.

[143] B. Tang, L. Xiang, L. Cheng, D. Liu, H. Kou, J. Li, The formation and evolution of shear bands in plane strain compressed nickel-base superalloy, Metals (Basel). 8 (2018). https://doi.org/10.3390/met8020141.

[144] C. Zener, J.H. Hollomon, Effect of strain rate upon plastic flow of steel, J Appl Phys. 15 (1944) 22–32. https://doi.org/10.1063/1.1707363.

[145] L. Qiang, M.N. Bassim, Effects of strain and strain-rate on the formation of the shear band in metals, Journal de Physique IV . 110 (2003) 87–91.

[146] R.W. Armstrongt, C.S. Coffey, W.L. Elban, ADIABATIC HEATING AT A DISLOCATION PILE-UP AVALANCHE, 1982.

[147] K. Huang, K. Marthinsen, Q. Zhao, R.E. Logé, The double-edge effect of second-phase particles on the recrystallization behaviour and associated mechanical properties of metallic materials, Prog Mater Sci. 92 (2018) 284–359. https://doi.org/10.1016/j.pmatsci.2017.10.004.

[148] L. Wang, G. Xie, J. Zhang, L.H. Lou, On the role of carbides during the recrystallization of a directionally solidified nickel-base superalloy, Scr Mater. 55 (2006) 457–460. https://doi.org/10.1016/j.scriptamat.2006.05.013.

[149] Y. Li, J. Cheng, X. Ma, Y. He, J. Guo, Effect of Boron and Carbon on the Hot Deformation Behavior of a Novel Third Generation Nickel-Based Powder Metallurgy Superalloy WZ-A3, J Mater Eng Perform. (2022). https://doi.org/10.1007/s11665-022-06674-y.

[150] H. Qin, Q. Tian, W. Zhang, Q. Du, H. Li, X. Liu, P. Yu, Microstructure Evolution of GH4742 Ni-Based Superalloy during Hot Forming, J Mater Eng Perform. (2022). https://doi.org/10.1007/s11665-022-06636-4.

[151] K. Sahithya, I. Balasundar, P. Pant, T. Raghu, Comparative study on the high temperature deformation behaviour of an as-cast Ni base superalloy subjected to different cooling rates after homogenization, J Alloys Compd. 849 (2020). https://doi.org/10.1016/j.jallcom.2020.156626.

[152] S.L. Semiatin, D.S. Weaver, P.N. Fagin, M.G. Glavicic, R.L. Goetz, N.D. Frey, R.C. Kramb, M.M. Antony, Deformation and Recrystallization Behavior during Hot Working of a Coarse-Grain, Nickel-Base Superalloy Ingot Material, Metallurgical and Materials Transactions A. 35 (2004) 679–693.

[153] E. Eriksson, J. Andersson, M. Hörnqvist Colliander, The Effect of Grain Boundary Carbides on Dynamic Recrystallization During Hot Compression of Ni-Based Superalloy Haynes 282 TM, Metall Mater Trans A Phys Metall Mater Sci. 53 (2022) 29–38. https://doi.org/10.1007/s11661-021-06524-x.

[154] R. Buerstmayr, F. Theska, R. Webster, M. Lison-Pick, S. Primig, Correlative analysis of grain boundary precipitates in Ni-based superalloy René 41, Mater Charact. 178 (2021). https://doi.org/10.1016/j.matchar.2021.111250.

[155] H.-J. Jou, G. Olson, T. Gabb, A. Garg, D. Miller, Characterization and Computational Modeling of Minor Precipitate Phases in Alloy LSHR, in: Superalloys 2012: 12th International Symposium on SUperalloys, The Minerals, Metals & Materials Society, 2012.

[156] X. Qin, X. Yan, D. Huang, X. Zhang, M. Qi, S. Yue, Evolution Behavior of M23C6 Carbides Under Different Hot Deformation Conditions in Alloy 602 CA, Metals and Materials International. 25 (2019) 1616–1625. https://doi.org/10.1007/s12540-019-00312-4.

[157] H. Wang, D. Liu, J. Wang, Y. Shi, Y. Zheng, Y. Hu, Investigation on the Thermal Deformation Behavior of the Nickel-Based Superalloy Strengthened by γ Phase, (2019). https://doi.org/10.3390/cryst9030125.

[158] Z. Shi, X. Yan, C. Duan, C. Tang, E. Pu, Characterization of the Hot Deformation Behavior of a Newly Developed Nickel-Based Superalloy, J Mater Eng Perform. 27 (2018) 1763–1776. https://doi.org/10.1007/s11665-018-3270-5.

[159] J. Wang, J. Dong, M. Zhang, X. Xie, Hot working characteristics of nickel-base superalloy 740H during compression, Material Science & Engineering A. 566 (2013) 61–70. https://doi.org/10.1016/j.msea.2012.12.077.

[160] M. Azarbarmas, M. Aghaie-Khafri, J.M. Cabrera, J. Calvo, Microstructural evolution and constitutive equations of Inconel 718 alloy under quasi-static and quasi-dynamic conditions, (2015). https://doi.org/10.1016/j.matdes.2015.12.157.

[161] A. Oradei-Basile, J.F. Radavich, A Current T-T-T Diagram for Wrought Alloy 718, in: Superalloys 718, 625 and Various Derivatives, The Minerals, Metals & Materials Society, 1991: pp. 325–335.

[162] H. Monajati, M. Jahazi, S. Yue, A.K. Taheri, Deformation Characteristics of Isothermally Forged UDIMET 720 Nickel-Base Superalloy, Metallurgical and Materials Transactions A. 36 (2005) 895–905.

[163] S.W. Hwang, K.-T. Park, J.S. Kim, Y. Kim, ANALYSIS OF HOT DEFORMATION OF NIMONIC 80A USING PROCESSING MAP, in: 23rd International Conference on Metallurgy and Metals, 2014.

[164] A.A. Guimaraes, J.J. Jonas, Recrystallization and Aging Effects Associated with the High Temperature Deformation of Waspaloy and Inconel 718, Metallurgical Transactions A. 12A (1981) 1655–1666.

[165] A.H. Cottrell, B.A. Bilby, Dislocation Theory of Yielding and Strain Ageing of Iron, in: Proceedings of the Physical Society. Section A, 1949: pp. 49–62.

[166] I. Philippart, H.J. Rack, High temperature dynamic yielding in metastable Ti-6.8Mo-4.5F-1.5Al, 1998.

[167] W.G. Johnston, J.J. Gilman, Dislocation velocities, dislocation densities, and plastic flow in lithium fluoride crystals, J Appl Phys. 30 (1959) 129–144. https://doi.org/10.1063/1.1735121.

[168] Z.L. Zhao, Y.Q. Ning, H.Z. Guo, Z.K. Yao, M.W. Fu, Discontinuous yielding in Ni-base superalloys during high-speed deformation, Materials Science and Engineering A. 620 (2015) 383–389. https://doi.org/10.1016/j.msea.2014.10.041.

[169] F. Margetan, E. Nieters, P. Haldipur, L. Brasche, T. Chiou, M. Keller, A. Degtyar, J. Umbach, W. Hassan, T. Patton, K. Smith, Fundamental Studies of Nickel Billet Materials-Engine Titanium Consortium Phase II, Ames, IA 50011, 2005.

[170] J. Cui, B. Li, Z. Liu, F. Qi, B. Zhang, J. Zhang, Numerical investigation of segregation evolution during the vacuum arc remelting process of ni-based superalloy ingots, Metals (Basel). 11 (2021). https://doi.org/10.3390/met11122046.

[171] Y. Tang, Y.Q. Zhang, N. v. Chawla, SVMs modeling for highly imbalanced classification, IEEE Transactions on Systems, Man, and Cybernetics, Part B: Cybernetics. 39 (2009) 281–288. https://doi.org/10.1109/TSMCB.2008.2002909.

[172] J.M. Jerez, I. Molina, P.J. García-Laencina, E. Alba, N. Ribelles, M. Martín, L. Franco, Missing data imputation using statistical and machine learning methods in a real breast cancer problem, Artif Intell Med. 50 (2010) 105–115. https://doi.org/10.1016/j.artmed.2010.05.002.

[173] B. Tang, L. Xiang, L. Cheng, D. Liu, H. Kou, J. Li, The formation and evolution of shear bands in plane strain compressed nickel-base superalloy, Metals (Basel). 8 (2018). https://doi.org/10.3390/met8020141.